\documentclass[useAMS,usenatbib,usegraphicx]{mn2e}
\usepackage[xcolor]{xcolor}

\def\deg{{$^{\circ}$}}

\def\kmsec{\mbox{km~s$^{\rm -1}$}}
\def\logg{\mbox{log~{\it g}}}
\def\teff{\mbox{$T_{\rm eff}$}}
\def\vt{\mbox{$v_{\rm t}$}}
\def\rpro{\mbox{$r$-process}}
\def\spro{\mbox{$s$-process}}
\def\ncap{\mbox{$n$-capture}}
\def\msun{$M_{\odot}$}
\def\ngc{\mbox{NGC~4833}}
\def\loggf{$\log gf$}

\def\apj{ApJ}
\def\aap{A\&A}
\def\aj{AJ}
\def\mnras{MNRAS}
\def\araa{ARA\&A}
\def\apjs{ApJS}
\def\apjl{ApJL}
\def\aaps{A\&AS}
\def\procspie{Proc.\ SPIE}
\def\jqsrt{J.\ Quant.\ Spectrosc.\ Rad.\ Trans.}
\def\pasp{PASP}
\def\pasj{PASJ}
\def\physscr{Phys.\ Scr.}

\title[Abundances in Globular Cluster NGC 4833]
{Detailed Abundances of 15~Stars in the Metal-Poor 
Globular Cluster NGC 4833\thanks{
This paper includes data gathered with the 6.5~meter 
Magellan Telescopes located at Las Campanas Observatory, Chile.}
}

\author[Ian U.\ Roederer and Ian B.\ Thompson]{%
Ian U.\ Roederer$^{1}$\thanks{Email:\ iur@umich.edu}
and
Ian B.\ Thompson$^{2}$\thanks{Email:\ ian@obs.carnegiescience.edu}\\
$^{1}$Department of Astronomy, University of Michigan,
1085 South University Avenue, Ann Arbor, MI 48109, USA\\
$^{2}$Carnegie Observatories, 
813 Santa Barbara Street, Pasadena, CA 91101, USA
}

\begin{document}

\pagerange{\pageref{firstpage}--\pageref{lastpage}} 
\pubyear{2014}
\maketitle
\label{firstpage}

\begin{abstract}

We have observed
15~red giant stars in the relatively massive,
metal-poor globular cluster NGC~4833
using the MIKE spectrograph at Magellan.
We calculate stellar parameters for each star and
perform a standard abundance analysis to derive abundances
of 43~species of 39~elements, including
20~elements heavier than the iron group.
We derive $\langle$[Fe/H]$\rangle = -$2.25~$\pm$~0.02 
from Fe~\textsc{i} lines and 
$\langle$[Fe/H]$\rangle = -$2.19~$\pm$~0.013 
from Fe~\textsc{ii} lines.
We confirm earlier results that found
no internal metallicity spread in NGC~4833,
and there are no significant star-to-star abundance dispersions
among any elements in the iron group
(19~$\leq Z \leq$~30).
We recover the usual abundance variations among the light elements
C, N, O, Na, Mg, Al, and possibly Si.
The heavy-element distribution reflects
enrichment by $r$-process nucleosynthesis
([Eu/Fe]~$= +$0.36~$\pm$~0.03),
as found in many other metal-poor globular clusters.
We investigate small star-to-star variations
found among the neutron-capture elements,
and we conclude that these are 
probably not real variations.
Upper limits on the Th abundance,
$\log \epsilon$~(Th/Eu)~$< -$0.47~$\pm$~0.09,
indicate that NGC~4833,
like other globular clusters where Th has been studied,
did not experience a so-called ``actinide boost.'' 

\end{abstract}

\begin{keywords}
globular clusters: individual (\mbox{NGC~4833}) ---
nuclear reactions, nucleosynthesis, abundances ---
stars: abundances ---
stars: Population II
\end{keywords}

\section{Introduction}
\label{intro}

Globular clusters are the remnants of
some of the most violent epochs of star formation
in the history of the Universe.
The compositions of the present-day stars 
provide a detailed chemical inventory 
to probe those ancient nucleosynthesis and
enrichment events.
Nevertheless, detailed chemical inventories for some
Galactic globular clusters are still relatively unknown.

Only recently were the chemical compositions 
of a large sample of stars examined in one 
such cluster, \ngc\
\citep{carretta14}.
The properties and orbital parameters of \ngc\ are listed 
in Table~\ref{basicdata}.
\ngc\ has one of the more extended, richly-populated
blue horizontal branches among 
Galactic globular clusters 
(e.g., \citealt{menzies72,samus95,melbourne00,piotto02}).
This metal-poor cluster
is a member of the ``old'' globular cluster population
(e.g., \citeauthor{melbourne00}; \citealt{marinfranch09}),
and it ranks in the top quartile of most luminous clusters
around the Milky Way \citep{harris96}.
The proper motion measurements and orbit calculations
of \citet{casettidinescu07} suggest that
\ngc\ and \mbox{NGC~5986}, another old and metal-poor cluster, 
share similar orbital characteristics.
\citeauthor{casettidinescu07}\ suggested 
that these two clusters
are dynamically associated and could be
an accreted pair with a common 
origin in a now-disrupted satellite galaxy.

\begin{table*}
\begin{minipage}{4.5in} 
\caption{Basic properties and orbital parameters of \ngc
\label{basicdata}}
\begin{tabular}{lccc}
\hline
Quantity & 
Symbol &
Value &
Reference \\
\hline
Right ascension               & $\alpha$ (J2000)   & 12:59:34            & 1 \\
Declination                   & $\delta$ (J2000)   & $-$70:52:35         & 1 \\
Galactic longitude            & $\ell$             & 303.6\deg           & 1 \\
Galactic latitude             & \textit{b}         & $-$8.0\deg          & 1 \\
Luminosity                    & $M_{V}$            & $-$8.17             & 1 \\
Mass-to-light ratio           & $M/L$      & 0.84~$\pm$~0.45     & 2 \\
Mass                          & $M_{*}$       & 1.2$\times 10^{5}$~\msun & 1 \\
Central concentration         & $c$                & 1.25                & 1 \\
Distance modulus              & $m-M$              & 15.05~$\pm$~0.06    & 3 \\
Distance from Sun             & R$_{\odot}$        & 6.5~kpc             & 4 \\
Distance from Galactic Center & R$_{\rm G.C.}$     & 7.0~kpc             & 4 \\
Perigalactic distance         & R$_{\rm peri}$     & 0.7~$\pm$~0.2~kpc   & 4 \\
Apogalactic distance          & R$_{\rm apo}$      & 7.7~$\pm$~0.7~kpc   & 4 \\
Maximum distance above Galactic plane & Z$_{\rm max}$ & 1.8~$\pm$~0.4~kpc& 4 \\
Orbital eccentricity          & $e$                & 0.84~$\pm$~0.03     & 4 \\
Orbital period                & $T_{\rm orbit}$    & 91~$\pm$~9~Myr      & 4 \\
\hline
\end{tabular}
\\
References:\
(1)~\citealt{harris96};
(2)~\citealt{carretta14};
(3)~\citealt{melbourne00};
(4)~\citealt{casettidinescu07}
\end{minipage}
\end{table*}

Until last year, the only high-resolution spectroscopic
observations of stars in \ngc\ had been conducted by
\citet{pilachowski83}, who observed two stars,
\citet{gratton89}, who observed two stars, and
\citet{minniti93,minniti96}, who observed one.
These studies derived abundances of $\approx$~15--20
species in individual stars in \ngc,
including O and Na abundances.
They recognized that [O/Fe] in some stars 
was depleted relative
to the maximum [O/Fe] ratios 
found in other clusters and 
field stars of similar metallicity.
The small sample sizes limited these authors'
abilities to characterize the
global chemical properties of \ngc.
\citet{carretta14} observed 78 stars in \ngc\ with the 
UVES and GIRAFFE spectrometers in the FLAMES instrument on the VLT.
This dataset enabled \citeauthor{carretta14}\ 
to characterize the pattern of light-element
(O, Na, Mg, Al, and Si) variations in \ngc.
That study also characterized the abundances of 18~species of
15~heavier elements.

\citet{carretta14} presented several forms of evidence that
suggest that \ngc\ may have been tidally stripped
more than the average cluster.
First, \ngc\ has a wide span of [Mg/Fe] ratios.
This property has only been identified in a few massive
or metal-poor clusters \citep{yong05,carretta09b,cohen12},
yet \ngc\ is only the 36$^{\rm th}$ 
most luminous Galactic globular cluster today.
Second, \ngc\ has an 
eccentric orbit, small perigalactic distance, and modest
central concentration.
These values predict a high destruction rate 
relative to other clusters \citep{casettidinescu07,allen08}.
Third, the inner-quartile range of the [O/Na] ratios
(IQR[O/Na]) in \ngc\
lies on the upper envelope of clusters in the 
IQR[O/Na] versus luminosity correlation.
Evidence discussed by \citet{carretta14} indicates
that less-concentrated clusters of a given luminosity 
may have larger values of IQR[O/Na] and have lost more mass
than more-concentrated ones.
Finally, the low mass-to-light ratio of \ngc\
hints that significant mass loss
may have occurred as low-mass stars have been
preferentially lost from the cluster.
\citet{carretta14} note that 
no attempt has been made to detect 
these escaped stars as tidal tails,
and potential investigators 
may have been intimidated by the stellar crowding and
high reddening at the low Galactic latitude of \ngc.
The preponderance of evidence suggests that \ngc\
has lost a larger fraction of its initial 
stellar mass than the average Galactic globular cluster.

Our sample includes only 15~stars in \ngc,
but our data cover most of the optical spectral range.
We detect 43~species of 39~elements
heavier than He,
and this complements the more limited 
chemical inventory derived from a larger sample of stars
studied by \citet{carretta14}.
These data enable us to focus new attention on the abundance pattern
exhibited by the heaviest elements in \ngc,
those produced by neutron ($n$) capture reactions.

Throughout this work we
adopt the standard definitions of elemental abundances and ratios.
For element X, the logarithmic abundance is defined
as the number of atoms of X per 10$^{12}$ hydrogen atoms,
$\log\epsilon$(X)~$\equiv \log_{10}(N_{\rm X}/N_{\rm H}) +$~12.0.
For elements X and Y, [X/Y] is 
the logarithmic abundance ratio relative to the solar ratio,
defined as $\log_{10} (N_{\rm X}/N_{\rm Y}) -
\log_{10} (N_{\rm X}/N_{\rm Y})_{\odot}$, using
like ionization states;
i.e., neutrals with neutrals and ions with ions.
We adopt the solar abundances listed in \citet{asplund09}.
Abundances or ratios denoted with the ionization state
indicate the total elemental abundance as derived from transitions of
that particular state.

\section{Observations}
\label{observations}

\begin{figure}
\centering
\includegraphics[angle=0,width=3.3in]{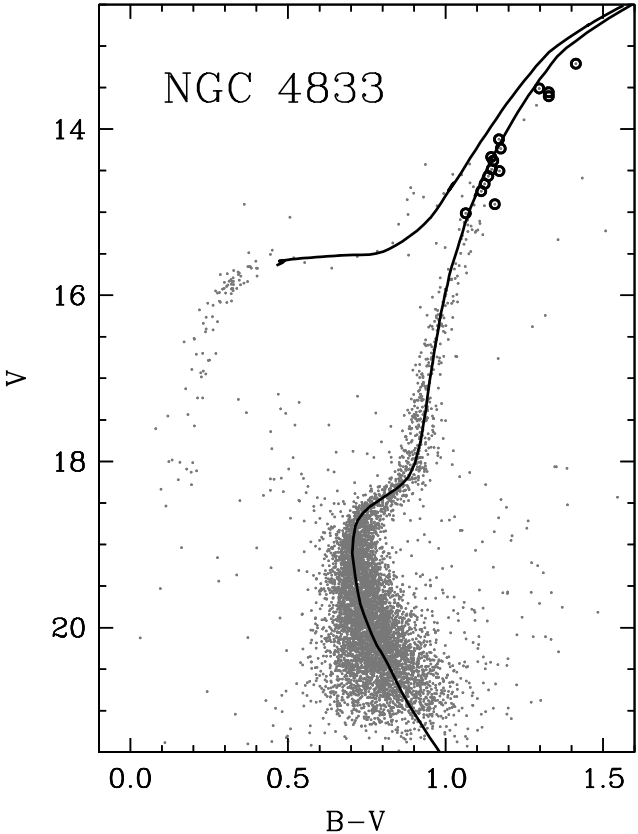}
\caption{
\label{cmdplot}
The $B-V$ versus $V$ colour-magnitude diagram
using photometry from \citet{piotto02}.
Stars observed with MIKE are marked with circles.
The line is an isochrone from the PARSEC database
(v.\ 1.2S; \citealt{bressan12,chen14})
computed for an age of 13~Gyr and
metallicity $Z =$~0.00025.
The isochrone is
shifted to a distance modulus of 15.05
and a reddening of 
$E(B-V) =$~0.32.
}
\end{figure}

We have observed 15 probable members of \ngc\ with the
Magellan Inamori Kyocera Echelle (MIKE)
spectrograph \citep{bernstein03} on the
6.5~m Landon Clay (Magellan~II) Telescope at Las Campanas Observatory.
Table~\ref{obstab} presents a log of these observations,
and Figure~\ref{cmdplot} highlights these 15~stars on a 
color-magnitude diagram of \ngc.
ThAr comparison lamp spectra have been taken immediately preceding or
following each observation.
The red and blue arms of MIKE are split 
by a dichroic at $\approx$~4950\AA.
This setup provides complete wavelength coverage from 
3350--9150\AA.
Data reduction, extraction, and wavelength calibration have been 
performed using 
the \textsc{CarPy} 
MIKE data reduction pipeline
written by D.\ Kelson (see also \citealt{kelson03}).
Continuum normalization and order stitching
have been performed within the 
\textsc{iraf} environment.

\begin{table}
\begin{minipage}{3.2in} 
\caption{Observing log
\label{obstab}}
\begin{tabular}{cccccc}
\hline
Star & 
Slit &
Date &
UT$_{\rm start}$ &
$t_{\rm exp}$ &
RV \\
 &
 &
 &
 &
(s) &
(\kmsec) \\
\hline
2-185   & 1\farcs0 & 2011 03 18 & 04:41 & 2600 & 211.0 \\
2-277   & 1\farcs0 & 2011 03 18 & 03:38 & 3600 & 201.7 \\
2-882   & 1\farcs0 & 2011 03 18 & 02:42 & 3200 & 209.8 \\
2-918   & 1\farcs0 & 2011 03 18 & 05:27 & 3000 & 209.8 \\
2-1578  & 0\farcs7 & 2011 03 17 & 02:52 & 5400 & 201.3 \\
2-1578  & 0\farcs7 & 2011 03 17 & 06:33 & 1800 & 201.2 \\
2-1664  & 0\farcs7 & 2011 03 17 & 04:30 & 7200 & 199.5 \\
3-742   & 0\farcs7 & 2011 01 31 & 07:43 & 2400 & 206.4 \\
3-772   & 1\farcs0 & 2011 03 17 & 23:40 & 2800 & 203.0 \\
3-1509  & 1\farcs0 & 2011 03 18 & 01:56 & 2400 & 207.4 \\
4-224   & 0\farcs7 & 2011 01 28 & 08:22 & 1200 & 201.5 \\
4-341   & 0\farcs7 & 2011 01 28 & 08:47 & 2400 & 202.2 \\
4-341   & 0\farcs7 & 2011 01 29 & 08:22 & 2400 & 202.5 \\
4-464   & 1\farcs0 & 2011 03 18 & 00:29 & 2600 & 297.6 \\
4-1255  & 0\farcs7 & 2011 01 31 & 08:29 & 2400 & 205.9 \\
4-1398  & 1\farcs0 & 2011 03 18 & 01:16 & 2200 & 200.1 \\
4-1706  & 1\farcs0 & 2011 03 18 & 06:20 & 3400 & 187.9 \\
\hline
\end{tabular}
\end{minipage}
\end{table}

\begin{table*}
\begin{minipage}{7.5in} 
\caption{Photometry, model atmosphere parameters, and S/N ratios
\label{startab}}
\begin{tabular}{ccccccccccccccc}
\hline
Star\footnote{This study, based on \citet{piotto02}} &
Star\footnote{\citet{menzies72}} &
Star\footnote{\citet{carretta14}} &
$V$ &
$(B-V)_{0}$ &
$(V-K)_{0}$ &
\teff\ &
\logg\ &
\vt\ &
{[M/H]} &
S/N &
S/N &
S/N &
S/N \\
 &
 &
 &
 &
 &
 &
(K) &
{[cgs]} &
(\kmsec) &
 &
(3950~\AA) &
(4550~\AA) &
(5200~\AA) &
(6750~\AA) \\
\hline
2-185   & D102 & \ldots& 14.57 & 0.85 & 2.35 & 4670 & 1.55 & 1.75 & $-$2.2 & 25 & 50 &  50 & 110 \\
2-277   & D109 & \ldots& 14.90 & 0.87 & 2.26 & 4752 & 1.72 & 1.80 & $-$2.2 & 25 & 50 &  50 & 110 \\
2-882   & D193 & \ldots& 14.66 & 0.84 & 2.45 & 4665 & 1.57 & 1.70 & $-$2.2 & 20 & 40 &  50 & 110 \\
2-918   &  C49 & 33027 & 14.75 & 0.83 & 2.40 & 4696 & 1.63 & 1.85 & $-$2.2 & 20 & 45 &  40 & 110 \\
2-1578  &  C91 & 31836 & 14.12 & 0.89 & 2.26 & 4595 & 1.35 & 1.75 & $-$2.2 & 30 & 50 &  60 & 140 \\
2-1664  &  C92 & \ldots& 14.34 & 0.86 & 2.82 & 4527 & 1.32 & 1.80 & $-$2.2 & 35 & 75 &  75 & 180 \\
3-742   & C129 & 31370 & 13.60 & 1.05 & 2.91 & 4357 & 0.93 & 2.00 & $-$2.2 & 25 & 65 &  75 & 200 \\
3-772   & D167 & \ldots& 14.50 & 0.89 & 2.37 & 4647 & 1.50 & 1.65 & $-$2.2 & 10 & 30 &  40 &  95 \\
3-1509  & C145 & 31929 & 14.38 & 0.87 & 2.52 & 4591 & 1.41 & 1.85 & $-$2.2 & 20 & 45 &  50 & 120 \\
4-224   & D136 & \ldots& 13.51 & 1.02 & 2.70 & 4385 & 0.96 & 2.05 & $-$2.2 & 20 & 45 &  55 & 120 \\
4-341   & D135 & \ldots& 13.56 & 1.05 & 2.69 & 4397 & 0.98 & 1.95 & $-$2.2 & 40 & 95 & 110 & 250 \\
4-464   & D153 & \ldots& 14.48 & 0.86 & 2.45 & 4628 & 1.48 & 1.90 & $-$2.2 & 10 & 40 &  45 & 115 \\
4-1255  &  D49 & \ldots& 13.21 & 1.14 & 2.60 & 4344 & 0.84 & 2.15 & $-$2.2 & 35 & 80 &  90 & 215 \\
4-1398  & C191 & \ldots& 14.23 & 0.89 & 2.56 & 4556 & 1.33 & 1.85 & $-$2.2 & 15 & 45 &  50 & 115 \\
4-1706  &  D41 & \ldots& 15.01 & 0.78 & 2.32 & 4764 & 1.76 & 1.85 & $-$2.2 & 25 & 45 &  50 & 105 \\
\hline
\end{tabular}
\end{minipage}
\end{table*}

Six stars were observed with the 0\farcs7$\times$5\farcs0 slit.
This setup yields a resolving power of $R \sim$~41,400 in the blue 
and $R \sim$~36,300 in the red, 
as measured from isolated ThAr lines in the 
extracted comparison lamp spectra.
Nine stars were observed during poorer seeing conditions 
with the 1\farcs0$\times$5\farcs0 slit,
which yields a resolving power of $R \sim$~30,500 in the blue and
$R \sim$~25,900 in the red.
Table~\ref{startab} presents 
signal-to-noise (S/N) estimates based on 
Poisson statistics for the number of photons collected 
per pixel in the continuum.
These values are measured 
from the co-added spectra.

We adopt a star naming convention based on the
\citet{piotto02} photometry files.
This convention has the form ``chip-ID.''
The first value (``chip'') indicates the 
WFPC2 chip number, and the second value (``ID'')
indicates the star identification number.

\begin{figure*}
\centering
\includegraphics[angle=270,width=6.0in]{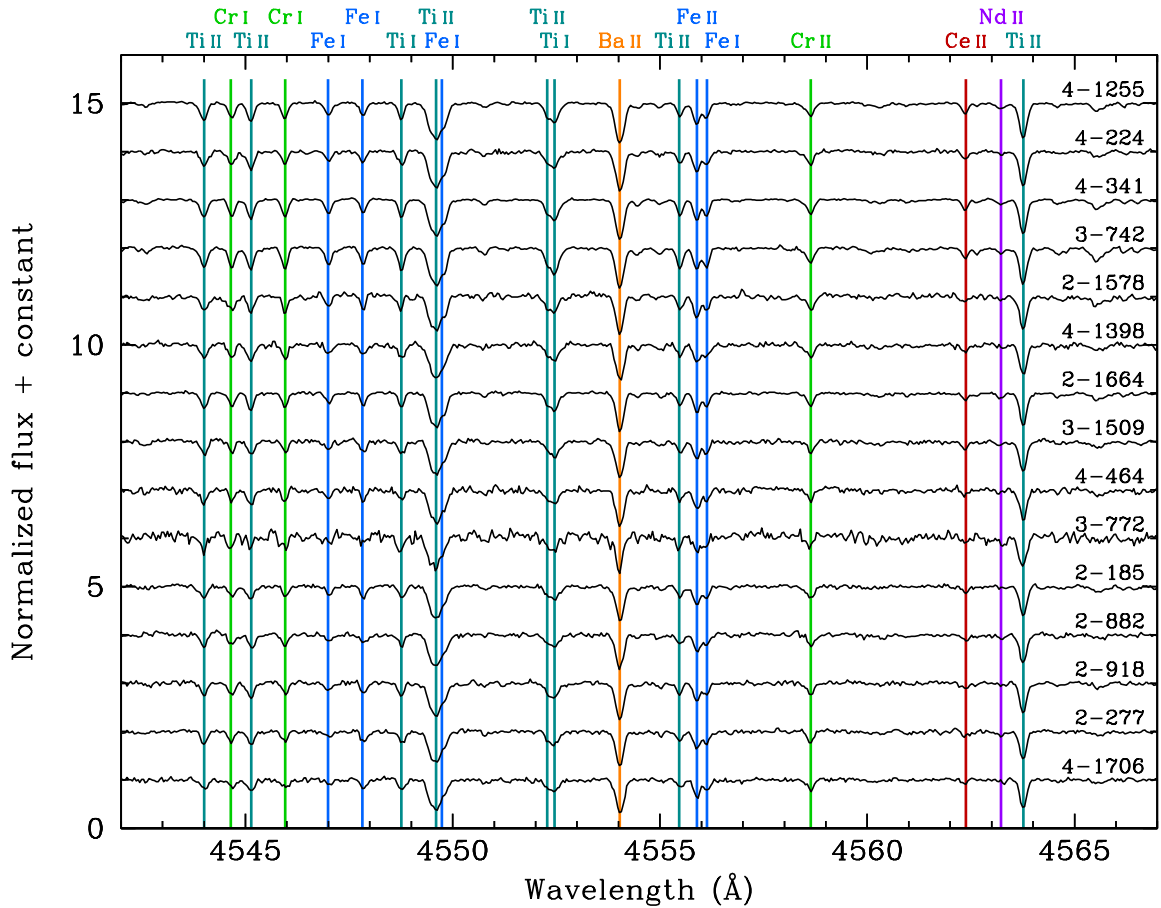}
\caption{
\label{specplot}
Small portions of the blue spectra of all 15~stars in our study.
Identifications are given for 19~absorption lines.
Velocity shifts have been removed, and 
the spectra have been shifted vertically 
for the sake of illustration.
The stars are sorted by increasing $V$-magnitude from top to bottom.
}
\end{figure*}

Figure~\ref{specplot} illustrates a 25~\AA\ portion of
the MIKE spectra of all 15~stars observed in our study.
The spectra are sorted by increasing $V$-magnitude
from top to bottom.
Lines of nine species of six elements are marked.
This illustrates the richness of blue spectra
of red giant stars,
even when the overall metallicity
is less than 100~times the solar metallicity.

\section{Radial velocities}
\label{rv}

We measure the stellar radial velocity (RV)
of each observation with respect to the ThAr lamp 
by cross-correlating the echelle order 
containing the Mg~\textsc{i} \textit{b} lines in each spectrum
against a template
using the \textit{fxcor} task in \textsc{iraf}.
We create the template by measuring the wavelengths of
unblended Fe~\textsc{i} lines in this order in star
\mbox{4-1255}, which has the highest S/N in a single exposure.
We compute velocity corrections to the Heliocentric rest frame
using the \textsc{iraf} \textit{rvcorrect} task.
This method yields a total uncertainty of $\approx$~0.6~\kmsec\ 
per observation (see \citealt{roederer14}).
Table~\ref{obstab} presents the Heliocentric RV
measured for each observation.

We derive a mean Heliocentric RV to \ngc\ of 203.0~$\pm$~1.5~\kmsec\
[s.e.m.]
($\sigma =$~5.9~\kmsec\ [s.d.]).
All stars are within 2' of the cluster center, and the 
RV of each star is within 2.5$\sigma$ of the systemic velocity.
There is a high probability that all 15~stars observed
are members of \ngc.

The mean heliocentric RV and dispersion of \ngc\
have been derived from high resolution GIRAFFE 
spectra by two previous studies, 
\citet{carretta14} and \citet{lardo15}.
\citeauthor{carretta14}\ reported a systemic velocity of
202.0~$\pm$~0.5~\kmsec\ and a dispersion of
4.1~$\pm$~0.3~\kmsec\ based on a sample of 78~stars.
\citeauthor{lardo15}\ reported a systemic velocity of
202.1~$\pm$~0.6~\kmsec\ and a dispersion of
3.9~\kmsec\ based on a sample of 82~stars.
Our values are in agreement.
These values are also in reasonable agreement
with previous measurements
derived from low-resolution spectra of smaller samples
of stars in \ngc\ by
\citet{geisler95} 
(201.0~$\pm$~1.3~\kmsec, $\sigma =$~4.6~\kmsec, 12~stars)
and \citet{rutledge97}
(194.1~$\pm$~7.5~\kmsec, $\sigma =$~3.3~\kmsec, 26~stars).
For completeness, we note that
no abundances have been presented
from the spectra presented by
\citeauthor{lardo15} 

\section{Equivalent Widths}
\label{ew}

We measure equivalent widths using a semi-automatic routine
that fits Voigt absorption line profiles to the
continuum-normalized spectra.
Following the same methods discussed in \citet{roederer14},
we inspect all equivalent width measurements by eye,
and we make manual adjustments to the fits when necessary.
Uncertainties in the equivalent widths are estimated
based on the S/N ratios that vary as a function of wavelength.
The complete list of equivalent widths is presented
in Table~\ref{ewtab},
which is
available in the Supplementary Information section
found in the online edition of the journal.
A short version is shown in the printed edition
to illustrate its form and content.

\begin{table}
\begin{minipage}{3.35in} 
\caption{Atomic data and equivalent widths
\label{ewtab}}
\begin{tabular}{ccccccc}
\hline
\hline
Wavelength &
Species\footnote{The number to the left of the decimal point indicates the atomic number, and the 
                     number to the right of the decimal point indicates the ionization state
                     (0~$=$~neutral, 1~$=$~first ion)} &
E.P.\ &
\loggf\ &
Ref.\ &
 2-185 &
\ldots \\
(\AA) &
 &
(eV) &
 &
 &
 & 
 \\
\hline
   6707.80 &  3.0 & 0.00 &  0.17 &  1 &  limit & \ldots \\
   6300.30 &  8.0 & 0.00 & -9.78 &  2 &  limit & \ldots \\
   7771.94 &  8.0 & 9.14 &  0.37 &  2 &  limit & \ldots \\
   7774.17 &  8.0 & 9.14 &  0.22 &  2 &  limit & \ldots \\
   7775.39 &  8.0 & 9.14 &  0.00 &  2 &  limit & \ldots \\
   5682.63 & 11.0 & 2.10 & -0.71 &  2 &   23.9 & \ldots \\
   5688.20 & 11.0 & 2.10 & -0.45 &  2 &   43.3 & \ldots \\
   6154.22 & 11.0 & 2.10 & -1.55 &  2 & \ldots & \ldots \\
   6160.75 & 11.0 & 2.10 & -1.25 &  2 &   13.1 & \ldots \\
   4702.99 & 12.0 & 4.33 & -0.38 &  2 &   88.9 & \ldots \\
   5528.40 & 12.0 & 4.34 & -0.50 &  2 &   95.7 & \ldots \\
   5711.09 & 12.0 & 4.34 & -1.72 &  2 &   20.9 & \ldots \\
 \vdots & \vdots & \vdots & \vdots & \vdots & \vdots & \\
\hline
\end{tabular}
\\
Notes.---
The full version of Table~\ref{ewtab} is available in the
supplementary material online. 
Here, ``synth'' denotes lines used to derive an abundance
via spectrum synthesis, and ``limit'' denotes lines
used to derive an upper limit. \\
References.---
 (1) \citealt{smith98}; 
 (2) \citealt{fuhr09};
 (3) \citealt{aldenius09};
 (4) \citealt{lawler89}, using HFS from \citealt{kurucz95};
 (5) \citealt{lawler13};
 (6) \citealt{wood13};
 (7) \citealt{doerr85a}, using HFS from \citealt{kurucz95};
 (8) \citealt{wood14b};
 (9) \citealt{sobeck07};
(10) \citealt{nilsson06};
(11) \citealt{denhartog11} for both \loggf\ value and HFS;
(12) \citealt{ruffoni14};
(13) \citealt{fuhr09}, using HFS from \citealt{kurucz95};
(14) \citealt{wood14};
(15) \citealt{roederer12a};
(16) \citealt{biemont11};
(17) \citealt{ljung06};
(18) \citealt{nilsson10};
(19) \citealt{palmeri05};
(20) \citealt{fuhr09}, using HFS/IS from \citealt{mcwilliam98} when available;
(21) \citealt{lawler01la}, using HFS from \citealt{ivans06};
(22) \citealt{roederer11c};  
(23) \citealt{lawler09}; 
(24) \citealt{li07}, using HFS from \citealt{sneden09};
(25) \citealt{ivarsson01}, using HFS from \citealt{sneden09};
(26) \citealt{denhartog03}, using HFS/IS from \citealt{roederer08} when available;
(27) \citealt{lawler06}, using HFS/IS from \citealt{roederer08} when available;
(28) \citealt{lawler01eu}, using HFS/IS from \citealt{ivans06};
(29) \citealt{denhartog06};
(30) \citealt{roederer12b};
(31) \citealt{lawler01tb}, using HFS from \citealt{lawler01tbhfs} when available;
(32) \citealt{wickliffe00}; 
(33) \citealt{lawler04} for both \loggf\ value and HFS;
(34) \citealt{lawler08};
(35) \citealt{wickliffe97};
(36) \citealt{sneden09} for both \loggf\ value and HFS/IS;
(37) \citealt{lawler07};
(38) \citealt{ivarsson03}, using HFS/IS from \citealt{cowan05}---see
       note on \loggf\ values there;
(39) \citealt{biemont00}, using HFS/IS from \citealt{roederer12b}; 
(40) \citealt{nilsson02}. 
\end{minipage}
\end{table}

\begin{figure}
\centering
\includegraphics[angle=0,width=3.3in]{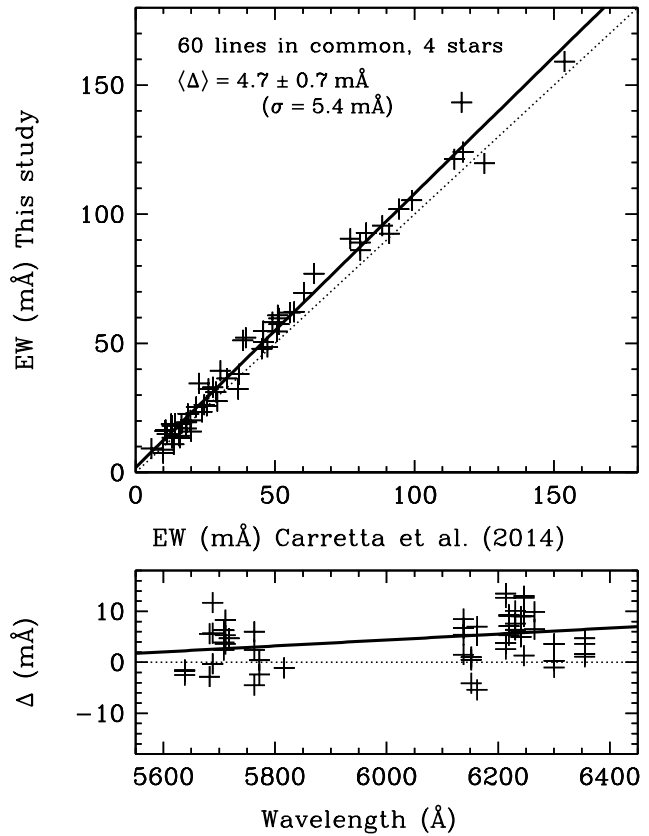}
\caption{
\label{ewplot}
Comparison of equivalent widths measured by \citet{carretta14}
and us for four stars in common.
The top panel illustrates the equivalent widths,
and the bottom panel illustrates the differences
as a function of wavelength.
The dotted lines represent a 1:1 correspondence, 
and the solid lines represent the linear fits.
}
\end{figure}

We observed four stars in common with the GIRAFFE sample
of \citet{carretta14}.
Those authors note that equivalent widths measured from their GIRAFFE
spectra are consistently larger than the equivalent widths
measured from their higher-resolution UVES spectra.
They adopt a linear transformation to convert their
GIRAFFE equivalent widths to the UVES scale.
We compare our measured equivalent widths 
with their corrected GIRAFFE equivalent widths.
As illustrated in Figure~\ref{ewplot},
we find a difference of
$+$4.7~$\pm$~0.7~m\AA\ ($\sigma =$~5.4~m\AA)
in the sense that our equivalent widths are larger.
There is a slight trend with wavelength,
but this is only significant at the 2$\sigma$ level.

\citet{carretta07} note that the transformation
from the GIRAFFE scale to the UVES scale is necessary
to ensure agreement with equivalent widths
measured independently by other investigators.
Equivalent widths measured from their UVES spectra of stars in
\mbox{NGC~6752} are in reasonable agreement with those
measured from $R \sim$~110,000 spectra obtained by \citet{yong03},
$\langle\Delta\rangle = +1.7 \pm 0.4$~m\AA\ ($\sigma =$~5.6~m\AA).
Our equivalent width measurements are made
with the same techniques and software 
used to examine hundreds of metal-poor stars observed
with MIKE \citep{roederer14}.
The \textsc{CarPy}
software package used to extract the MIKE observations
is also identical.
Our previous work found excellent agreement among the 
equivalent widths measured from MIKE spectra and those 
measured from echelle data collected at McDonald Observatory.
There was a difference of
$+$0.7~$\pm$~0.2~m\AA\ ($\sigma =$~4.0~m\AA)
for the HRS on the
Hobby-Eberly Telescope
and a difference of 
$+$1.0~$\pm$~0.1~m\AA\ ($\sigma =$~3.7~m\AA)
for the Tull Coud\'{e} Spectrograph on the Harlan J.\ Smith Telescope.
Furthermore, \citeauthor{roederer14}\ made extensive
comparisons between the equivalent widths for weak lines
in stars in common with \citet{carretta02}, \citet{johnson02},
\citet{ivans03}, \citet{cayrel04}, \citet{honda04}, and \citet{lai08}.
The offsets were all
$\leq$~1.3~m\AA.~
Thus we conclude that equivalent widths measured
from MIKE spectra taken with the 0\farcs7 slit
are not systematically in error.

We also investigate whether the wider MIKE slit used
for some of our observations may be the source of this difference.
Two of the four stars in \ngc\ in common with \citet{carretta14}
were observed using the 0\farcs7 slit, and two were observed
using the 1\farcs0 slit.
The mean difference
is $+4.5 \pm 1.1$~m\AA\ ($\sigma =$~6.2~m\AA)
for the two stars observed with the 0\farcs7 slit
and 
$+4.8 \pm 0.8$~m\AA\ ($\sigma =$~4.6~m\AA)
for the two stars observed with the 1\farcs0 slit.
The difference is not significant,
and we conclude that the different slits are
not the source of the discrepancy.

In principle, poor sky subtraction could result in a 
systematic bias in the equivalent widths measured from a single night.
The four stars in common with \citet{carretta14} were observed
on three different nights.
Star \mbox{3-742} was observed on 2011 01 31, when the moon
was only 6~per cent illuminated and 69\deg\ away from \ngc.
The other three stars were observed on 2011 03 17 and 2011 03 18,
when the moon was $>$90~per cent illuminated and $>$78\deg\
away from \ngc.
The mean difference is
$+5.6 \pm 1.6$~m\AA\ ($\sigma =$~6.7~m\AA)
for the observations near new moon and
$+4.2 \pm 0.7$~m\AA\ ($\sigma =$~4.6~m\AA)
for the observations near full moon.
The difference is not significant, 
and we conclude that poor sky subtraction is not the
source of the discrepancy.

Finally, we note that Roederer, Marino, \& Sneden (\citeyear{roederer11c})
examined spectra of several stars in globular cluster \mbox{M22}
that were taken on the same two nights in March, 2011.
That study compared the equivalent widths measured in those 
stars with those measured by \citet{marino11}.
Again, the offset is small, 
$\langle\Delta\rangle = +1.8 \pm 0.3$~m\AA\ ($\sigma =$~3.7~m\AA).
There is no compelling evidence that our MIKE observations
of \mbox{M22} or \ngc\ 
conducted in March, 2011, are contaminated
by the solar spectrum reflected by the full moon.

In conclusion, 
we cannot explain the 4.7~m\AA\ offset 
in the mean equivalent width differences
between our data and that of \citet{carretta14}.
In Section~\ref{compare},
we find that the cumulative effect of the 
choice of Fe~\textsc{i} lines used for the analysis,
the \loggf\ values, and this equivalent width
difference is 0.04~dex.
This value is relatively small,
and we do not pursue the matter further.

\section{Broadband photometry and reddening}
\label{photometry}

Table~\ref{startab} presents the photometry and colours.
$V$ and $(B-V)$ are taken from \citet{piotto02}\footnote{
Downloaded from http://www.astro.unipd.it/globulars/ on 2010 October 22}
and have been transformed
by those authors to the Johnson system from their 
\textit{Hubble Space Telescope} (\textit{HST})
WFPC2 $F439W$ and $F555W$
broadband photometry.
$K$ magnitudes are taken from the Two-Micron All Sky Survey
(2MASS; \citealt{skrutskie06}).
The $(B-V)$ and $(V-K)$ colours 
are dereddened according to the differential 
reddening maps of \citet{melbourne00}
and the extinction coefficients of \citet{cardelli89}.
All 15~stars that we have observed are within 2' of the cluster center,
and the full range of reddening for these stars
is 0.310~$\leq E(B-V) \leq$~0.322 
according to the \citeauthor{melbourne00}\ maps.
The differential reddening is a small
effect compared to the relatively large extinction
toward \ngc\ at low Galactic latitude
($b = -$8\deg).

\section{Model atmospheres}
\label{atmpar}

We interpolate model atmospheres from
the \textsc{atlas9} $\alpha$-enhanced grid \citep{castelli03}.
We use a 
recent version of the line analysis code \textsc{moog} 
(\citealt{sneden73}; see also \citealt{sobeck11}),
operated in batch mode,
to perform the analysis.
In Section~\ref{models}, we discuss our methods
to estimate the appropriate model atmosphere parameters.
We compare these values with those derived 
by \citet{carretta14} in Section~\ref{compare}.

\subsection{Model parameters}
\label{models}

\begin{figure}
\centering
\includegraphics[angle=0,width=3.3in]{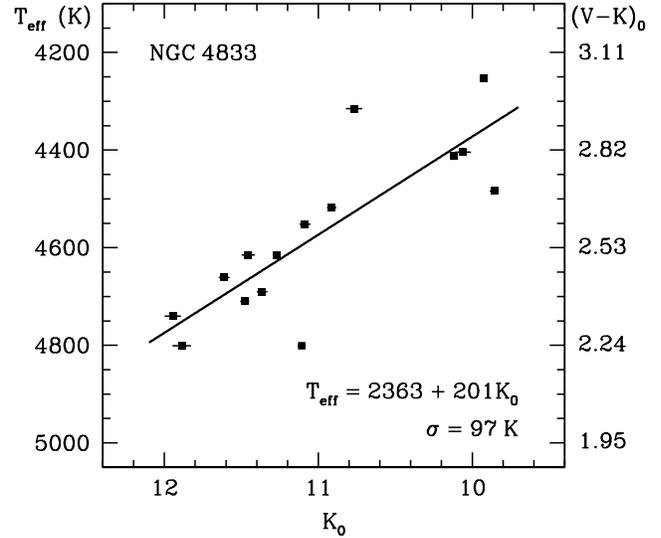}
\caption{
\label{kmagteff}
Relationship between dereddened $K$ magnitudes 
and \teff\ predicted by dereddened $(V-K)$ colours,
which are approximated on the
right axis for reference.
The linear least-squares fit is shown.
}
\end{figure}

We first estimate the stellar effective temperature (\teff)
using the metallicity-dependent $V-K$ colour-\teff\ relations 
of \citet{alonso99b}.  
We transform the $K$~magnitudes
from the 2MASS system to the TCS system
according to Equation~5 of \citet{ramirez05a}.
We refine the initial estimate by 
fitting a line to the relation between
dereddened $K$ magnitude, $K_{0}$, and \teff\
calculated from $(V-K)_{0}$.
Figure~\ref{kmagteff} illustrates this relationship.
The standard deviation of the residuals 
with respect to this fit is 97~K,
which we adopt as the 
uncertainty in \teff\ for each star.

The two stars that deviate most from the fit
are \mbox{2-1578} and \mbox{2-1664}.
Both stars are within 2$'$ of the cluster center,
as are most other stars in our sample.
No neighboring stars within 5~magnitudes and 10~pixels
are found in the \citet{piotto02} 
photometry files,
and no faint contaminants 
are apparent in the WFPC2 images.
No photometry error flags are reported for either star.
The $(V-K)_{0}$ colors of these stars
would need to be different by about $+$0.2~mag
and $-$0.2~mag, respectively,
to bring each star to the best-fit line
in Figure~\ref{kmagteff}.
These stars are only about 11$''$ apart,
so it is highly unlikely that 
differential reddening 
is underestimated for one and overestimated for the other.
We derive metallicities of these two stars
([Fe~\textsc{i}/H]~$= -$2.23 and $-$2.24,
[Fe~\textsc{ii}/H]~$= -$2.23 and $-$2.18,
respectively; Section~\ref{results})
that are in excellent agreement with the cluster means.
In summary, we find no reason to exclude
either star when computing the standard deviation
of the residuals to the fit shown in Figure~\ref{kmagteff}.

Uncertainties in $K_{0}$ translate to uncertainties
of only 33~K in \teff\ on average.
Small changes in the differential reddening 
across the face of \ngc\
have a minimal effect on \teff\ when derived 
this way.
$A(K)/E(B-V) \approx$~0.35 \citep{cardelli89},
so a differential reddening of $\pm$~0.012~mag in $E(B-V)$ 
\citep{melbourne00}
translates to $\pm$~0.004~mag in $K$.
The slope of the \teff-$K_{0}$ relation shown in 
Figure~\ref{kmagteff} is 201~K/mag.
Thus differential reddening contributes an average of
$\approx$~1~K to the \teff\ uncertainty budget,
which is negligible.
We assume
that all giants in \ngc\
reside along a single red giant branch (RGB) fiducial
with [Fe/H]~$= -$2.2 
(Section~\ref{iron}).
A change in [Fe/H] of 0.1~dex affects the
calculated \teff\ values by $<$~1~K,
which is also negligible.

\citet{melbourne00} derived the distance to \ngc\ using
the absolute magnitudes of RR~Lyrae variables along the 
horizontal branch (HB).  
We use this distance, $(m-M) =$~15.05~$\pm$~0.06,
to derive the surface gravities (\logg)
for all 15~stars observed from the fundamental relation 
$\log g = 4\log(T_{\rm eff,\star})
-\log(M/M_{\odot})
-0.4(M_{\rm bol,\sun} - M_{\rm bol,\star})
-10.61$.
Here, $M_{\rm bol,\star} = 
{\rm BC}_{K} + m_{K} - (m-M)$,
and the constant 10.61 is derived from the 
solar values given in \citet{cox00}.
We assume a mass of 0.8~\msun\ for all stars on the RGB in \ngc.
We interpolate BC$_{K}$ from
the grid of bolometric corrections presented by \citet{alonso99a}.
Uncertainties of 0.05 in magnitude, 97~K in \teff, 
0.1~\msun\ in mass, and 0.06 in the distance modulus 
only affect the derived \logg\ values by
0.02, 0.04, 0.05, and 0.03~dex, respectively.
Their collective uncertainties
contribute a total of 0.07~dex uncertainty in \logg.

We derive the microturbulent velocity, \vt,
by minimizing correlations between the equivalent width
(expressed as the $\log$ of the equivalent width
divided by $\lambda$)
of Fe~\textsc{i} lines and the abundances derived from them.
The internal uncertainty associated with this
method is $\approx$~0.05~dex.

Finally, we assign a single metallicity
to all model atmospheres,
[Fe/H]~$= -$2.2.
This is justified by our results
indicating that \ngc\ is mono-metallic
and has a mean metallicity 
of [Fe~\textsc{ii}/H]~$= -$2.19~$\pm$0.013
(Section~\ref{iron}).
Our use of the $\alpha$-enhanced grid of models
is justified because the 
most abundant electron-donating metals 
are enhanced by factors of a few relative to Fe in \ngc\
when compared with the solar ratios.

\subsection{Comparison with Carretta et al.}
\label{compare}

\citet{carretta14} used a procedure similar to ours to derive 
their \teff\ values for stars in \ngc.
Their procedure is discussed in detail by \citet{gratton07}.
There are four stars in common between our study and theirs,
and we derive a mean difference in \teff\ of
$-$28~$\pm$~1~K ($\sigma =$~2~K).
Both studies use the same 2MASS $K$ magnitudes
to calculate \teff\ after computing a relation
like that shown in Figure~\ref{kmagteff}.
There appears to be a zeropoint offset in the $V$ magnitudes
used to compute the initial relationship between
$V-K$ and \teff.
We adopt $V$ magnitudes from \citet{piotto02},
who calibrated their \textit{HST} $F555W$ magnitudes to $V$ in 
the standard Johnson system.
\citeauthor{carretta14}\ used their own ground-based
$V$ magnitudes
calibrated with an absolute zeropoint uncertainty of 0.03~mag.
A systematic zeropoint difference of $\pm$~0.03~mag translates
to an average difference in \teff\ of
$\mp$~27~K,   
exactly the difference found for these four stars.
The \teff\ offset between our study
and \citeauthor{carretta14}\
can be explained by the stated uncertainties
in the $V$-band zeropoints.

On average, our \logg\ values are 
lower than those calculated by \citet{carretta14} by
0.21~$\pm$~0.02~dex ($\sigma =$~0.04~dex).
Small differences in the adopted solar constants,
mass of stars on the RGB in \ngc, 
and distance modulus to \ngc\ 
only affect this offset by 0.04~dex.
Adopting the \citeauthor{carretta14}\
\teff\ values changes the \logg\ values by only 0.01~dex.
We adopt 0.21~dex as the systematic uncertainty
in \logg.

Our \vt\ values are greater by 0.18~$\pm$~0.04~\kmsec\
($\sigma =$~0.08~\kmsec)
for the four stars in common with \citet{carretta14}.
We adopt 0.18~\kmsec\ as the systematic uncertainty
in \vt.

Our metallicities (derived from Fe~\textsc{i} lines)
are lower by 0.20~$\pm$~0.02~dex ($\sigma =$~0.04~dex)
for the four stars in common with \citet{carretta14}.
The stated offset accounts for the different
adopted solar Fe abundances.
The discrepancy can be attributed to the
model atmosphere grids and line analysis codes,
as we now show.
We rederive metallicities for these four stars
using the \citeauthor{carretta14} model atmosphere parameters,
their line list and equivalent widths, 
our adopted grid of models (\textsc{atlas9}),
and our adopted line analysis code (\textsc{moog}).
The metallicities are lower by 
0.20~$\pm$~0.01~dex ($\sigma =$~0.02~dex),
which fully accounts for the discrepancy.

\section{Abundance analysis}
\label{analysis}

Abundances of lighter elements ($Z \leq$~30) that do not exhibit
broad isotope shifts (IS) or
hyperfine splitting structure (HFS)
are derived from equivalent widths.
We use \textsc{moog} to compute theoretical equivalent
widths that are forced to match the
measured values by adjusting the abundance.
Abundances of all other elements are derived
by matching synthetic spectra
with the observed spectra.
Lines of these elements are denoted by the word ``synth'' 
in Table~\ref{ewtab}.
When a line is not detected, 
we derive a 3$\sigma$ abundance upper limit using
a version of the formula presented on p.\ 590 of
\citet{frebel08}, which is derived from
equation A8 of \citet{bohlin83}.
These lines are denoted by the word ``limit''
in Table~\ref{ewtab}.

The atomic data for all lines examined
are presented in Table~\ref{ewtab}.
References for the \loggf\ values, IS, and HFS
(when available) are also given in Table~\ref{ewtab}.
We derive C abundances from the 
CH $A^2\Delta - X^2\Pi$ G band ($\approx$~4290--4330~\AA)
using a line list kindly provided by B.\ Plez (2007, private communication).
We derive N abundances from the 
CN $B^2\Sigma - X^2\Sigma$ violet band ($\approx$~3875--3885~\AA)
using the line list from \citet{kurucz95}
after setting the C abundance using the CH G band.

We assume local thermodynamic equilibrium (LTE)
holds for the line-forming layers of the atmosphere
for all species except for those described below.
We adopt non-LTE corrections for 
Li~\textsc{i} \citep{lind09},
O~\textsc{i} \citep{fabbian09}, 
Na~\textsc{i} \citep{lind11}, and 
K~\textsc{i} \citep{takeda02}.
We use the
correction for the nearest point on the grid
when the stellar parameters
extend beyond the edges of the grids of 
pre-computed non-LTE corrections.
These corrections are included in the abundances
presented in Tables~\ref{linetab}, \ref{abundtab}, and \ref{meantab}.
We include an additional 0.1~dex statistical uncertainty
in our error estimates to account for uncertainties
in the corrections.
We do not detect the Li~\textsc{i} 6707~\AA\ line;
corrections are based on the 3$\sigma$ upper limits
and range from $+$0.10 to $+$0.21~dex.
The non-LTE corrections for each of the
three high-excitation O~\textsc{i} 
7771, 7774, and 7775~\AA\ triplet lines
range from $-$0.05 to $-$0.08~dex.
The non-LTE corrections for the Na~\textsc{i} 
5682, 5688, 6154, and 6160~\AA\ lines
range from $-$0.02 to $-$0.14~dex.
The K~\textsc{i} 7698~\AA\ line is 
contaminated by an atmospheric O$_{2}$ line;
fortunately, however, the K~\textsc{i} 7664~\AA\ line is not
contaminated.
\citet{roederer14} found that these two lines
yield consistent abundances (to better than 0.02~dex)
after non-LTE corrections are applied.
The non-LTE corrections for the K~\textsc{i} 7664~\AA\ line 
range from $-$0.58 to $-$0.61~dex.

We apply one additional correction to the O abundance
derived from the high-excitation O~\textsc{i} triplet lines.
The [O~\textsc{i}]~6300~\AA\ line
is generally considered to be a reliable abundance
indicator formed under conditions of LTE
\citep{kiselman01}, but 
the O~\textsc{i} 7771, 7774, and 7775~\AA\ lines
are not.
Both abundance indicators are detected in seven stars.
After applying the non-LTE corrections for the O~\textsc{i} triplet lines,
these lines still yield abundances higher
by $+$0.40~$\pm$~0.05~dex ($\sigma =$~0.13~dex)
than the [O~\textsc{i}] 6300~\AA\ line.
This offset between the O~\textsc{i} and [O~\textsc{i}] lines
is comparable to those found by
\citet{garciaperez06} and \citet{roederer14}
in other metal-poor giants.
We apply a downward correction to all non-LTE-corrected O~\textsc{i} 
triplet line abundances by 0.40~dex.
We include an additional 0.05~dex uncertainty 
to account for the uncertainty in this correction.

\section{Results}
\label{results}

Table~\ref{linetab} lists
the abundances or upper limits derived from each line examined
in each star.
Table~\ref{abundtab} lists the weighted mean abundances
of each species examined in each star.
We adopt the upper limit that provides the
strongest constraint on the abundance
when multiple lines of the same species are not detected.
The complete versions of 
Tables~\ref{linetab} and \ref{abundtab} are 
available in the Supplementary Information available online.
Table~\ref{meantab} lists the mean abundance ratios
of all species examined in \ngc.
These are computed using the
formalism presented in \citet{mcwilliam95}.
The statistical uncertainty, $\sigma_{\rm statistical}$, 
is given by equation~A17 of \citeauthor{mcwilliam95} 
This includes uncertainties in the equivalent width measurement or
line profile fitting, \loggf\ values, and any non-LTE corrections.
The total uncertainty, $\sigma_{\rm total}$, 
is given by equation~A16 of \citeauthor{mcwilliam95} 
This includes the statistical uncertainty and 
uncertainties in the model atmosphere parameters.
We recommend that $\sigma_{\rm neutrals}$ for element A 
be added in quadrature with $\sigma_{\rm statistical}$ for
element B when computing the ratio [A/B] when B is 
derived from lines of the neutral species.
We recommend a similar procedure,
utilizing $\sigma_{\rm ions}$ instead of $\sigma_{\rm neutrals}$, 
when element B 
is derived from lines of the ionized species.
The adopted solar reference abundances
are listed in Table~\ref{solartab}.

\begin{table}
\begin{minipage}{2.6in} 
\caption{Abundances derived from individual lines
\label{linetab}}
\begin{tabular}{ccccc}
\hline
\hline
Star &
Species &
Wavelength &
$\log \epsilon$ &
$\sigma$ \\
 &
 &
(\AA) &
 &
 \\
\hline
 2-185           &     Li~\textsc{i}  & 6707.80 & $<$ 0.53 &  \ldots \\
 2-185           &    [O~\textsc{i}]  & 6300.30 & $<$ 7.16 &  \ldots \\
 2-185           &      O~\textsc{i}  & 7771.94 & $<$ 6.83 &  \ldots \\
 2-185           &      O~\textsc{i}  & 7774.17 & $<$ 7.24 &  \ldots \\
 2-185           &      O~\textsc{i}  & 7775.39 & $<$ 7.40 &  \ldots \\
 2-185           &     Na~\textsc{i}  & 5682.63 &     4.51 &    0.20 \\
 2-185           &     Na~\textsc{i}  & 5688.20 &     4.60 &    0.20 \\
 2-185           &     Na~\textsc{i}  & 6160.75 &     4.74 &    0.21 \\
 2-185           &     Mg~\textsc{i}  & 4702.99 &     5.38 &    0.32 \\
 2-185           &     Mg~\textsc{i}  & 5528.40 &     5.60 &    0.29 \\
 2-185           &     Mg~\textsc{i}  & 5711.09 &     5.53 &    0.15 \\
\vdots & \vdots & \vdots & \vdots & \vdots \\
\hline
\end{tabular}
\\
Notes.--- 
The full version of Table~\ref{linetab} is available in the 
Supplementary Information section online.  
\end{minipage}
\end{table}

\begin{table*}
\begin{minipage}{4.6in} 
\caption{Mean abundances in each star
\label{abundtab}}
\begin{tabular}{ccccccccc}
\hline
\hline
Star &
Species &
$N_{\rm lines}$ &
$\log\epsilon$ &
[X/Fe]\footnote{[Fe/H] is given for Fe~\textsc{i} and Fe~\textsc{ii}} &
$\sigma_{\rm statistical}$ &
$\sigma_{\rm total}$ &
$\sigma_{\rm neutrals}$ &
$\sigma_{\rm ions}$ \\
\hline
2-185 & Fe~\textsc{i}  & 102 &     5.33 &$-$2.17&  0.06 &  0.14 &  0.00 &  0.00  \\
2-185 & Fe~\textsc{ii} &   6 &     5.27 &$-$2.23&  0.07 &  0.10 &  0.00 &  0.00  \\
2-185 & Li~\textsc{i}  &   1 & $<$ 0.53 & \ldots& \ldots& \ldots& \ldots& \ldots \\
2-185 & C~(CH)         &   1 &     5.62 &$-$0.64&  0.10 &  0.22 &  0.15 &  0.15  \\
2-185 & N~(CN)         &   1 &     7.29 &  1.63 &  0.20 &  0.28 &  0.23 &  0.23  \\
2-185 & O~\textsc{i}   &   4 & $<$ 6.83 &  0.31 & \ldots& \ldots& \ldots& \ldots \\
2-185 & Na~\textsc{i}  &   3 &     4.61 &  0.55 &  0.12 &  0.17 &  0.13 &  0.17  \\
2-185 & Mg~\textsc{i}  &   3 &     5.52 &  0.09 &  0.07 &  0.14 &  0.09 &  0.15  \\
\vdots & \vdots & \vdots & \vdots & \vdots & \vdots & \vdots & \vdots & \vdots \\
\hline
\end{tabular}
\\
Notes.--- 
The full version of Table~\ref{abundtab} is available in the 
Supplementary Information section online.  
\end{minipage}
\end{table*}

\begin{table*}
\begin{minipage}{5.5in} 
\caption{Mean abundance ratios in NGC~4833
\label{meantab}}
\begin{tabular}{cccccccccc}
\hline
 & 
 &
\multicolumn{5}{c}{This study} &
 &
\multicolumn{2}{c}{\citet{carretta14}} \\
\cline{3-7} \cline{9-10}
Ratio &
Species or &
Mean &
Std.\ err.\ &
Std.\ dev.\ &
Num.\ &
Notes &
 &
Mean &
Std.\ err.\ \\
 &
molecule &
 &
 &
 &
stars &
 & 
 &
 \\
\hline
{[C/Fe]}  & CH          & $-$0.44 & 0.06 & 0.24 & 15 &   & & \ldots  & \ldots \\
{[N/Fe]}  & CN          & $+$1.28 & 0.09 & 0.36 & 15 &   & & \ldots  & \ldots \\
{[O/Fe]}  & \textsc{i}  & $+$0.81 & 0.08 & 0.22 &  8 & 1 & & $+$0.17 & 0.22   \\
{[Na/Fe]} & \textsc{i}  & $+$0.39 & 0.06 & 0.22 & 14 & 1 & & $+$0.52 & 0.29   \\
{[Mg/Fe]} & \textsc{i}  & $+$0.35 & 0.06 & 0.24 & 15 &   & & $+$0.27 & 0.22   \\
{[Al/Fe]} & \textsc{i}  & $+$1.09 & 0.07 & 0.20 &  8 & 1 & & $+$0.90 & 0.34   \\
{[Si/Fe]} & \textsc{i}  & $+$0.74 & 0.02 & 0.09 & 15 &   & & $+$0.47 & 0.04   \\
{[K/Fe]}  & \textsc{i}  & $+$0.40 & 0.02 & 0.10 & 15 &   & & \ldots  & \ldots \\
{[Ca/Fe]} & \textsc{i}  & $+$0.49 & 0.011& 0.04 & 15 &   & & $+$0.35 & 0.01   \\
{[Sc/Fe]} & \textsc{ii} & $+$0.19 & 0.03 & 0.11 & 15 &   & & $-$0.11 & 0.01   \\
{[Ti/Fe]} & \textsc{i}  & $+$0.08 & 0.014& 0.06 & 15 &   & & $+$0.18 & 0.02   \\
{[Ti/Fe]} & \textsc{ii} & $+$0.39 & 0.02 & 0.08 & 15 &   & & $+$0.23 & 0.01   \\
{[V/Fe]}  & \textsc{i}  & $-$0.44 & 0.03 & 0.10 & 15 &   & & $-$0.08 & 0.01   \\
{[V/Fe]}  & \textsc{ii} & $+$0.05 & 0.05 & 0.19 & 14 &   & & \ldots  & \ldots \\
{[Cr/Fe]} & \textsc{i}  & $-$0.26 & 0.02 & 0.07 & 15 &   & & $-$0.24 & 0.02   \\
{[Cr/Fe]} & \textsc{ii} & $+$0.29 & 0.03 & 0.12 & 12 &   & & $+$0.01 & 0.05   \\
{[Mn/Fe]} & \textsc{i}  & $-$0.58 & 0.03 & 0.11 & 15 &   & & $-$0.54 & 0.01   \\
{[Fe/H]}  & \textsc{i}  & $-$2.25 & 0.02 & 0.09 & 15 &   & & $-$2.02 & 0.01   \\
{[Fe/H]}  & \textsc{ii} & $-$2.19 & 0.013& 0.05 & 15 &   & & $-$2.01 & 0.02   \\
{[Co/Fe]} & \textsc{i}  & $-$0.39 & 0.04 & 0.14 & 14 & 1 & & $-$0.03 & 0.03   \\
{[Ni/Fe]} & \textsc{i}  & $-$0.06 & 0.015& 0.06 & 15 &   & & $-$0.18 & 0.01   \\
{[Cu/Fe]} & \textsc{i}  & $-$0.65 & 0.05 & 0.14 &  9 & 1 & & $-$0.80 & 0.09   \\
{[Zn/Fe]} & \textsc{i}  & $+$0.19 & 0.02 & 0.10 & 15 &   & & $+$0.07 & 0.03   \\
{[Rb/Fe]} & \textsc{i}  &$<+$0.85 &\ldots&\ldots& 15 &   & & \ldots  & \ldots \\
{[Sr/Fe]} & \textsc{ii} & $+$0.18 & 0.03 & 0.12 & 14 &   & & \ldots  & \ldots \\
{[Y/Fe]}  & \textsc{ii} & $-$0.22 & 0.03 & 0.11 & 15 &   & & $-$0.15 & 0.06   \\
{[Zr/Fe]} & \textsc{ii} & $+$0.25 & 0.03 & 0.11 & 14 &   & & \ldots  & \ldots \\
{[Nb/Fe]} & \textsc{ii} &$<+$0.93 &\ldots&\ldots&  2 &   & & \ldots  & \ldots \\
{[Mo/Fe]} & \textsc{i}  & $+$0.19 & 0.11 & 0.23 &  4 & 1 & & \ldots  & \ldots \\
{[Ba/Fe]} & \textsc{ii} & $-$0.02 & 0.016& 0.06 & 15 &   & & $-$0.06 & 0.07   \\
{[La/Fe]} & \textsc{ii} & $+$0.08 & 0.03 & 0.13 & 14 &   & & $+$0.05 & 0.03   \\
{[Ce/Fe]} & \textsc{ii} & $+$0.03 & 0.03 & 0.11 & 15 &   & & \ldots  & \ldots \\
{[Pr/Fe]} & \textsc{ii} & $+$0.19 & 0.09 & 0.29 & 10 & 1 & & \ldots  & \ldots \\
{[Nd/Fe]} & \textsc{ii} & $+$0.18 & 0.03 & 0.11 & 15 &   & & \ldots  & \ldots \\
{[Sm/Fe]} & \textsc{ii} & $+$0.29 & 0.03 & 0.11 & 14 &   & & \ldots  & \ldots \\
{[Eu/Fe]} & \textsc{ii} & $+$0.36 & 0.03 & 0.13 & 15 &   & & \ldots  & \ldots \\
{[Gd/Fe]} & \textsc{ii} & $+$0.32 & 0.09 & 0.21 &  6 & 1 & & \ldots  & \ldots \\
{[Tb/Fe]} & \textsc{ii} & $+$0.19 & 0.04 & 0.05 &  2 & 1 & & \ldots  & \ldots \\
{[Dy/Fe]} & \textsc{ii} & $+$0.43 & 0.05 & 0.17 & 13 & 1 & & \ldots  & \ldots \\
{[Ho/Fe]} & \textsc{ii} & $+$0.16 & 0.18 & 0.18 &  1 & 1 & & \ldots  & \ldots \\
{[Er/Fe]} & \textsc{ii} & $+$0.33 & 0.08 & 0.20 &  7 & 1 & & \ldots  & \ldots \\
{[Tm/Fe]} & \textsc{ii} & $+$0.19 & 0.07 & 0.10 &  2 & 1 & & \ldots  & \ldots \\
{[Yb/Fe]} & \textsc{ii} & $+$0.01 & 0.08 & 0.25 &  9 & 1 & & \ldots  & \ldots \\
{[Hf/Fe]} & \textsc{ii} & $+$0.27 & 0.04 & 0.06 &  2 & 1 & & \ldots  & \ldots \\
{[Ir/Fe]} & \textsc{i}  & $+$0.66 & 0.11 & 0.11 &  1 & 1 & & \ldots  & \ldots \\
{[Pb/Fe]} & \textsc{i}  &$<+$0.02 &\ldots&\ldots& 15 &   & & \ldots  & \ldots \\
{[Th/Fe]} & \textsc{ii} &$<+$0.32 &\ldots&\ldots& 14 &   & & \ldots  & \ldots \\
\hline
\end{tabular}
\\
Notes--- 1.\ Does not include one or more upper limits \\
\end{minipage}
\end{table*}

\begin{table}
\begin{minipage}{2.0in} 
\caption{Adopted solar abundances
\label{solartab}}
\begin{tabular}{cccc}
\hline
Element &
$Z$ &
$\log \epsilon_{\odot}$ &
Photospheric (P) \\
 &
 &
 &
or meteoritic (M) \\
\hline
C  &  6 &  8.43  & P \\
N  &  7 &  7.83  & P \\
O  &  8 &  8.69  & P \\
Na & 11 &  6.24  & P \\
Mg & 12 &  7.60  & P \\
Al & 13 &  6.45  & P \\
Si & 14 &  7.51  & P \\
K  & 19 &  5.03  & P \\
Ca & 20 &  6.34  & P \\
Sc & 21 &  3.15  & P \\
Ti & 22 &  4.95  & P \\
V  & 23 &  3.93  & P \\
Cr & 24 &  5.64  & P \\
Mn & 25 &  5.43  & P \\
Fe & 26 &  7.50  & P \\
Co & 27 &  4.99  & P \\
Ni & 28 &  6.22  & P \\
Cu & 29 &  4.19  & P \\
Zn & 30 &  4.56  & P \\
Rb & 37 &  2.52  & P \\
Sr & 38 &  2.87  & P \\
Y  & 39 &  2.21  & P \\
Zr & 40 &  2.58  & P \\
Nb & 41 &  1.46  & P \\
Mo & 42 &  1.88  & P \\
Ba & 56 &  2.18  & P \\
La & 57 &  1.10  & P \\
Ce & 58 &  1.58  & P \\
Pr & 59 &  0.72  & P \\
Nd & 60 &  1.42  & P \\
Sm & 62 &  0.96  & P \\
Eu & 63 &  0.52  & P \\
Gd & 64 &  1.07  & P \\
Tb & 65 &  0.30  & P \\
Dy & 66 &  1.10  & P \\
Ho & 67 &  0.48  & P \\
Er & 68 &  0.92  & P \\
Tm & 69 &  0.10  & P \\
Yb & 70 &  0.92  & M \\
Hf & 72 &  0.85  & P \\
Ir & 77 &  1.38  & P \\
Pb & 82 &  2.04  & M \\
Th & 90 &  0.06  & M \\
\hline
\end{tabular}
\end{minipage}
\end{table}

\begin{figure*}
\centering
\includegraphics[angle=0,width=3.4in]{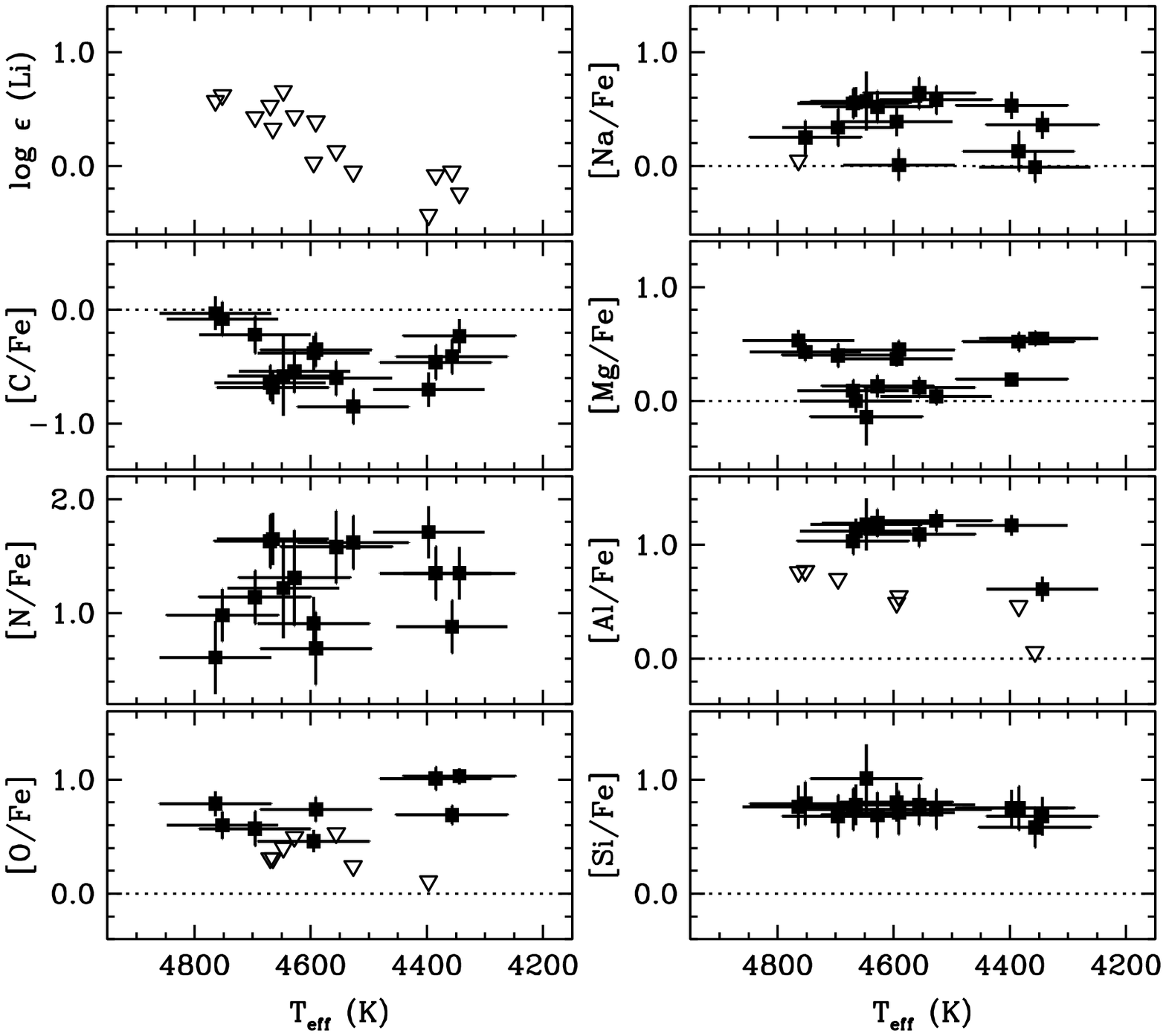}
\hspace*{0.0in}
\includegraphics[angle=0,width=3.4in]{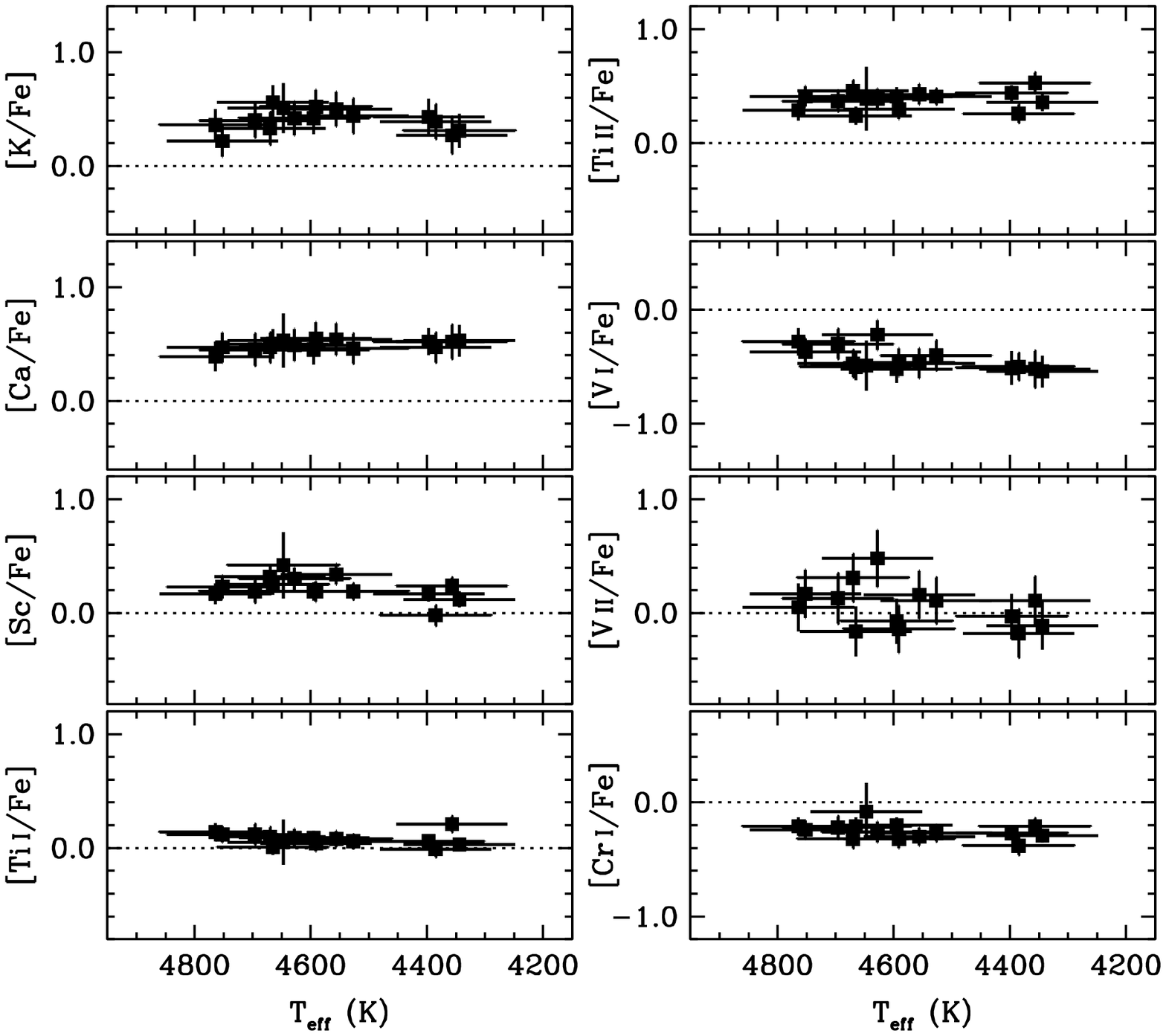}
\caption{
\label{teffplot1}
Derived [X/Fe] ratios ($\log \epsilon$ for Li)
as a function of \teff\ for Li through Cr~\textsc{i}.
The dotted lines represent the solar ratios,
and the downward facing triangles represent upper limits.
}
\end{figure*}

\begin{figure*}
\centering
\includegraphics[angle=0,width=3.4in]{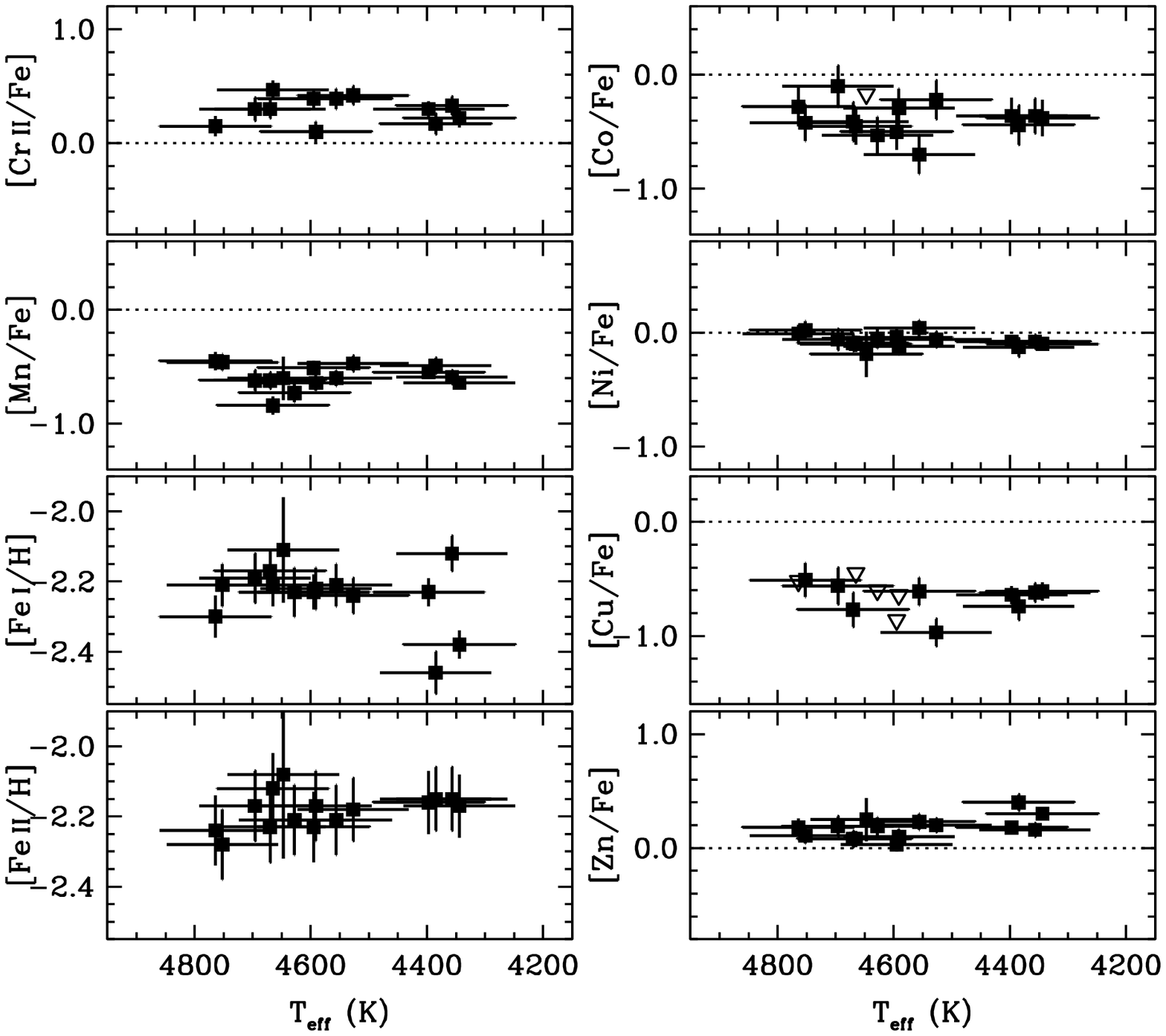}
\hspace*{0.0in}
\includegraphics[angle=0,width=3.4in]{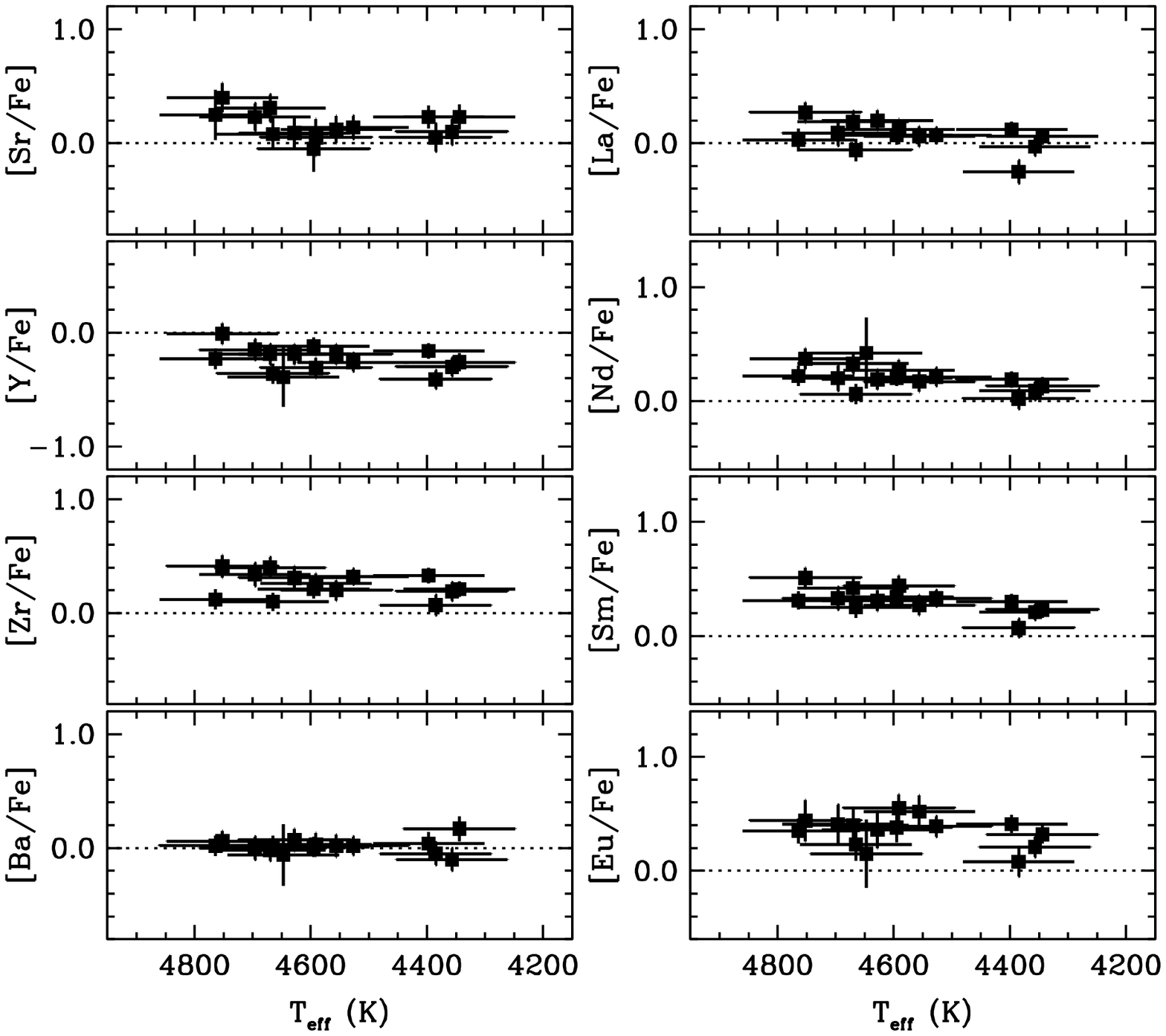}
\caption{
\label{teffplot2}
Derived [X/Fe] ratios ([Fe/H] for Fe) 
as a function of \teff\ for Cr~\textsc{ii} through Yb.
The dotted lines represent the solar ratios,
and the downward facing triangles represent upper limits.
}
\end{figure*}

Figures~\ref{teffplot1} and \ref{teffplot2} illustrate
abundance ratios as a function of \teff.
In most cases, no trends are found.
Subtle trends may be present in cases where
one of the elements in question 
is known to vary with evolutionary status.
A small downward trend of [C/Fe]
and a corresponding increase in [N/Fe] 
are found with decreasing \teff.
The Li upper limits get progressively lower
with decreasing \teff,
which is simply a reflection of relatively constant equivalent width
upper limits (from relatively constant S/N levels)
applied to stars with decreasing \teff.
There are also subtle differences at the $\approx$~1--2$\sigma$ level
($\approx$~0.10--0.15~dex)
in the mean [X/Fe] ratios between the
warm (\teff~$>$~4500~K) and cool (\teff~$<$~4500~K) stars,
where ``X'' represents
Sc, V~\textsc{i} and \textsc{ii}, 
Zn, Nd, Ce, Sm, Eu, and Dy.
We investigate these differences in detail
in Sections~\ref{irongroup} and \ref{dispersion}.

Table~\ref{atmvar} lists the variations in 
abundance ratios that result from
changes in the model atmosphere parameters
for one representative star
with high S/N ratios, 
\mbox{4-1255}.
The final column of Table~\ref{atmvar} lists the
total variation, calculated by adding the 
four individual variations in quadrature.

\begin{table*}
\begin{minipage}{4.3in} 
\caption{Variations in Abundance Ratios Resulting from Changes in the 
Model Atmosphere Parameters for Star 4-1255
\label{atmvar}}
\begin{tabular}{ccccccc}
\hline
Ratio &
Ion &
$\delta$[X/Y]/$\delta$\teff\ &
$\delta$[X/Y]/$\delta$\logg\ &
$\delta$[X/Y]/$\delta$\vt\ &
$\delta$[X/Y]/$\delta$[M/H] &
$\sqrt{\Sigma\delta_{i}^{2}}$ \\
\hline
{[O/Fe]}  & \textsc{i}  & $-$0.27 & $+$0.10 & $+$0.06 & $+$0.04 & 0.30\\
{[Na/Fe]} & \textsc{i}  & $-$0.06 & $+$0.00 & $+$0.07 & $+$0.01 & 0.09\\
{[Mg/Fe]} & \textsc{i}  & $-$0.06 & $-$0.03 & $+$0.01 & $+$0.00 & 0.07\\
{[Al/Fe]} & \textsc{i}  & $-$0.08 & $+$0.01 & $+$0.07 & $+$0.01 & 0.11\\
{[Si/Fe]} & \textsc{i}  & $-$0.11 & $+$0.04 & $+$0.07 & $+$0.03 & 0.14\\
{[K/Fe]}  & \textsc{i}  & $+$0.03 & $+$0.01 & $-$0.03 & $-$0.01 & 0.04\\
{[Ca/Fe]} & \textsc{i}  & $-$0.01 & $+$0.02 & $-$0.03 & $-$0.01 & 0.04\\
{[Sc/Fe]} & \textsc{ii} & $+$0.01 & $+$0.05 & $+$0.07 & $+$0.05 & 0.10\\
{[Ti/Fe]} & \textsc{i}  & $+$0.08 & $-$0.01 & $+$0.02 & $-$0.02 & 0.09\\
{[Ti/Fe]} & \textsc{ii} & $-$0.06 & $-$0.02 & $-$0.04 & $-$0.01 & 0.08\\
{[V/Fe]}  & \textsc{i}  & $+$0.11 & $+$0.02 & $+$0.07 & $+$0.05 & 0.14\\
{[V/Fe]}  & \textsc{ii} & $-$0.02 & $-$0.05 & $+$0.00 & $-$0.07 & 0.09\\
{[Cr/Fe]} & \textsc{i}  & $+$0.04 & $-$0.01 & $+$0.02 & $-$0.01 & 0.05\\
{[Cr/Fe]} & \textsc{ii} & $-$0.12 & $+$0.00 & $+$0.03 & $-$0.02 & 0.13\\
{[Mn/Fe]} & \textsc{i}  & $+$0.03 & $+$0.04 & $+$0.10 & $+$0.05 & 0.12\\
{[Fe/H]}  & \textsc{i}  & $+$0.15 & $-$0.03 & $-$0.07 & $-$0.02 & 0.17\\
{[Fe/H]}  & \textsc{ii} & $-$0.06 & $+$0.08 & $-$0.04 & $+$0.05 & 0.12\\
{[Co/Fe]} & \textsc{i}  & $+$0.13 & $+$0.00 & $+$0.01 & $+$0.02 & 0.13\\
{[Ni/Fe]} & \textsc{i}  & $-$0.03 & $+$0.02 & $+$0.05 & $+$0.01 & 0.06\\
{[Cu/Fe]} & \textsc{i}  & $-$0.02 & $+$0.03 & $+$0.08 & $+$0.02 & 0.09\\
{[Zn/Fe]} & \textsc{i}  & $-$0.18 & $+$0.07 & $+$0.04 & $+$0.05 & 0.20\\
{[Sr/Fe]} & \textsc{ii} & $+$0.01 & $-$0.06 & $+$0.02 & $-$0.07 & 0.09\\
{[Y/Fe]}  & \textsc{ii} & $+$0.09 & $-$0.01 & $+$0.02 & $+$0.00 & 0.09\\
{[Zr/Fe]} & \textsc{ii} & $+$0.05 & $-$0.01 & $+$0.02 & $+$0.00 & 0.05\\
{[Mo/Fe]} & \textsc{i}  & $+$0.03 & $+$0.00 & $+$0.05 & $-$0.01 & 0.06\\
{[Ba/Fe]} & \textsc{ii} & $+$0.16 & $-$0.02 & $-$0.08 & $+$0.00 & 0.18\\
{[La/Fe]} & \textsc{ii} & $+$0.10 & $-$0.02 & $+$0.05 & $+$0.01 & 0.11\\
{[Ce/Fe]} & \textsc{ii} & $+$0.07 & $+$0.00 & $+$0.05 & $+$0.02 & 0.09\\
{[Pr/Fe]} & \textsc{ii} & $+$0.13 & $+$0.02 & $+$0.08 & $+$0.03 & 0.16\\
{[Nd/Fe]} & \textsc{ii} & $+$0.13 & $+$0.00 & $+$0.04 & $+$0.02 & 0.14\\
{[Sm/Fe]} & \textsc{ii} & $+$0.11 & $-$0.01 & $+$0.03 & $+$0.03 & 0.12\\
{[Eu/Fe]} & \textsc{ii} & $+$0.09 & $+$0.02 & $+$0.07 & $+$0.01 & 0.12\\
{[Gd/Fe]} & \textsc{ii} & $+$0.09 & $+$0.05 & $+$0.04 & $-$0.01 & 0.11\\
{[Tb/Fe]} & \textsc{ii} & $+$0.10 & $+$0.03 & $+$0.07 & $+$0.05 & 0.14\\
{[Dy/Fe]} & \textsc{ii} & $+$0.07 & $-$0.02 & $-$0.02 & $-$0.02 & 0.08\\
{[Er/Fe]} & \textsc{ii} & $+$0.00 & $+$0.03 & $+$0.00 & $+$0.01 & 0.03\\
{[Tm/Fe]} & \textsc{ii} & $+$0.09 & $-$0.04 & $+$0.00 & $-$0.08 & 0.13\\
{[Yb/Fe]} & \textsc{ii} & $+$0.21 & $-$0.09 & $+$0.04 & $-$0.08 & 0.25\\
{[Hf/Fe]} & \textsc{ii} & $+$0.12 & $+$0.07 & $+$0.09 & $-$0.02 & 0.17\\
\hline
\end{tabular}
\\
Notes--- 
$\delta$\teff~$= +$97~K, 
$\delta$\logg~$= +$0.21~dex, 
$\delta$\vt~$= +$0.18~\kmsec, 
$\delta$[M/H]~$= +$0.20;
see Sections~\ref{models} and \ref{compare} for details.
\end{minipage}
\end{table*}

\subsection{Comparison of abundance ratios}
\label{abundcompare}

Table~\ref{meantab} also lists
the mean abundance ratios derived by \citet{carretta14}
from their UVES spectra of stars in \ngc.
In many cases the two sets of mean abundances are 
in reasonable (2$\sigma$) agreement.
At first glance,
several [X/Fe] ratios are not in agreement,
where ``X'' represents
O~\textsc{i},
Si~\textsc{i},
Ca~\textsc{i},
Sc~\textsc{ii},
Ti~\textsc{i} and \textsc{ii},
V~\textsc{i},
Cr~\textsc{ii},
Co~\textsc{i}, 
Ni~\textsc{i},
Zn~\textsc{i}, and
Nd~\textsc{ii}.
Most of these
differences can be explained by the
sets of lines used, the \loggf\ values,
and the adopted solar abundances,
as we now show.

\citet{carretta14} used the 
atomic data and solar reference abundances
presented in \citet{gratton03}.
Their approach 
ensures a uniform abundance scale
for the dozens of globular clusters observed
in the last decade using GIRAFFE and UVES.~
Our analysis incorporates atomic data from
more recent laboratory studies and
adopts the \citet{asplund09} solar abundance scale.
The appeal of using updated \loggf\ values is that
recent laboratory studies
regularly report uncertainties 
on absolute transition probabilities as small as 5~per cent,
which is effectively negligible
when compared with other sources of uncertainty.
The choice of \loggf\ values is irrelevant 
if only relative differences are of interest and
if the same set of lines 
is employed for each star.
This condition is not met
since the S/N and stellar parameters vary,
thus we use updated \loggf\ values 
whenever possible.

The [Ca/Fe] ratios provide a good example
of the consequences of this choice.
There are seven Ca~\textsc{i} lines 
studied in the GIRAFFE spectra obtained by \citet{carretta14}.
The Ca~\textsc{i} \loggf\ values
used by \citeauthor{carretta14}\ and those we adopt from NIST
or \citet{aldenius09}
differ by $-$0.26~dex to $+$0.27~dex
for these lines.
Only one of these absolute differences is smaller than 0.17~dex.
In our study, we use three of these lines and
several others at shorter wavelengths.
Thus the exact set of lines measured significantly
affects the derived [Ca/Fe] ratios.
Substantial differences ($\geq$~0.05~dex) are also present
in the adopted \loggf\ values of lines of Na~\textsc{i},
Si~\textsc{i}, Sc~\textsc{ii}, Ti~\textsc{i},
Fe~\textsc{ii}, and Ni~\textsc{i}.
It is impossible to assess the exact differences
since no single source of \loggf\ values
covers all of the lines used by \citeauthor{carretta14}\ and us.

Our use of the \citet{asplund09} solar abundance scale
introduces several substantial differences in the [X/Fe] ratios
relative to \citet{carretta14}.
Our [O~\textsc{i}/Fe] ratios would change 
by $-$0.06~dex on their scale;
[Na~\textsc{i}/Fe], $+$0.07~dex; 
[Mg~\textsc{i}/Fe], $+$0.21~dex;
[Ca~\textsc{i}/Fe], $+$0.11~dex;
[Ti~\textsc{ii}/Fe], $-$0.13~dex;
[Cr~\textsc{ii}/Fe], $-$0.08~dex;
[Mn~\textsc{i}/Fe], $+$0.13~dex; and 
[Zn~\textsc{i}/Fe], $+$0.07~dex.

We use our software, model atmospheres, and solar reference
abundances to rederive the [X/Fe] ratios using 
the \citet{carretta14} line lists and equivalent widths
for the four stars in common.
On our scale, the \citeauthor{carretta14}\ ratios change by
$+$0.15~$\pm$~0.04~dex for [O~\textsc{i}/Fe],
$+$0.10~$\pm$~0.03~dex for [Si~\textsc{i}/Fe], 
$-$0.05~$\pm$~0.05~dex for [Ca~\textsc{i}/Fe], 
$+$0.05~$\pm$~0.03~dex for [Ti~\textsc{i}/Fe], and
$+$0.07~$\pm$~0.03~dex for [Ni~\textsc{i}/Fe].
These account for
about half of the total discrepancy
in O~\textsc{i}, Si~\textsc{i}, and Ni~\textsc{i},
but they widen the discrepancy for
Ca~\textsc{i} and Ti~\textsc{i}.
V~\textsc{i} and Co~\textsc{i} cannot be assessed
due to the corrections for HFS.~
\citeauthor{carretta14}\ only derived
abundances for Ti~\textsc{ii}, Cr~\textsc{ii},
Zn~\textsc{i}, and Nd~\textsc{ii} from their UVES spectra,
so we have no data in common.

To summarize,
some mean [X/Fe] ratios we have derived
are in disagreement with those found by \citet{carretta14}.
We can explain some of these discrepancies
in part by the different solar reference abundances,
and we attribute others to the different
sets of lines used and the \loggf\ values of those lines.
Finally,
the exact sets of stars observed could affect the mean values
of the light elements known to vary star-by-star.

\section{Discussion}
\label{discussion}

\subsection{Iron}
\label{iron}

We derive a mean metallicity, 
$\langle$[Fe/H]$\rangle$, of 
$-$2.25~$\pm$~0.02 ($\sigma =$~0.09)
from Fe~\textsc{i} lines in the 15~stars observed in \ngc.
We derive
$\langle$[Fe/H]$\rangle = -$2.19~$\pm$~0.013 ($\sigma =$~0.05)
from Fe~\textsc{ii} lines.
The standard deviation associated with
each of these values does not exceed
that expected from the model atmosphere uncertainties
listed in Table~\ref{atmvar}.
We find no evidence for an intrinsic dispersion
in the Fe abundances from one star to another within \ngc,
which reaffirms the results of \citet{carretta14}.

As noted in Section~\ref{compare},
we find a metallicity lower by
0.20~$\pm$~0.02~dex ($\sigma =$~0.04~dex)
relative to \citet{carretta14} for four stars in common.
We have identified the causes of this discrepancy previously,
and now we compare the metallicity of \ngc\
to an external metallicity standard.
The temperature of the K-giant Arcturus ($\alpha$~Boo)
is known to better than 30~K
from measurements of its angular diameter and bolometric flux.
\citet{koch08} performed a differential abundance analysis
between Arcturus and
red giants in the moderately metal-poor 
globular cluster \mbox{47~Tuc}
([Fe/H]~$= -$0.76~$\pm$~0.01 [statistical]\ $\pm$~0.04 [systematic]).
\citet{koch11} extended this differential analysis
to the very metal-poor cluster \mbox{NGC~6397}
([Fe/H]~$= -$2.10~$\pm$~0.02 [statistical]\ $\pm$~0.07 [systematic]),
which we use for our comparison.

We use the Fe~\textsc{i} line list from
\citet{koch11} to identify 19~lines in common
to three red giants in \mbox{NGC~6397}
(stars 7230, 8958, and 13414)
and three red giants in \ngc\
(stars \mbox{3-742}, \mbox{4-341}, \mbox{4-1255})
with similar \teff.
We measure equivalent widths for each of these lines 
in \ngc,
and we perform a differential abundance analysis between
\mbox{NGC~6397} and \ngc.
We find a mean difference in [Fe/H] of 
$+$0.09~$\pm$~0.04~dex between \mbox{NGC~6397}
and \ngc,
In other words, if we adopt 
[Fe/H]~$= -$2.10~$\pm$~0.02 for \mbox{NGC~6397}
as derived by \citet{koch11},
we infer
[Fe/H]~$= -$2.19~$\pm$~0.04 for \ngc.
This value is in fair agreement with
our derived mean metallicity,
[Fe/H]~$= -$2.25~$\pm$~0.02.
On the basis of the differential globular cluster abundance scale
established by \citeauthor{koch08}\ relative to Arcturus,
we prefer the lower metallicity value for \ngc.

\subsection{Lithium through Silicon}
\label{light}

We do not detect Li in any of the stars in our sample.
The upper limits we derive
are illustrated in Figure~\ref{teffplot1}.
These limits rule out 
substantial enhancement of the surface Li abundances
in these stars,
as has been found in a few red giants in other globular clusters
(e.g., \citealt{kraft99,smith99}).

\begin{figure}
\centering
\includegraphics[angle=0,width=3.3in]{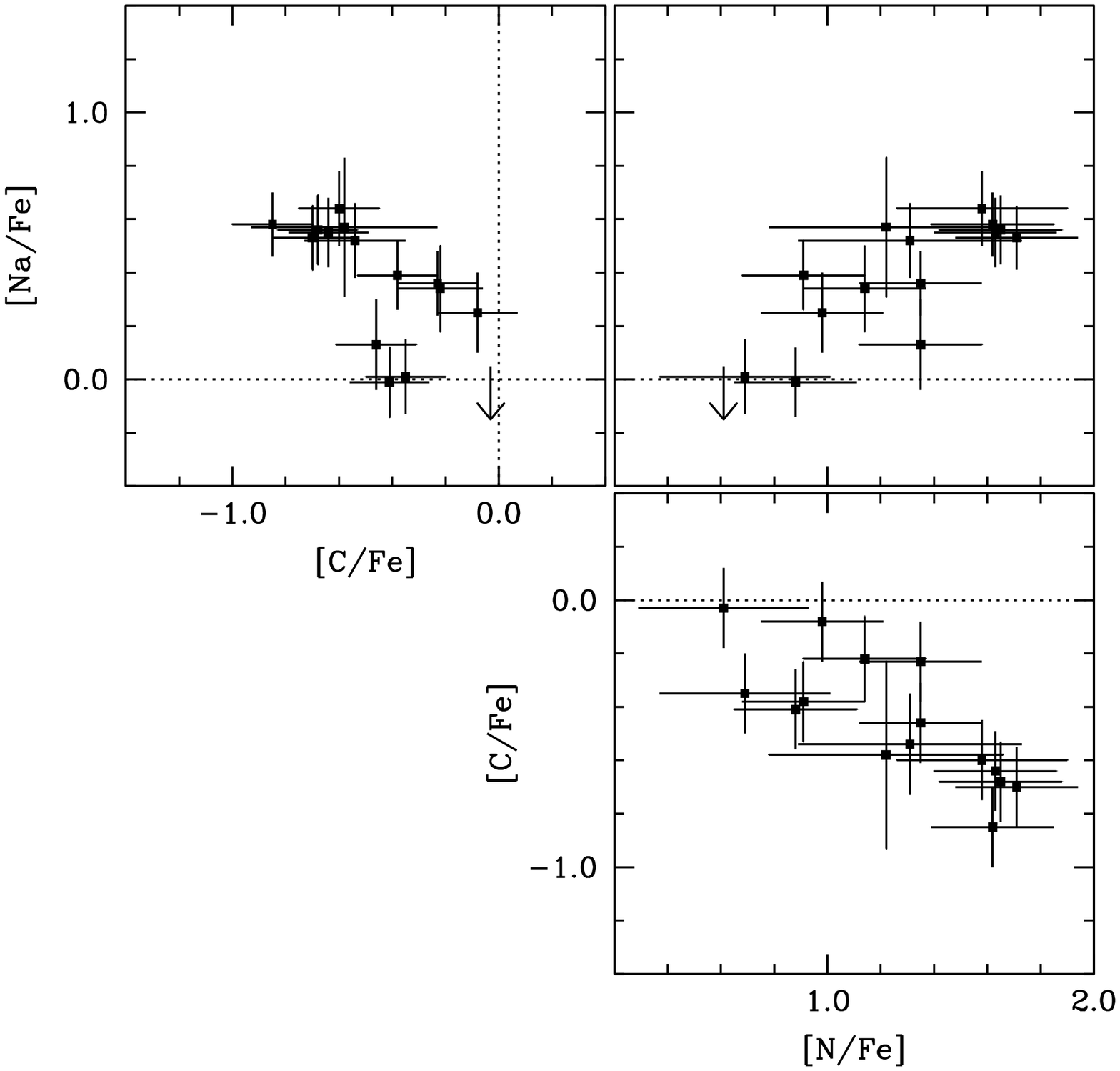}
\caption{
\label{lightplot1}
Relationships among C, N, and Na in \ngc.
Black symbols mark data from this study.
The dotted lines indicate the solar ratios.
}
\end{figure}

Figure~\ref{lightplot1} shows the relationships among the
light elements C, N, and Na in \ngc.
There is a clear decrease in the [C/Fe] ratio with 
increasing [N/Fe].
[C/Fe] also anti-correlates with [Na/Fe], and
[N/Fe] correlates with [Na/Fe].
There is no compelling evidence for a bi-modality
in either [C/Fe] or [N/Fe], but 
this tentative conclusion should
be checked using larger sample sizes.

\begin{figure*}
\centering
\includegraphics[angle=0,width=6.5in]{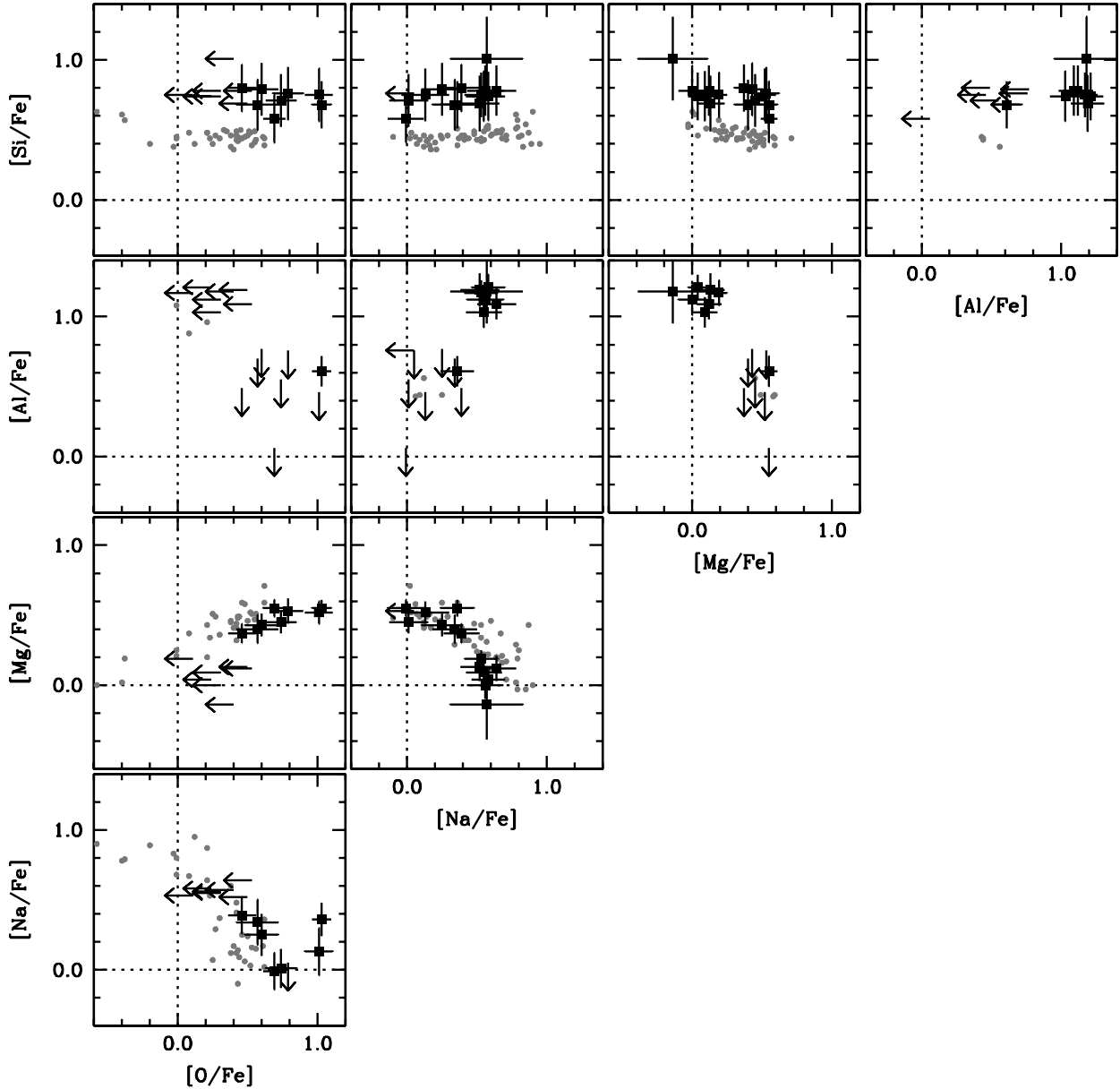}
\caption{
\label{lightplot2}
Relationships among O, Na, Mg, Al, and Si in \ngc.
Black symbols mark data from this study,
and small gray circles mark data from \citet{carretta14}.
The dotted lines indicate the solar ratios.
}
\end{figure*}

Figure~\ref{lightplot2} reveals a 
prominent anti-correlation between [O/Fe] and
[Na/Fe] in our sample of 15~stars in \ngc.
\citet{carretta14} detected O and Na in 51~stars
in their sample, and these data are shown for comparison
in Figure~\ref{lightplot2}.
The offset between our mean [O/Fe] and that of \citeauthor{carretta14}\
is apparent.
This offset is strongly influenced by two factors.
First, we only derive upper limits for [O/Fe]
in stars with [O/Fe]~$< +$0.4,
while \citeauthor{carretta14}\ detected O
in $\approx$~20 such stars.
These upper limits are not included in the mean
[O/Fe] value reported in Table~\ref{meantab}.
Second, two of the eight stars with O detections in our sample
show [O/Fe]~$\approx +$1.0,
which is $\approx$~0.2~dex higher than
the next-highest [O/Fe] ratio in any other star in our sample.
Neither of these stars, \mbox{4-224} and \mbox{4-1225},
was included in the \citeauthor{carretta14}\ sample.
The remaining stars in our sample are found to overlap the 
\citeauthor{carretta14}\ data relatively well
in the [O/Fe] versus [Na/Fe] plane.

Our data and that of \citet{carretta14}
independently exhibit a broad anti-correlation between
[O/Fe] and [Na/Fe].
This includes stars with [O/Fe] and [Na/Fe] ratios
consistent with normal halo field stars
and stars with depleted [O/Fe] and enhanced [Na/Fe] ratios.
\citeauthor{carretta14}\ calculated that 31~$\pm$~8 per cent
of stars in \ngc\ belong to the primordial component
(i.e., those with normal [O/Fe] and [Na/Fe] ratios).
Another 59~$\pm$~11 per cent 
of stars belong to the intermediate component
(i.e., those with depleted [O/Fe] ratios and enhanced [Na/Fe] ratios),
and 10~$\pm$~4 per cent of stars belong to the extreme component
(i.e., those with [O/Na]~$< -$0.9; \citealt{carretta09a}).
Our sample is more than three times smaller than that of
\citeauthor{carretta14},
so we do not attempt to rederive these fractions.
The fraction of stars in the primordial component, one-third,
is typical of most globular clusters that have been studied.
The fraction of stars in the extreme component is
larger than in most clusters, 
which have typically 2--3 per cent or less \citep{carretta09a}.

Figure~\ref{lightplot2} also reveals 
correlations between 
[O/Fe] and [Mg/Fe],
[Na/Fe] and [Al/Fe], possibly
[Na/Fe] and [Si/Fe], and possibly
[Al/Fe] and [Si/Fe].
Anti-correlations are also present among
[O/Fe] and [Al/Fe],
[Na/Fe] and [Mg/Fe], and
[Mg/Fe] and [Al/Fe]. 
These features were also present
in the sample of \citet{carretta14}.
Collectively, these abundance patterns
indicate that proton-capture reactions
operating at high temperatures occurred
in the stars that enriched 
some of the present-day members of \ngc.

\citet{carretta14}
identified a bi-modal distribution
among the Mg and Al abundances
derived from the UVES spectra.
Our data confirm this separation, which
is easily discerned in the
[O/Fe] versus [Al/Fe],
[Na/Fe] versus [Mg/Fe],
[Na/Fe] versus [Al/Fe], and
[Mg/Fe] versus [Al/Fe]
planes in Figure~\ref{lightplot2}.

Trends involving [Si/Fe] are more subtle.
\citeauthor{carretta14}\ found evidence of such trends
in their sample,
which can be seen by simple 
inspection of their Figure~8.
Similar inspection of our Figure~\ref{lightplot2} 
only hints that [Si/Fe] may correlate with 
[Na/Fe] and [Al/Fe] and anti-correlate with [Mg/Fe].
We divide our sample into two groups
to test whether these correlations are significant.
The [Mg/Fe] ratios are well-separated into two groups
divided at [Mg/Fe]~$\approx +$0.3,
and we compute the weighted mean [Si/Fe] ratio within each group.
In the Mg-rich group of stars,
$\langle$[Si/Fe]$\rangle = +$0.72~$\pm$~0.03;
in the Mg-poor group,
$\langle$[Si/Fe]$\rangle = +$0.76~$\pm$~0.04.
The [Si/Fe] ratios 
are not significantly different in the Mg-rich and Mg-poor groups of stars.

We also compute $p$-values for the
linear correlation coefficient $r$
between [Si/Fe] and other light-element ratios.
The $p$-value 
gives the probability that a random sample of
$N$ uncorrelated points would yield an experimentally-derived value
$\geq |r|$.  
For example, 
the $p$-value 
for the [Na/Fe] and [Mg/Fe] ratios
shown in Figure~\ref{lightplot2} is 0.0002.
The $p$-values between [Si/Fe] and [O/Fe],
[Na/Fe], [Mg/Fe], and [Al/Fe]
are 0.645, 0.077, 0.019, and 0.356, respectively.
Only the anti-correlation between [Mg/Fe] and [Si/Fe]
is significant at the 2$\sigma$ level in our data.

Censored (``survivor'') statistical tests
can be applied to check whether the [Si/Fe] correlations
may be significant if upper limits are included.
We apply
the generalized version of Kendall's $\tau$ rank correlation test
to our data using the \textsc{asurv} code
(Rev.\ 1; \citealt{lavalley92})
as described in \citet{isobe86}.
This test evaluates whether
a correlation may be present between two variables
when upper limits are found in either variable.
We find that
correlations among [Na/Fe], [Mg/Fe], and [Al/Fe] are all
highly significant ($p <$~0.004 in each case).
We also find that
[Si/Fe] does not exhibit a significant correlation
with either [Na/Fe] or [Al/Fe] if the upper limits
are considered
($p >$~0.12 in each case).

We are unable to reproduce correlations between
[Si/Fe] and any other ratios,
except perhaps [Mg/Fe], in our data.
Our sample size is considerably smaller than that of
\citet{carretta14},
which may explain the difference in our results.

\subsection{Potassium through Zinc}
\label{irongroup}

\begin{figure*}
\centering
\includegraphics[angle=0,width=5.0in]{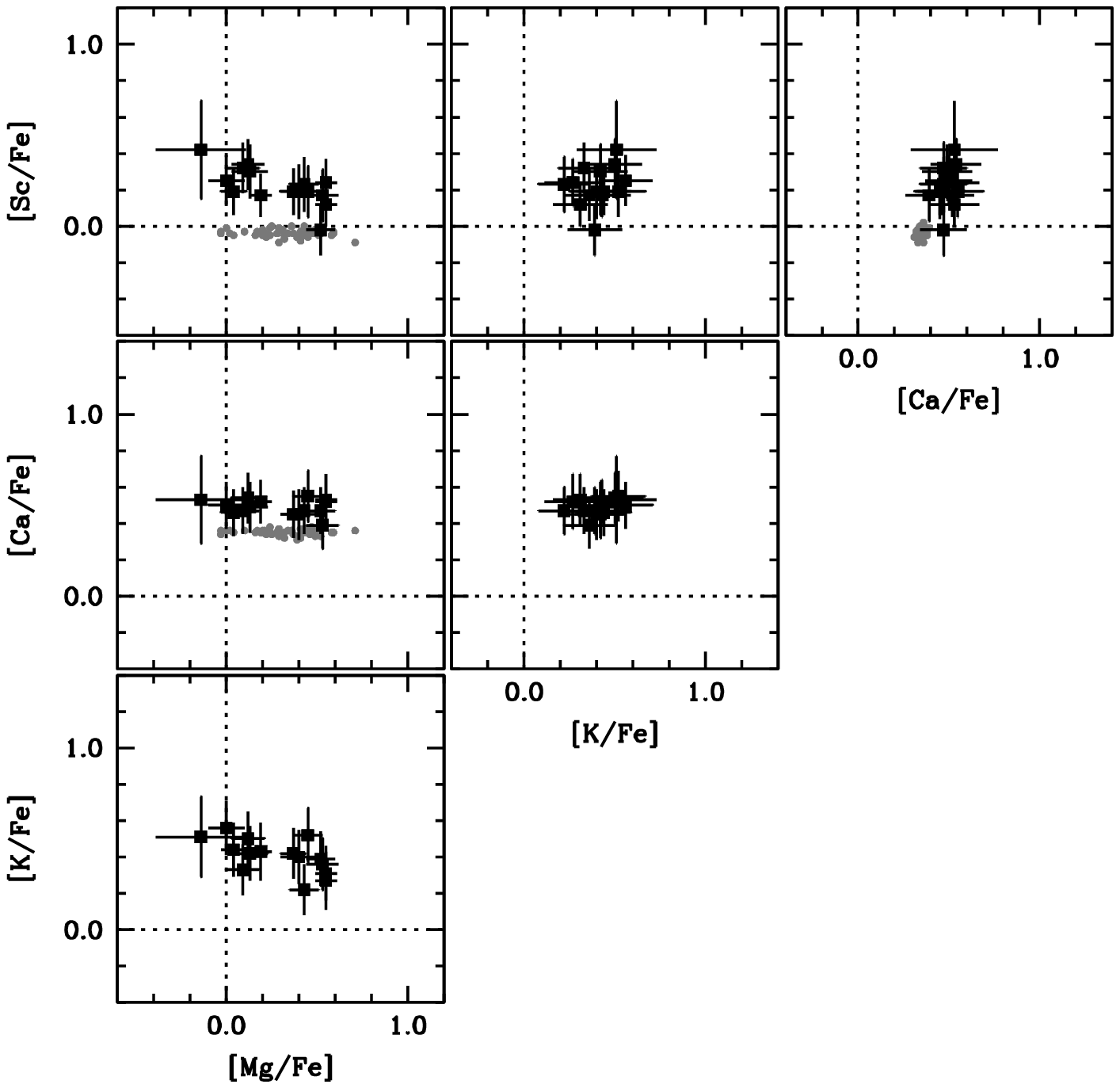}
\caption{
\label{lightplot3}
Relationships among Mg, K, Ca, and Sc in \ngc.
Black symbols mark data from this study,
and small gray circles mark data from \citet{carretta14}.
The dotted lines indicate the solar ratios.
}
\end{figure*}

Figure~\ref{lightplot3} illustrates the
relationships among [Mg/Fe], [K/Fe], [Ca/Fe], and [Sc/Fe].
Visual inspection of Figure~\ref{lightplot3} suggests that
[K/Fe] might anti-correlate with [Mg/Fe].
We calculate the mean [K/Fe] ratios
in the Mg-rich and Mg-poor groups of stars.
We find
$\langle$[K/Fe]$\rangle = +$0.36~$\pm$~0.03
for the Mg-rich group.
We find
$\langle$[K/Fe]$\rangle = +$0.45~$\pm$~0.03
for the Mg-poor group.
These ratios are only
distinct at the $\sim$~1.5$\sigma$ level,
which is not significant.
The $p$-value for the linear correlation between
[Mg/Fe] and [K/Fe] is 0.018, which is significant.
If [Mg/Fe] and [K/Fe] are anti-correlated,
we might also expect correlations between
[K/Fe] and [Na/Fe], [Al/Fe], or [Si/Fe];
however, 
the $p$-values for the correlations between [K/Fe]
and other light elements are not significant at the
2$\sigma$ level.
We conclude that 
the evidence for an anti-correlation between
[Mg/Fe] and [K/Fe] is curious but not compelling.

Hints of
an anti-correlation 
between [Mg/Fe] and [Sc/Fe] are 
seen in Figure~\ref{lightplot3}, but a
closer analysis suggests otherwise.
Figure~\ref{teffplot1} demonstrates that
the three lowest [Sc/Fe] ratios are found in 
three of the four coolest stars in our sample,
signaling that these low ratios 
could be an artifact of the analysis.
This correlation appears weaker
if only the 11~stars with \teff~$>$~4500~K
are considered.
When dividing the sample of all 15~stars
into Mg-rich and Mg-poor groups,
we find 
$\langle$[Sc/Fe]$\rangle = +$0.16~$\pm$~0.03
in the Mg-rich group and
$\langle$[Sc/Fe]$\rangle = +$0.25~$\pm$~0.04
in the Mg-poor group.
If only the 11~stars with \teff~$>$~4500~K
are considered,
we find
$\langle$[Sc/Fe]$\rangle = +$0.19~$\pm$~0.01
in the Mg-rich group and
$\langle$[Sc/Fe]$\rangle = +$0.27~$\pm$~0.04
in the Mg-poor group.
The [Sc/Fe] ratios
are not distinct at the 
2$\sigma$ level.
We also conduct a line-by-line differential analysis
of the [Sc/Fe] ratios 
in two stars 
(\mbox{2-185} and \mbox{2-918})
with similar stellar parameters
(\teff~$=$~4670~K and 4696~K, respectively)
and different [Mg/Fe] ratios
([Mg/Fe]~$= +$0.09 and $+$0.40, respectively).
An unweighted average of these differentials
yields $\delta$[Sc/Fe]~$= +$0.018~$\pm$~0.043~dex,
which indicates no difference.
Finally, we note that \citet{carretta14} found
no evidence for an anti-correlation between
[Mg/Fe] and [Sc/Fe] in their data,
which are also shown in Figure~\ref{lightplot2}.
We conclude that 
there is no compelling evidence
for variations in the [Sc/Fe] ratios 
within \ngc.

\citet{cohen11b}, 
\citet{cohen12}, and 
\citet{mucciarelli12}
detected an
anti-correlation between [Mg/Fe] and [K/Fe] in
the massive, metal-poor globular cluster \mbox{NGC~2419}.
\citet{mucciarelli15}
found a smaller, but significant, anti-correlation
between [Mg/Fe] and [K/Fe] in \mbox{NGC~2808}.
These two massive clusters are the only ones
where such relations have been identified
\citep{carretta13}.
Our results indicate that such extreme
variations in K do not occur in \ngc.
\citet{cohen12} also detected an anti-correlation
between [Mg/Fe] and [Sc/Fe] in their \mbox{NGC~2419} data,
and there is no compelling evidence for 
a similar relation in \ngc.

\begin{figure*}
\centering
\includegraphics[angle=270,width=6.5in]{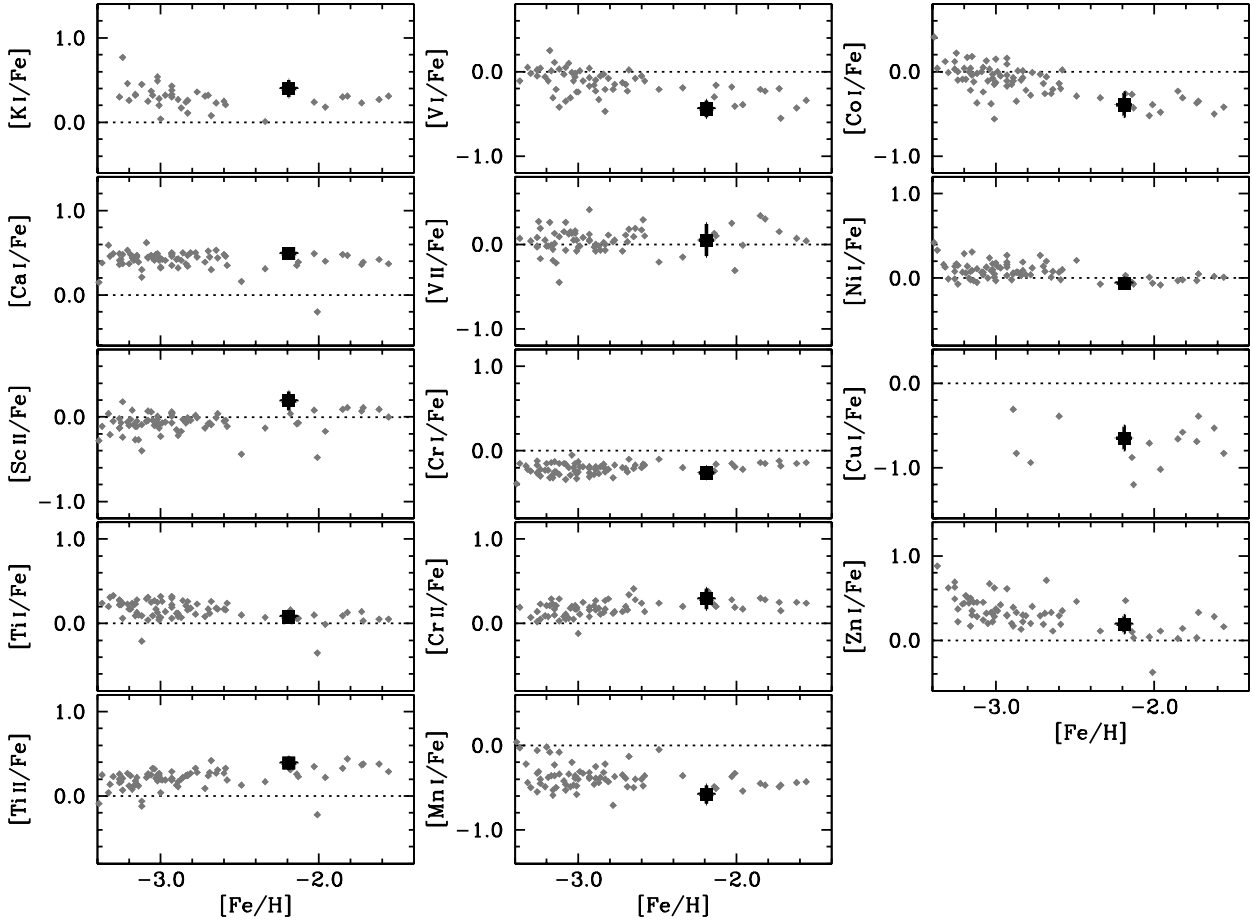}
\caption{
\label{irongroupplot}
[X/Fe] ratios for elements, X, in the iron group.
The large black square indicates the mean 
abundances in \ngc, and the small
gray diamonds represent individual red giant halo field
stars analyzed by \citet{roederer14}.
The dotted lines indicate the solar ratios.
}
\end{figure*}

Figure~\ref{irongroupplot}
compares the mean [X/Fe] ratios for elements 
in and near the iron group (loosely considered
here as K through Zn; 19~$\leq Z \leq$~30)
in \ngc\ and normal metal-poor halo field stars.
The comparison stars are drawn from the 
sample of 98~red giants examined by \citet{roederer14}.
Upper limits in the comparison sample have been omitted
from Figure~\ref{irongroupplot} for clarity.
Overall, there is excellent agreement between
the mean [X/Fe] ratios for elements in the iron group
in \ngc\ and normal halo field stars.

\citet{gratton04} pointed out that the [Ca/Fe] 
in two stars in \ngc\
reported by \citet{gratton89} 
is higher by $\approx$~0.2--0.3~dex than 
the [Ca/Fe] ratio in other metal-poor globular clusters.
The \citeauthor{gratton89}\ result was already at odds with the 
data of Pilachowski et al.\ (\citeyear{pilachowski83}).
Neither \citet{carretta14} nor we reproduce
this high [Ca/Fe] ratio in \ngc.

\subsection{Neutron-capture elements}
\label{ncap}

\begin{figure*}
\centering
\includegraphics[angle=0,width=2.2in]{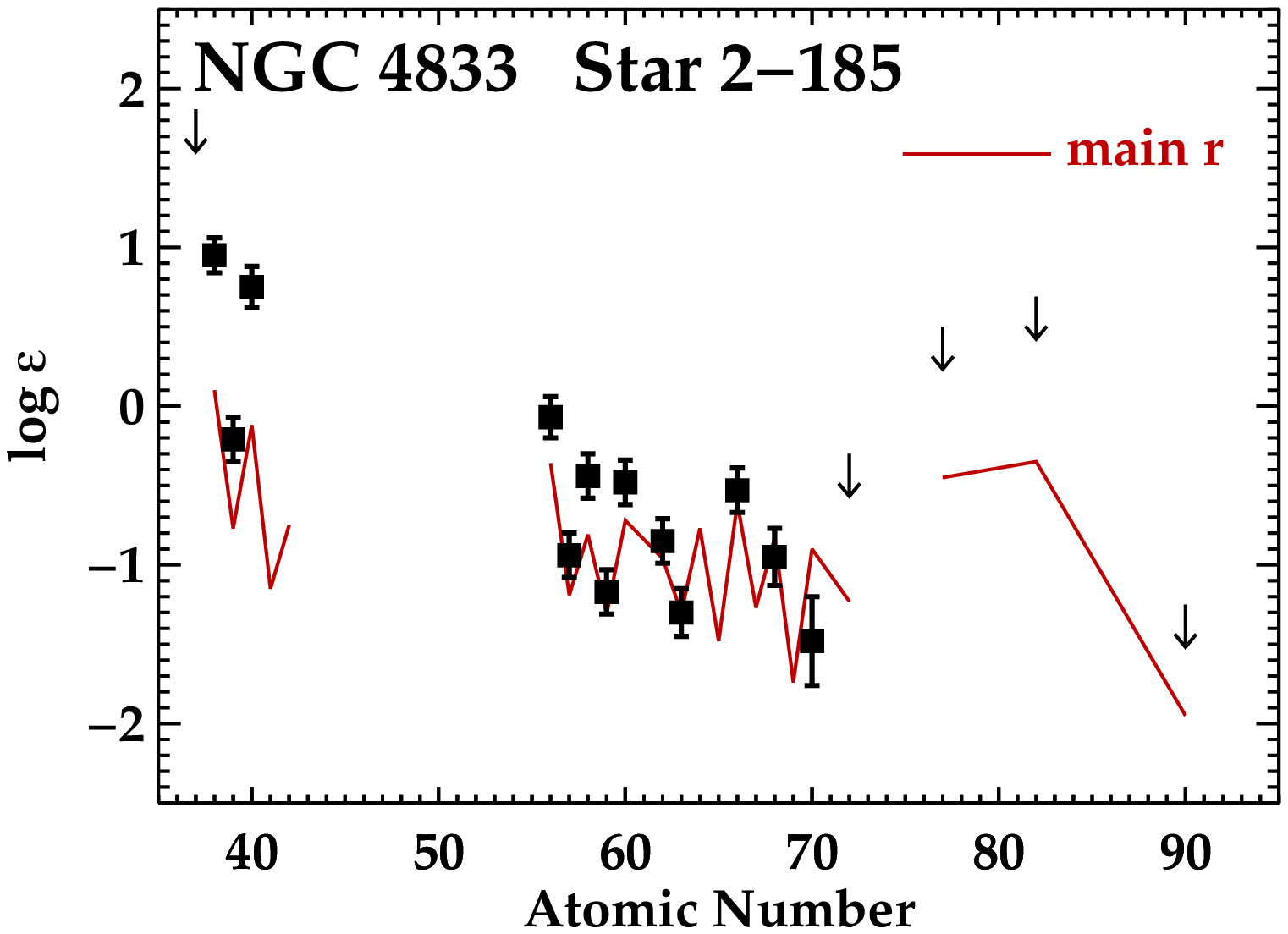}
\hspace*{0.1in}
\includegraphics[angle=0,width=2.2in]{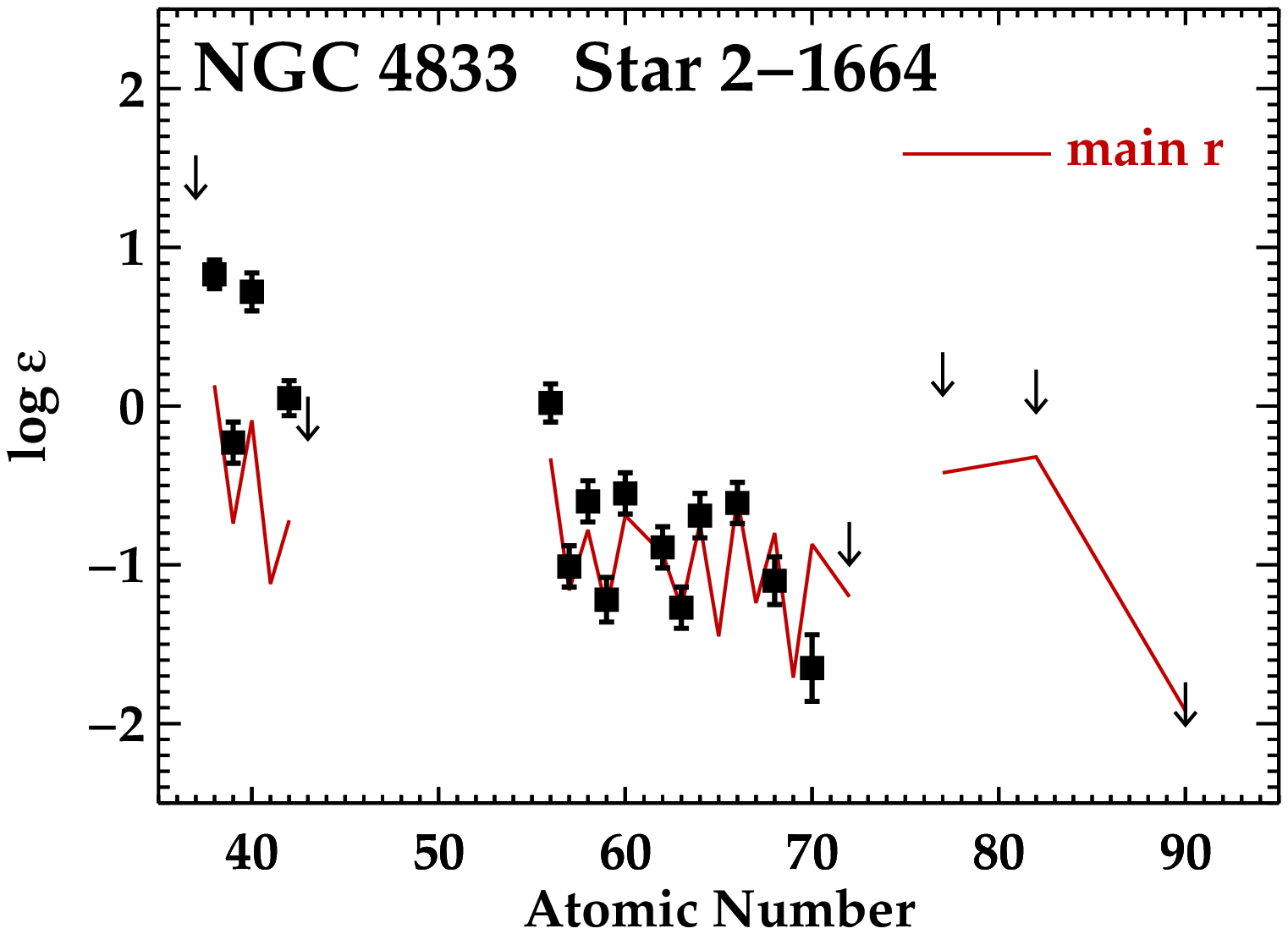}
\hspace*{0.1in}
\includegraphics[angle=0,width=2.2in]{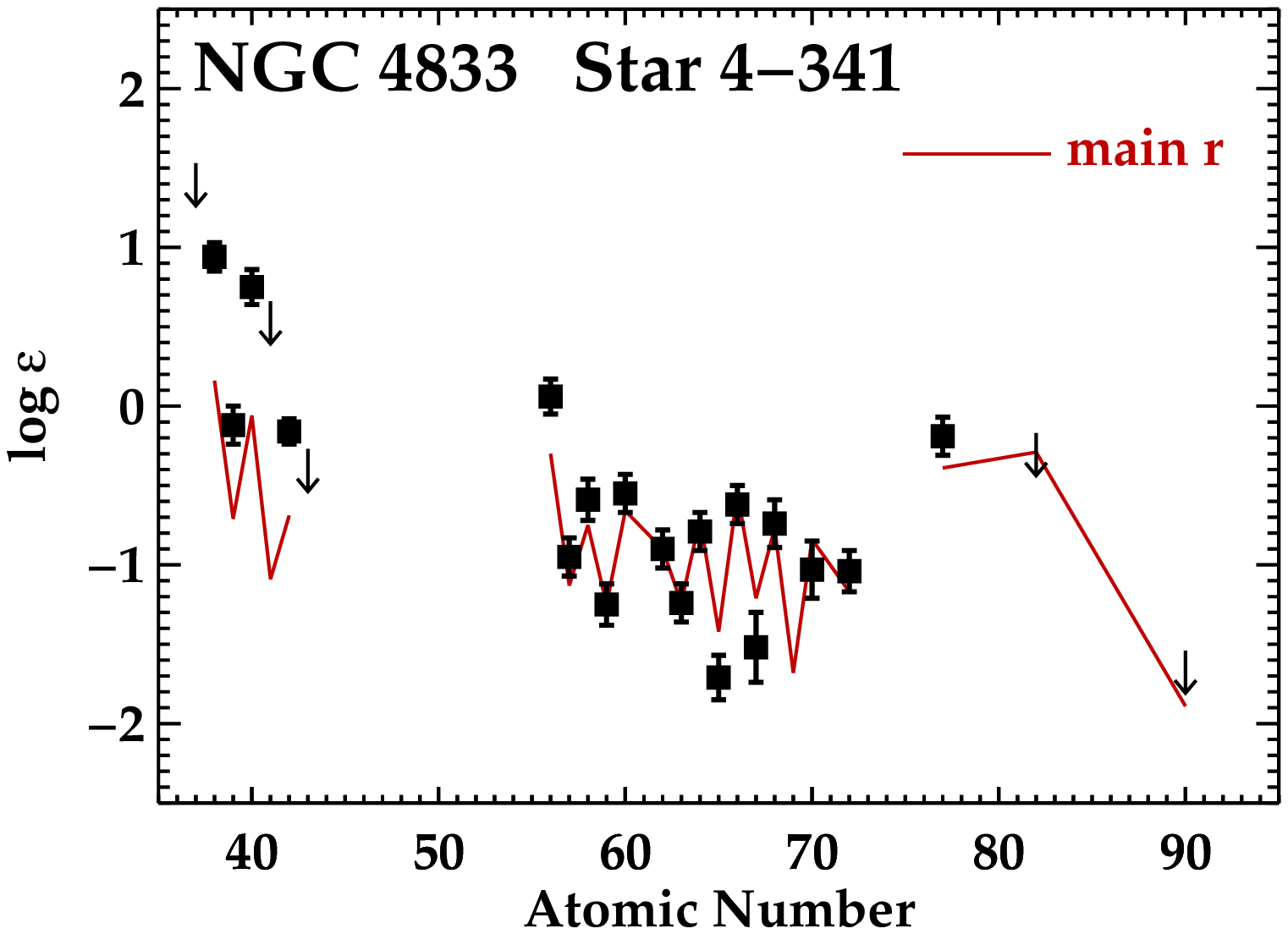} \\
\vspace*{0.1in}
\includegraphics[angle=0,width=2.2in]{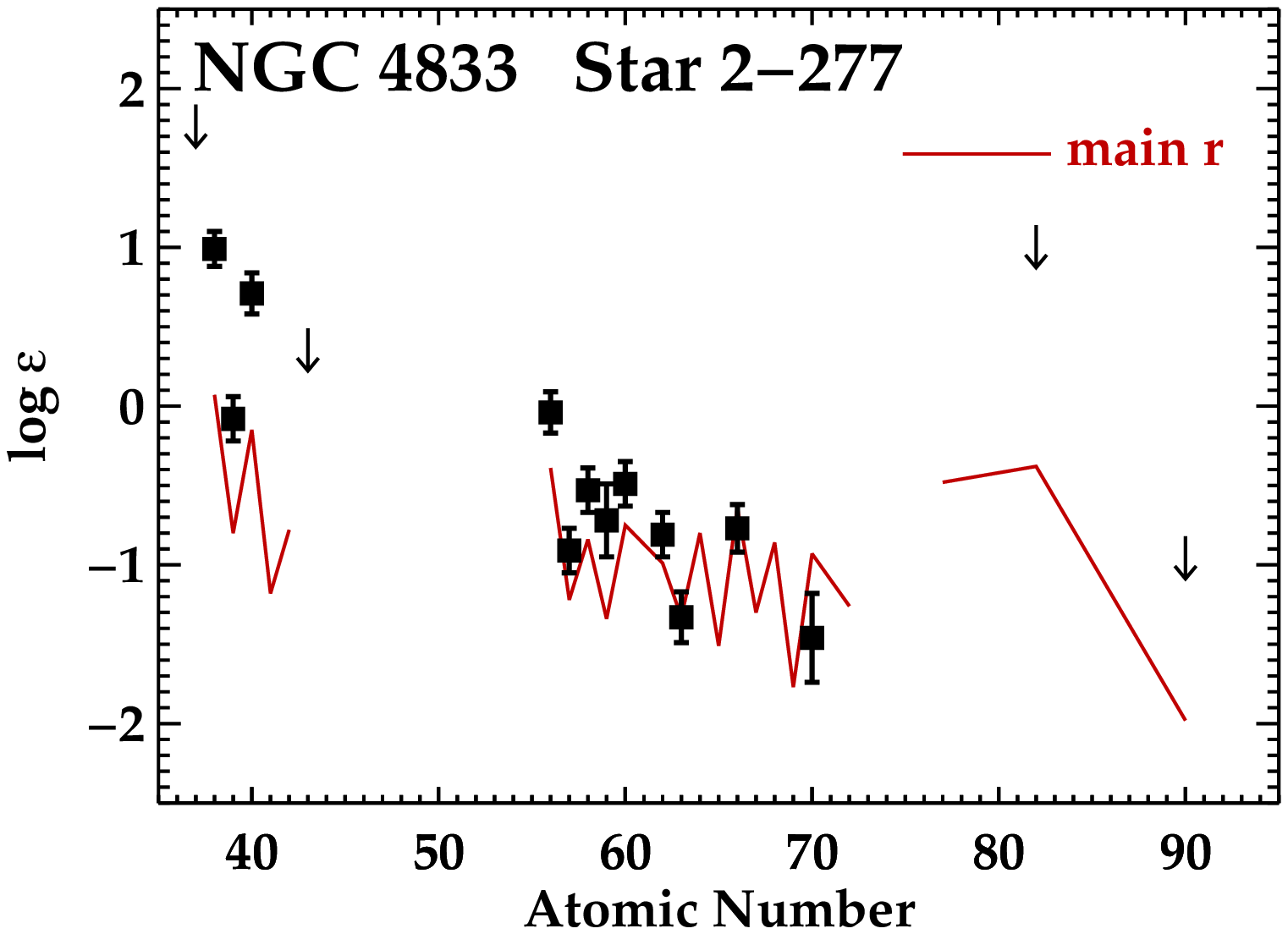}
\hspace*{0.1in}
\includegraphics[angle=0,width=2.2in]{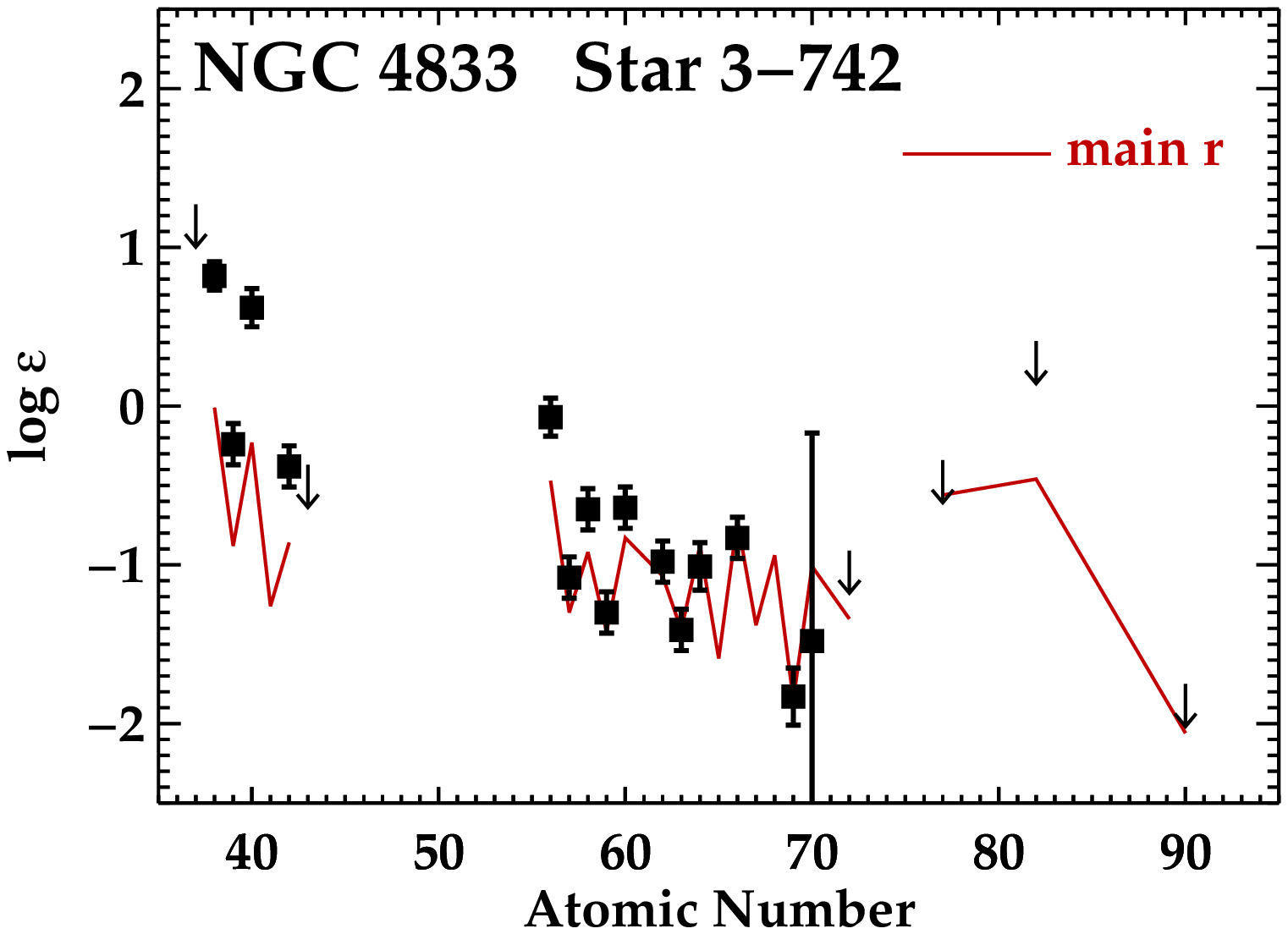}
\hspace*{0.1in}
\includegraphics[angle=0,width=2.2in]{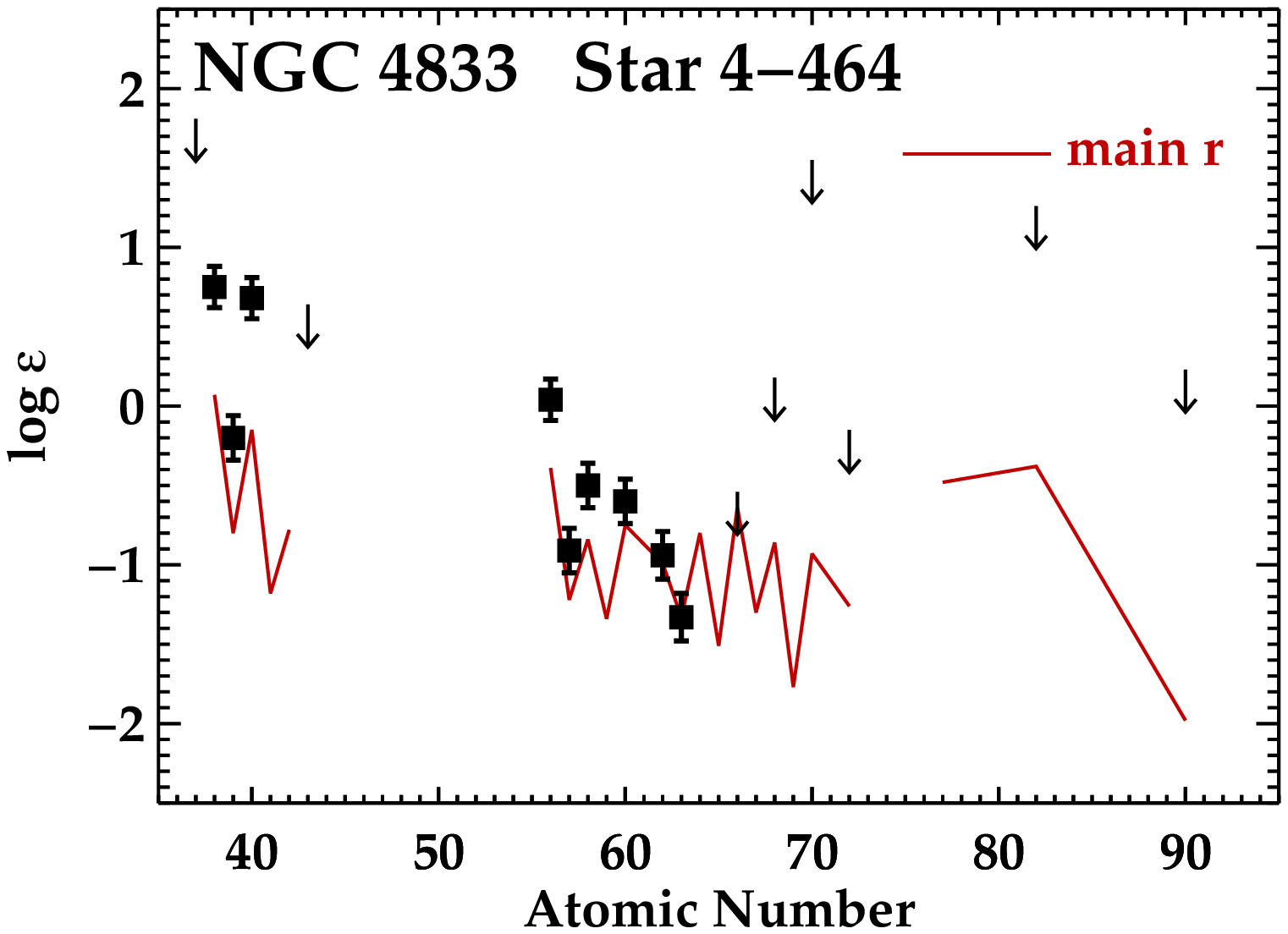} \\
\vspace*{0.1in}
\includegraphics[angle=0,width=2.2in]{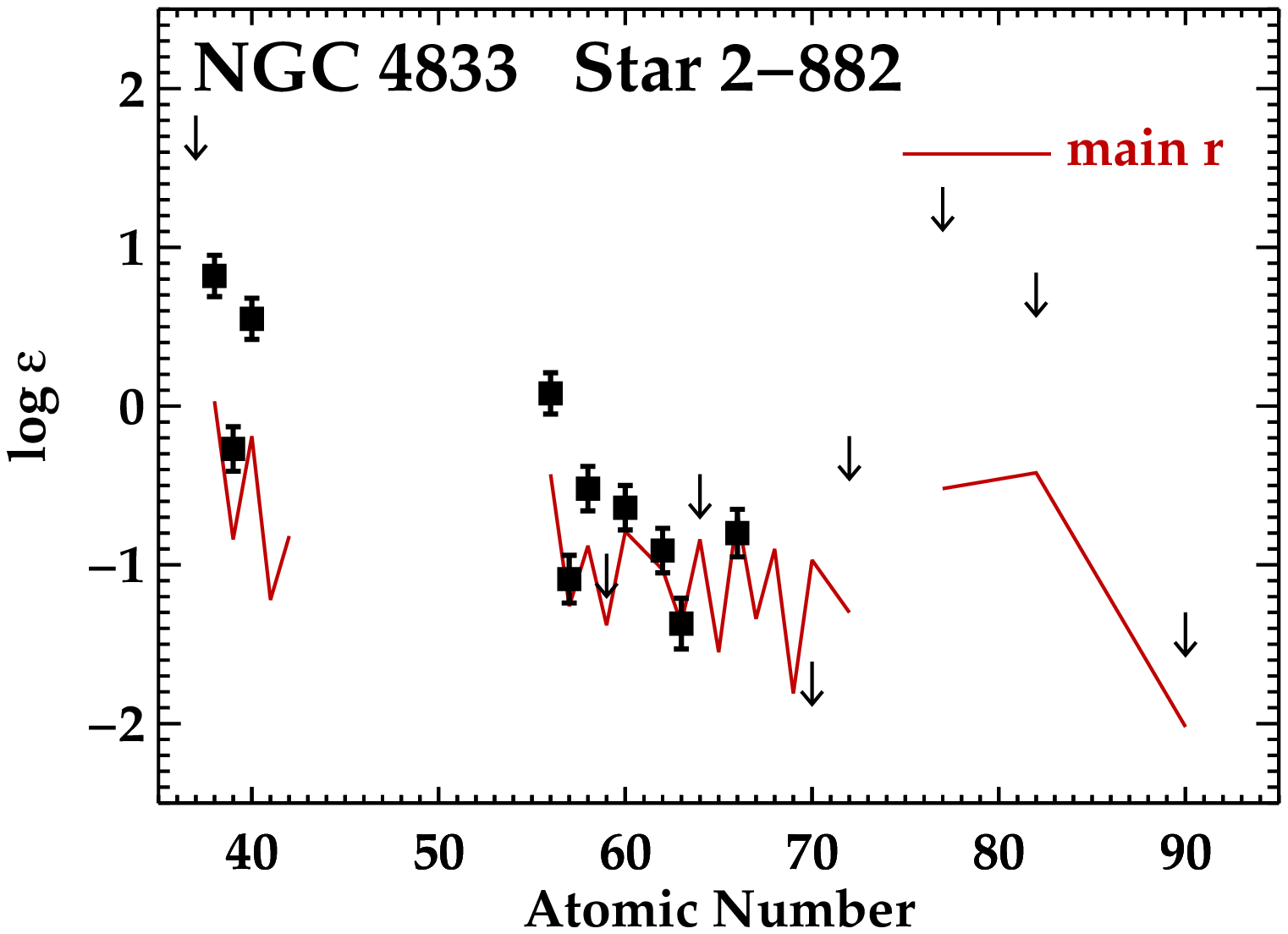}
\hspace*{0.1in}
\includegraphics[angle=0,width=2.2in]{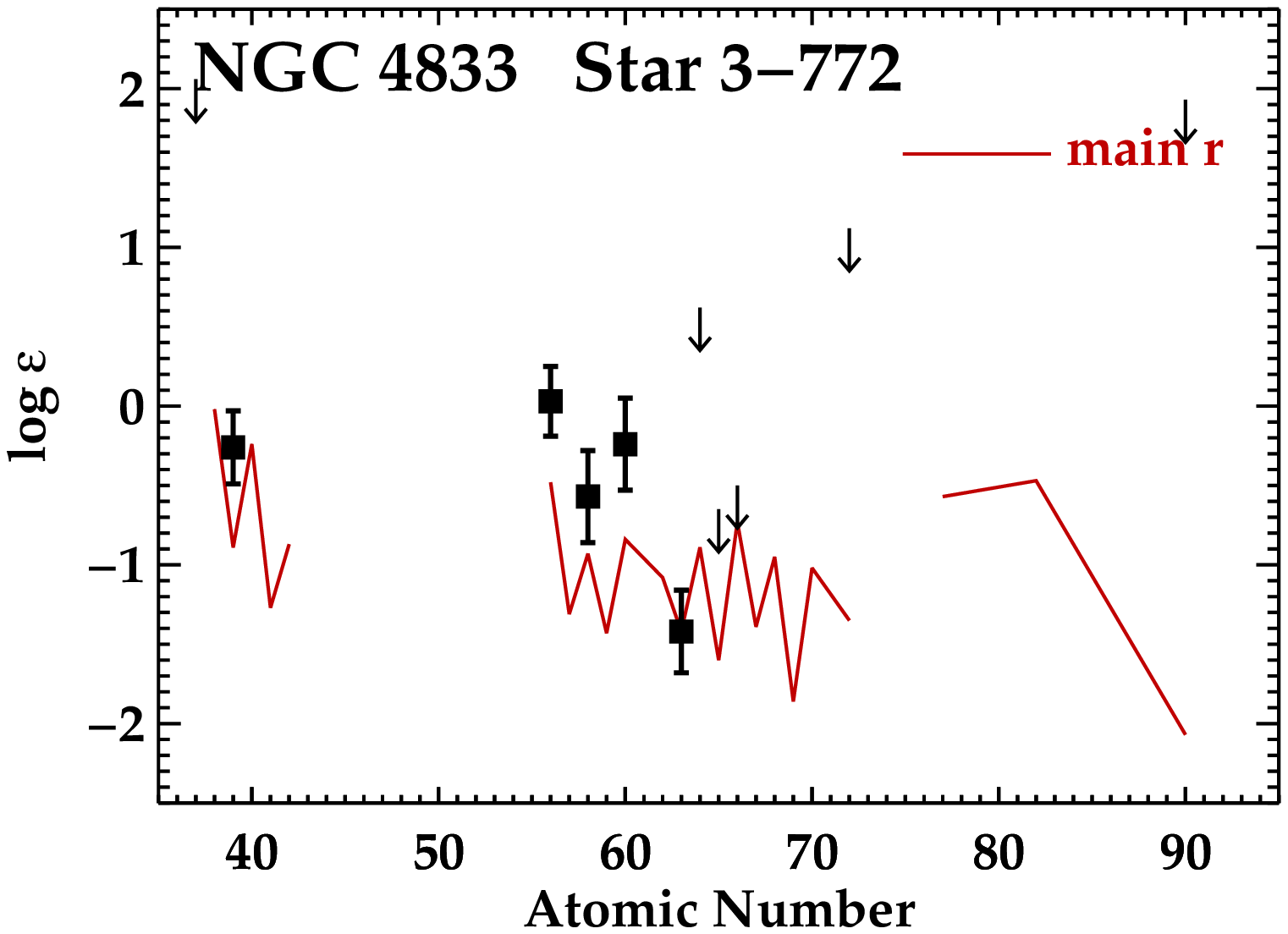}
\hspace*{0.1in}
\includegraphics[angle=0,width=2.2in]{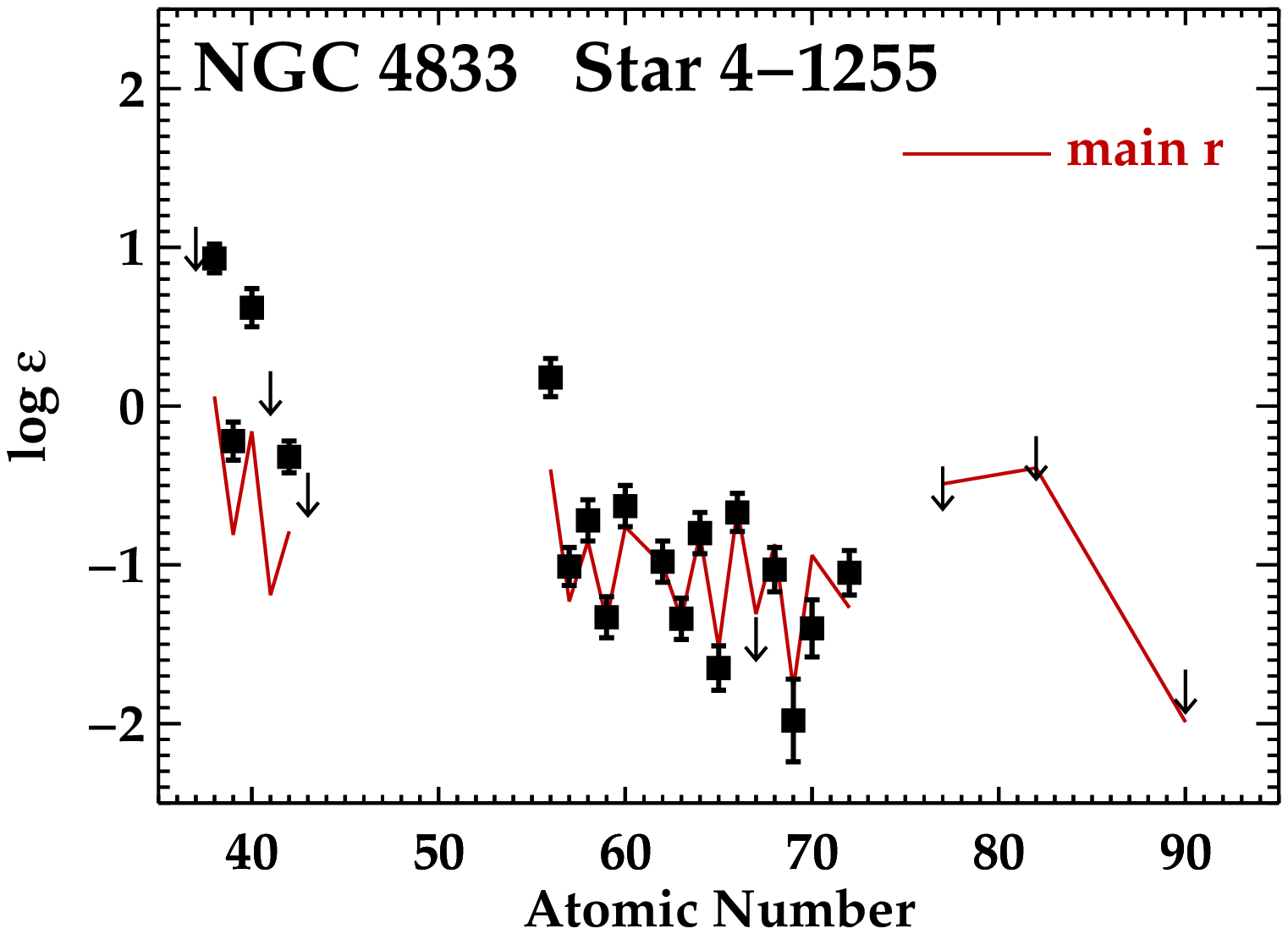} \\
\vspace*{0.1in}
\includegraphics[angle=0,width=2.2in]{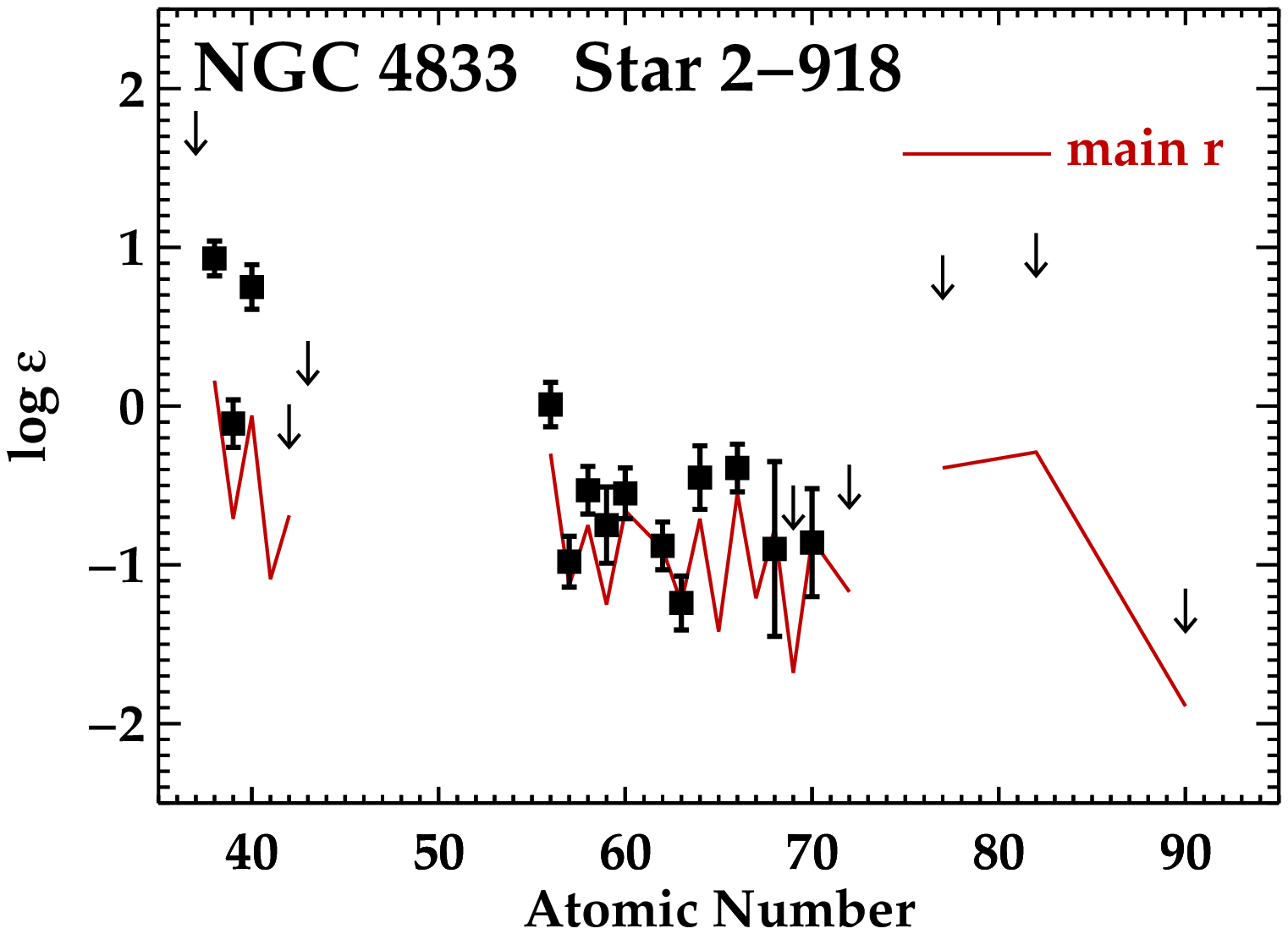}
\hspace*{0.1in}
\includegraphics[angle=0,width=2.2in]{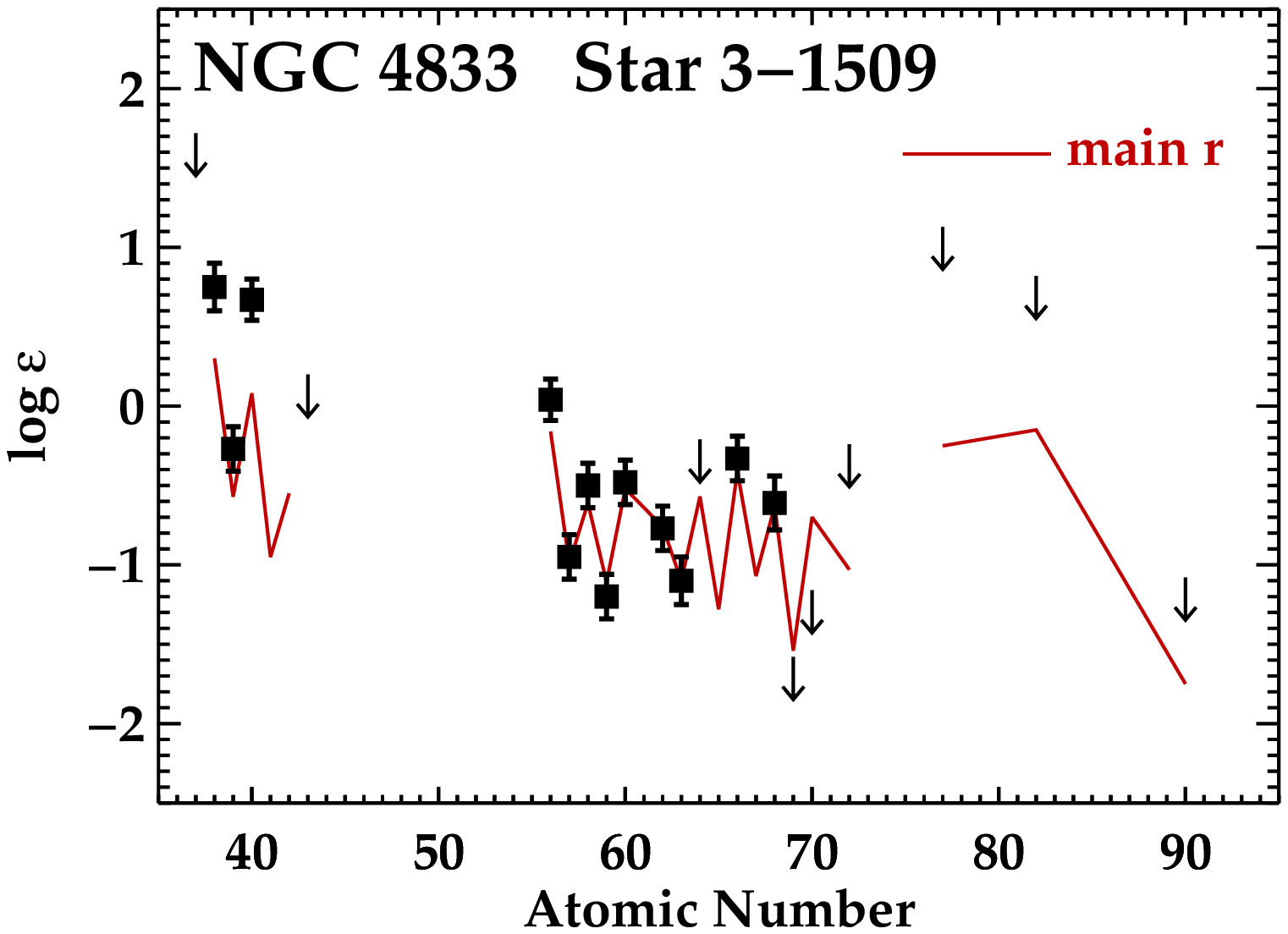}
\hspace*{0.1in}
\includegraphics[angle=0,width=2.2in]{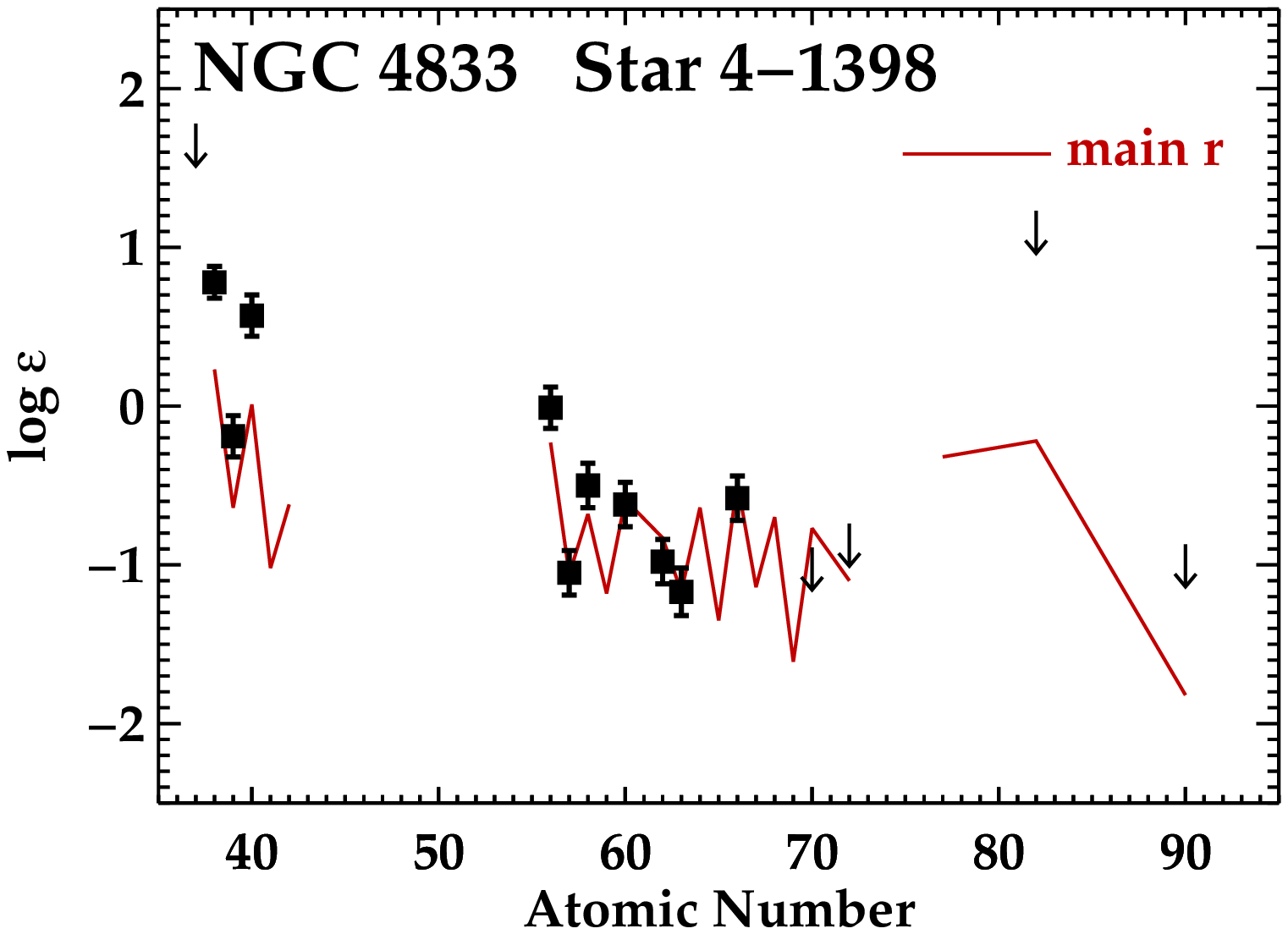} \\
\vspace*{0.1in}
\includegraphics[angle=0,width=2.2in]{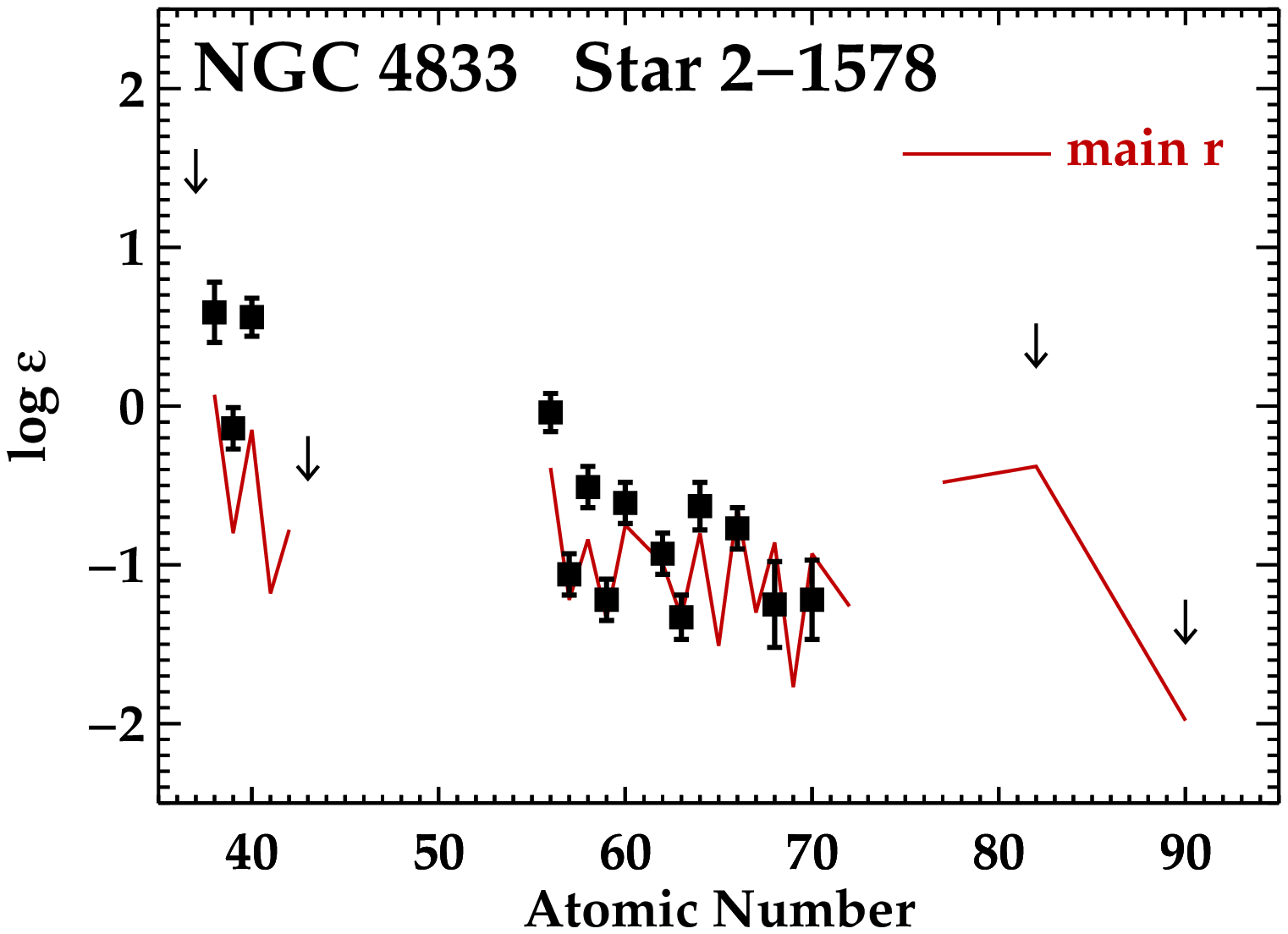}
\hspace*{0.1in}
\includegraphics[angle=0,width=2.2in]{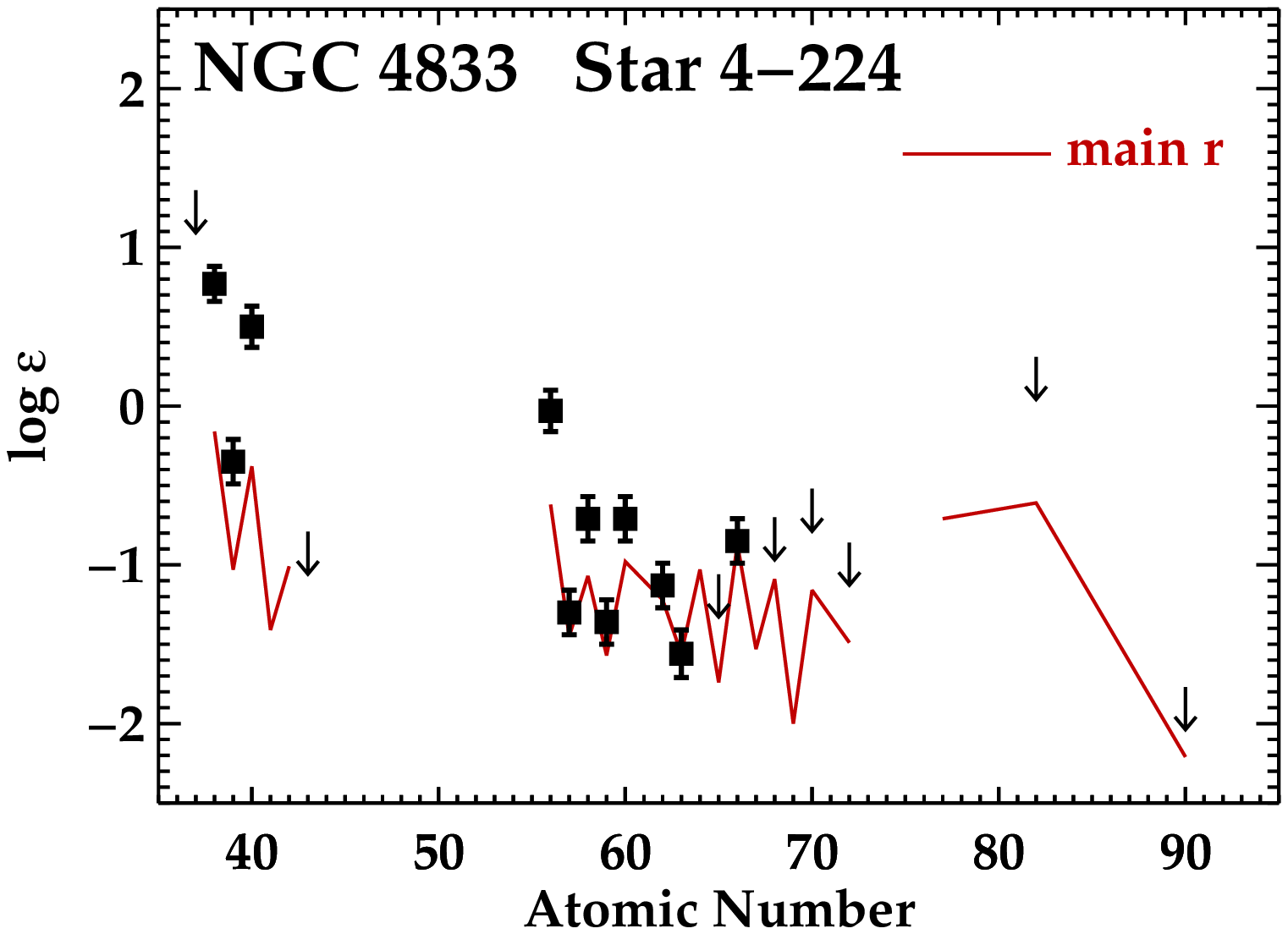}
\hspace*{0.1in}
\includegraphics[angle=0,width=2.2in]{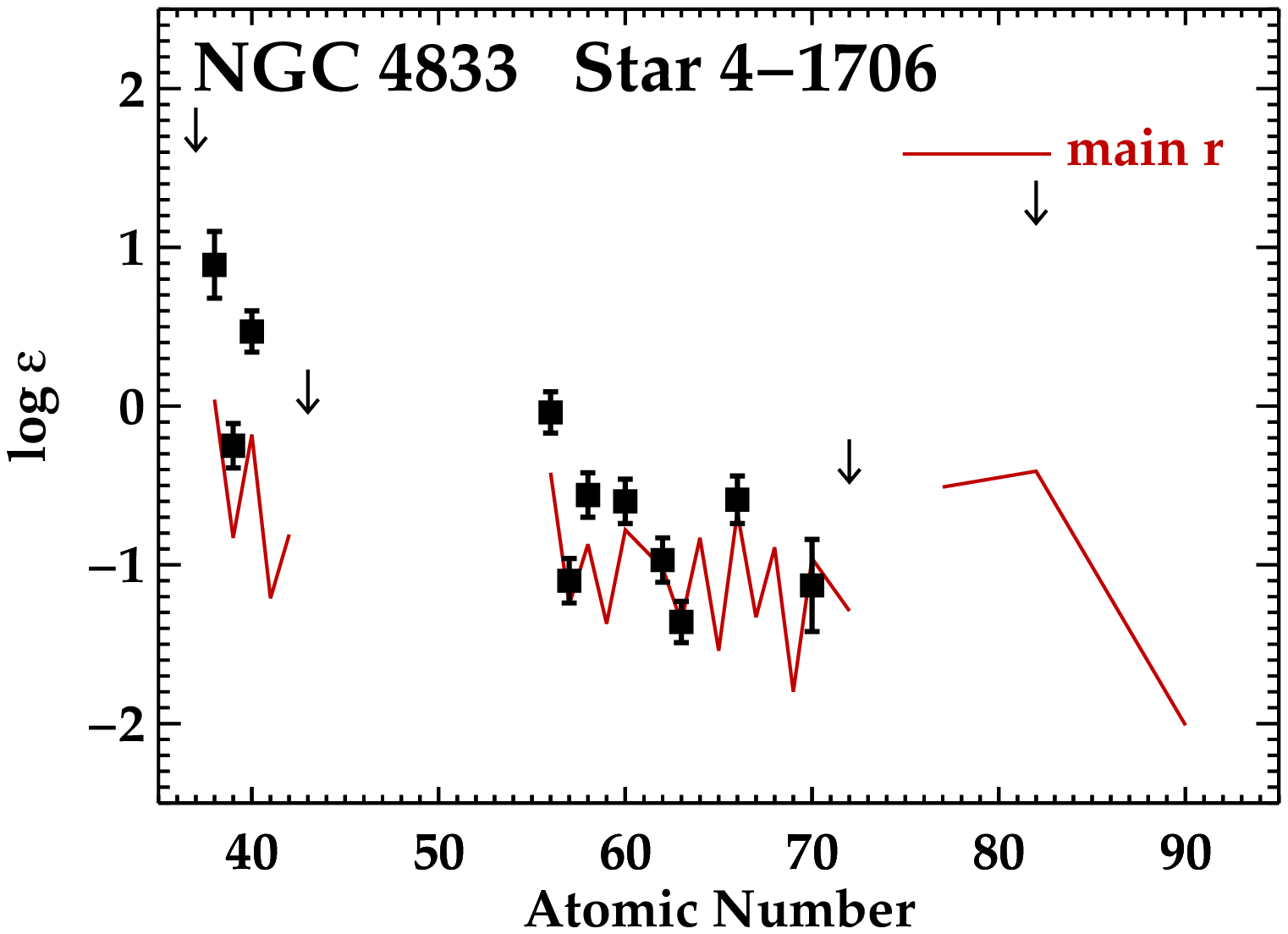} 
\caption{
\label{ncapplots}
Heavy element abundance pattern for each star observed in \ngc.
The red lines mark the template of the main component
of the \rpro\ (\mbox{CS~22892--052}).
Filled symbols mark detections, and downward-facing triangles
mark 3$\sigma$ upper limits.
}
\end{figure*}

Figure~\ref{ncapplots} illustrates the heavy-element
abundance distributions in each star observed in \ngc.
For comparison,
Figure~\ref{ncapplots} also shows
the abundance distribution of the
main component of the \rpro\
as found 
in the metal-poor giant \mbox{CS~22892--052}
\citep{sneden03,sneden09,roederer09}.
There is generally good agreement among the heavy-element
abundance distribution from one star to the next in \ngc.
Large variations in the heavy-element 
distributions, like those found in 
$\omega$~Cen (e.g., \citealt{norris95,smith00}),
M2 \citep{yong14},
M22 \citep{marino09,roederer11c}, or
\mbox{NGC~1851} \citep{yong08,carretta11}
are excluded.

\begin{figure}
\centering
\includegraphics[angle=0,width=3.3in]{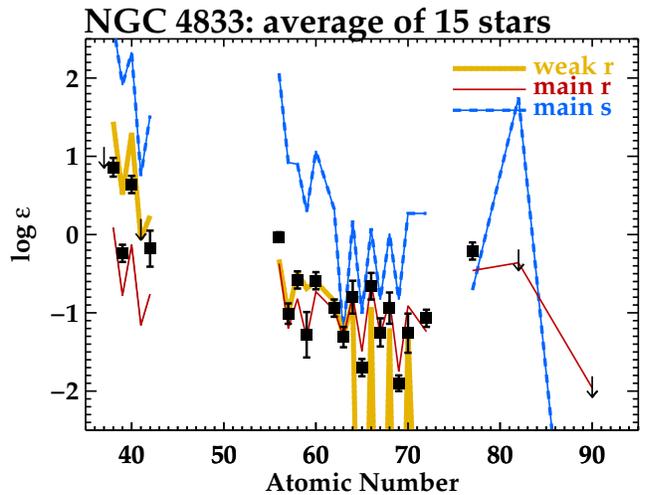}
\caption{
\label{ncapavgplot}
Average heavy element abundance pattern in \ngc.
Filled squares mark detections, and arrows mark 3$\sigma$ upper limits.
The red line, 
thick gold line, and long-dashed blue line
represent the main component of the \rpro,
the weak component of the \rpro, and the
main and strong components of the \spro\
as described in the text.
Each of the three curves has been renormalized
to the mean Eu abundance in \ngc.
}
\end{figure}

Figure~\ref{ncapavgplot} shows the mean heavy element abundance
distribution for all 15~stars observed in \ngc.
Three templates are shown for comparison.
The red line marks the main component of the \rpro,
and this line is identical to that found in Figure~\ref{ncapplots}.
The 
thick gold line marks the 
distribution found in the metal-poor giant \mbox{HD~122563}
\citep{honda06,roederer12b},
frequently referred to as the distribution
produced by the weak component of the \rpro.
The long-dashed blue line marks the 
distribution predicted by the main and strong components
of the \spro\ \citep{sneden08,bisterzo11}.
These curves 
illustrate the general characteristics of
$r$- and \spro\ nucleosynthesis
and are not intended to be rigid representations.
For example,
the abundance distributions that result from
\spro\ nucleosynthesis 
depend on the parameters of stars passing through the
thermally-pulsing asymptotic giant branch phase of evolution
(e.g., \citealt{busso99,bisterzo10}).
The mean heavy element abundance distribution in \ngc\
traces the main component of the \rpro\ 
within $\approx$~2$\sigma$ for the elements
with $Z \geq$~57.
Ba ($Z =$~56) is enhanced by $\approx$~0.3~dex
relative to the template for the main component of the \rpro\ when
normalized to Eu.
Sr, Y, Zr, and Mo ($Z =$~38, 39, 40, and 42, respectively)
are enhanced by $\approx$~0.9, 0.5, 0.9, and 0.6~dex.

One possible explanation of the 
Sr, Y, Zr, Mo, and Ba enhancements could
be a small contribution from some form of \spro\ nucleosynthesis.
If so, the results of \citet{roederer10} 
suggest that we might expect enhancement of 
other elements with a substantial \spro\ component
in solar system material,
such as Yb, Hf, and Pb ($Z =$~70, 72, and 82).
No such enhancements are observed.
Another possible explanation is that the
weak component of the \rpro\ has contributed 
to the heavy elements in \ngc.
Halo field stars, like \mbox{HD~122563}, whose heavy elements
have been produced by this mechanism do not typically
show Ir ($Z =$~77), which is detected in \ngc.
Our data suggest that
the most likely explanation
for the heavy element abundance distribution
in \ngc\
is a combination of the main and weak components 
of the \rpro.
A single \rpro\ with physical characteristics
intermediate between those of the main and weak
components of the \rpro\
could also be responsible.

\begin{figure}
\centering
\includegraphics[angle=0,width=3.3in]{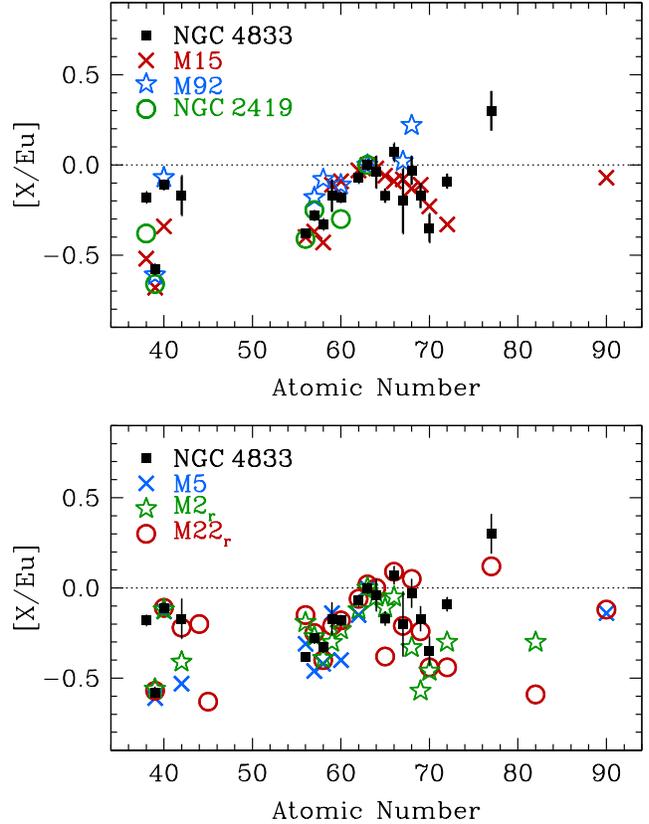}
\caption{
\label{ncapgcplot}
Average heavy element abundance pattern in \ngc\
and six other metal-poor globular clusters.
All ratios have been normalized to Eu.
The three comparison clusters shown in the top panel have
$\langle$[Fe/H]$\rangle < -$2.0, 
and the three comparison clusters
shown in the bottom panel have 
$\langle$[Fe/H]$\rangle > -$2.0.
References are given in the text.
}
\end{figure}

Figure~\ref{ncapgcplot} compares the heavy element
abundance distribution in \ngc\ with six other
metal-poor globular clusters.
These clusters are selected for comparison
because large numbers of heavy elements have been
studied in multiple red giants within each cluster.
The top panel of Figure~\ref{ncapgcplot} 
compares \ngc\ with three clusters with [Fe/H]~$< -$2.0:\
M15 ([Fe/H]~$= -$2.53, RGB stars only; \citealt{sobeck11}),
M92 ([Fe/H]~$= -$2.70; \citealt{roederer11b}),
and
\mbox{NGC~2419} ([Fe/H]~$= -$2.06; Cohen et al.\ \citeyear{cohen11b}).
The bottom panel of Figure~\ref{ncapgcplot} 
compares \ngc\ with three other clusters with [Fe/H]~$> -$2.0:\
M2  ([Fe/H]~$= -$1.68, $r$-only group of stars; \citealt{yong14}),
M5  ([Fe/H]~$= -$1.40, RGB stars only; \citealt{lai11}),
and
M22 ([Fe/H]~$= -$1.81, $r$-only group of stars; 
Roederer et al.\ \citeyear{roederer11c}).
The distributions are normalized to the Eu abundance
to eliminate differences in the overall amount of
heavy elements from one cluster to another.
Regardless, these differences are small:\
$\langle$[Eu/Fe]$\rangle = +$0.36 (\ngc),
$+$0.55 (M15),
$+$0.54 (M92),
$+$0.38 (M2),
$+$0.46 (M5), 
$+$0.35 (M22), and
$+$0.30 (\mbox{NGC~2419}).

A few general trends emerge from this comparison.
First, 
the [X/Eu] ratios are consistently sub-solar
when ``X'' represents the lighter \ncap\ elements.
The [Y/Eu] and [Zr/Eu] ratios are similar
in all of the
clusters where they have been examined.
Small enhancements 
in Sr, Y, Zr, Mo, and Ba
relative to the main component of the \rpro\
are not unusual in metal-poor globular clusters.
Second, there is a consistent upward trend 
in the [X/Eu] ratios when ``X'' represents
Ba through Sm.
Third, there is a consistent 
downward trend in [X/Eu] when ``X'' represents
Gd through Yb.
In summary,
the heavy element distribution in \ngc\ 
closely resembles
that in other well-studied metal-poor globular clusters.

\subsection{Heavy Element Dispersion in NGC~4833}
\label{dispersion}

In many globular clusters,
the heavy elements appear to be homogeneous
at the limit of observations.
M15 is unique among all known clusters
in that 
(1) it has a large range
(spanning $>$~0.7~dex, \citealt{sneden97,worley13})
of heavy-element abundances relative to Fe,
(2) the star-to-star abundance distribution
remains relatively constant and 
appears to have originated predominantly 
via \rpro\ nucleosynthesis, and
(3) the heavy-element dispersion shows
no correlation with the light element dispersion.
These characteristics of M15 have been
verified by 
\citet{sneden97,sneden00}, \citet{preston06},
\citet{otsuki06}, \citet{sobeck11}, and \citet{worley13}.

\citet{roederer11a} reanalyzed literature data
and presented evidence 
that several other metal-poor clusters 
(M3, M5, M13, and \mbox{NGC~3201})
may also exhibit a less-extreme \rpro\ dispersion.
\citet{roederer11b} reported
an \rpro\ dispersion in M92, which was not
found in the higher-quality data
obtained by \citet{cohen11a}.
The [La/Fe] and [Eu/Fe] ratios showed a high degree
of correlation for stars within each of these clusters.
This would not be expected if the La and Eu abundances
were independent measures of a sample 
with no cosmic dispersion in La or Eu.
The [La/Eu] ratios were constant, indicating
that the distributions were constant
and only the total amount of material
(relative to, e.g., H or Fe) was changing.
In principle, errors in the choice of model atmosphere could also
produce the observed correlation since the 
La~\textsc{ii} and Eu~\textsc{ii} lines
form similarly.
This possibility was dismissed because no correlation
was found among other species 
that might also form similarly, including
Sc~\textsc{ii} and Ti~\textsc{ii}.
In these clusters,
the \rpro\ dispersion did not correlate with the
[Na/Fe] ratio, which was chosen to represent
the light-element dispersion commonly found in globular clusters.

\begin{figure*}
\centering
\includegraphics[angle=0,width=3.4in]{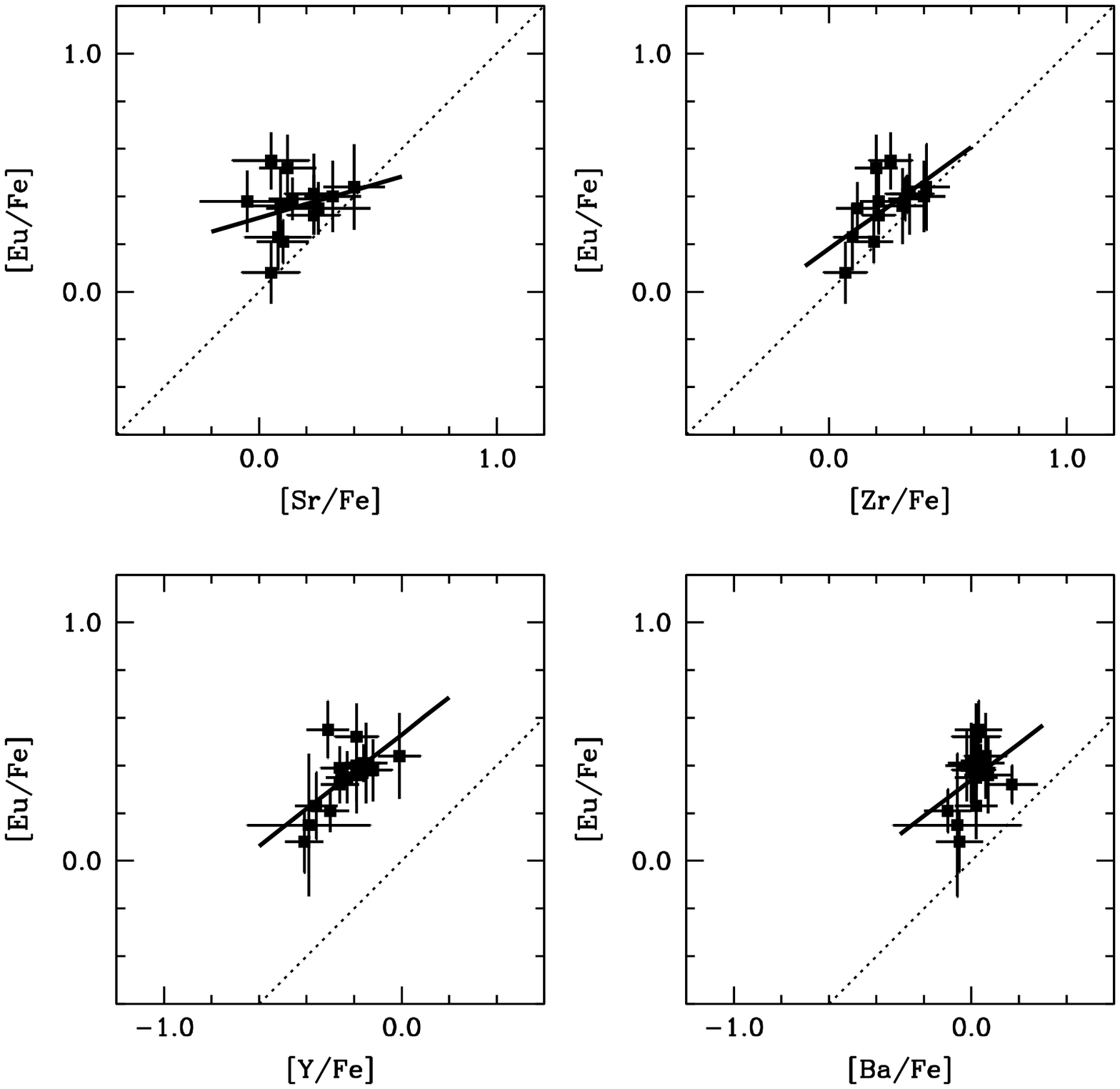}
\hspace*{0.0in}
\includegraphics[angle=0,width=3.4in]{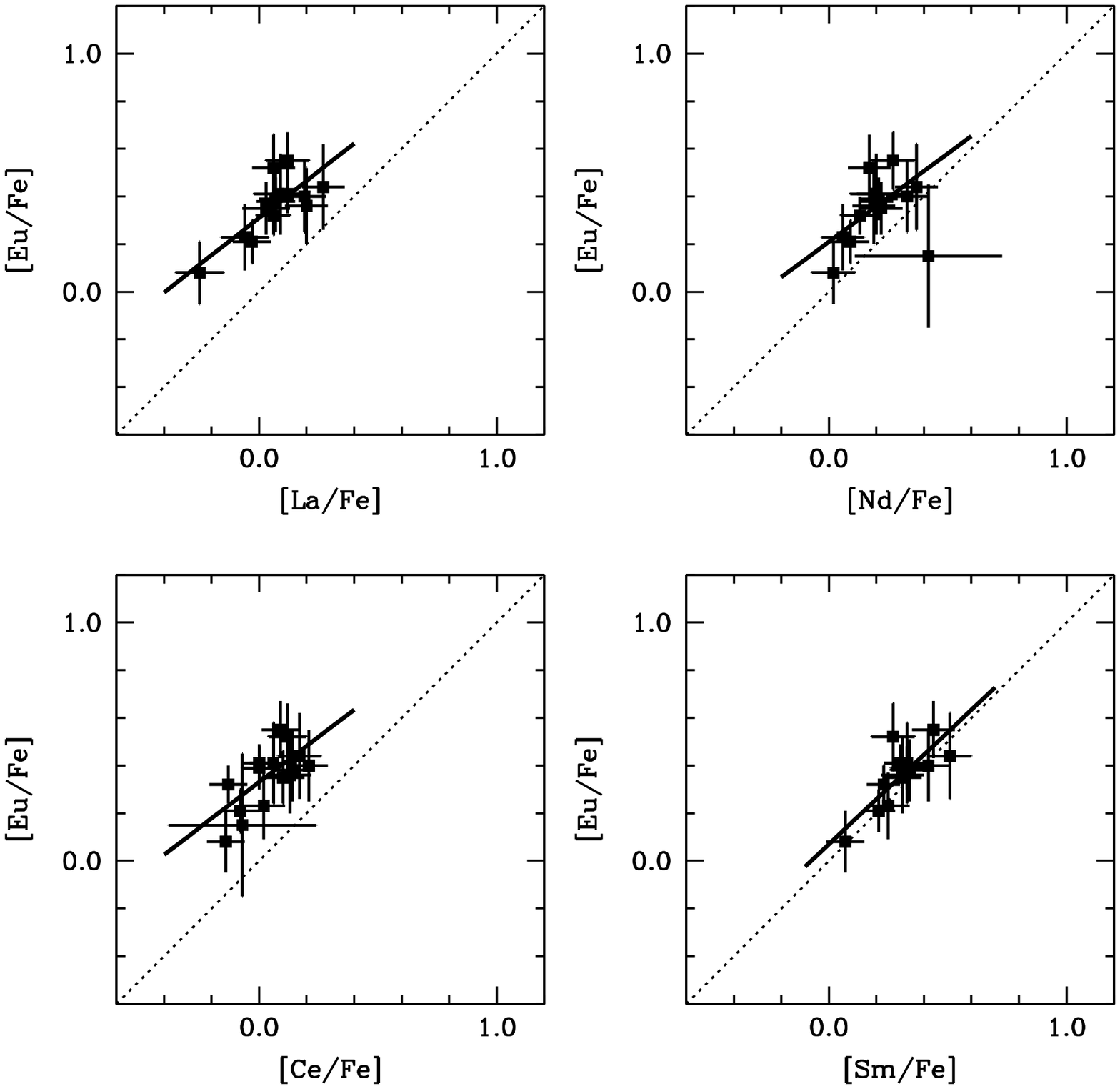} \\
\caption{
\label{rprodispplot}
Correlations among heavy element ratios in \ngc\
for elements measured in 14 or more stars.
The red lines mark the weighted least-squares fits.
Dotted lines mark the solar values.
We have checked, and confirmed, the weighted fit for [Ba/Fe] versus [Eu/Fe],
which appears to deviate from what might be expected from a by-eye fit.
The uncertainty on this fit is considerably larger than for the
other rare earth elements, as can be seen in Figure~\ref{euslopes}.
}
\end{figure*}

We calculate a $p$-value of 0.001
for the relation between
[La/Fe] and [Eu/Fe] in 15~stars in \ngc.
The null hypothesis
that [La/Fe] and [Eu/Fe] are not correlated is rejected
at a $>$~3$\sigma$ significance level.
Figure~\ref{rprodispplot} illustrates the correlations
between [Eu/Fe] and eight heavy-element abundance ratios in \ngc.
These eight pairs of ratios are selected
because they have been measured in 14 or 15
of the 15~stars studied.
The correlations bear strong resemblance to those 
identified by \citet{roederer11a}.
Of these eight ratios, only [Sr/Fe], [Ba/Fe], and [Nd/Fe]
do not correlate with [Eu/Fe] at a $>$~2$\sigma$ level.

\begin{figure}
\centering
\includegraphics[angle=0,width=3.3in]{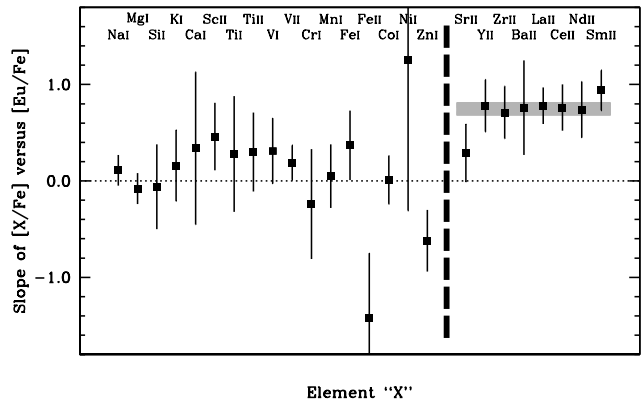} \\
\caption{
\label{euslopes}
Slopes in the [X/Fe] versus [Eu/Fe] relationships,
where ``X'' represents a given species.
Recall that all ratios are constructed using
ions with ions (e.g., [La~\textsc{ii}/Fe~\textsc{ii}])
and neutrals with neutrals (e.g., [Na~\textsc{i}/Fe~\textsc{i}].
The vertical dashed line separates the light from the heavy elements.
The shaded region represents the mean $\pm$~1$\sigma$ of the slopes
of the eight heavy elements shown to the right of the vertical dashed line.
The horizontal dotted line marks a slope of zero.
}
\end{figure}

Weighted linear least-squares fits to the data are also shown
in each panel in Figure~\ref{rprodispplot}.
A slope of $+$1 corresponds to a direct correlation,
$-$1 corresponds to a direct anti-correlation,
and 0 corresponds to no correlation.
The values of the slopes and 1$\sigma$ uncertainties
are illustrated in Figure~\ref{euslopes}.
The slopes 
are remarkably consistent and non-zero, and the mean
slope (0.75~$\pm$~0.07, $\sigma =$~0.19) is
illustrated by the shaded box in Figure~\ref{euslopes}.
Only the slope between [Eu/Fe] and [Sr/Fe] 
is not strongly positive
among the elements produced by \ncap\ reactions.

Slopes of the relations between [Eu/Fe] and lighter element ratios
($Z \leq$~30)
are shown in Figure~\ref{euslopes} for comparison.
Most of these slopes are consistent with zero,
in contrast to the slopes between [Eu/Fe] and other
elements produced by \ncap\ reactions.
There is no coherent, significant pattern among them.
This suggests that none of the lighter elements
correlate with Eu or other \ncap\ elements.

\begin{figure*}
\centering
\includegraphics[angle=0,width=3.4in]{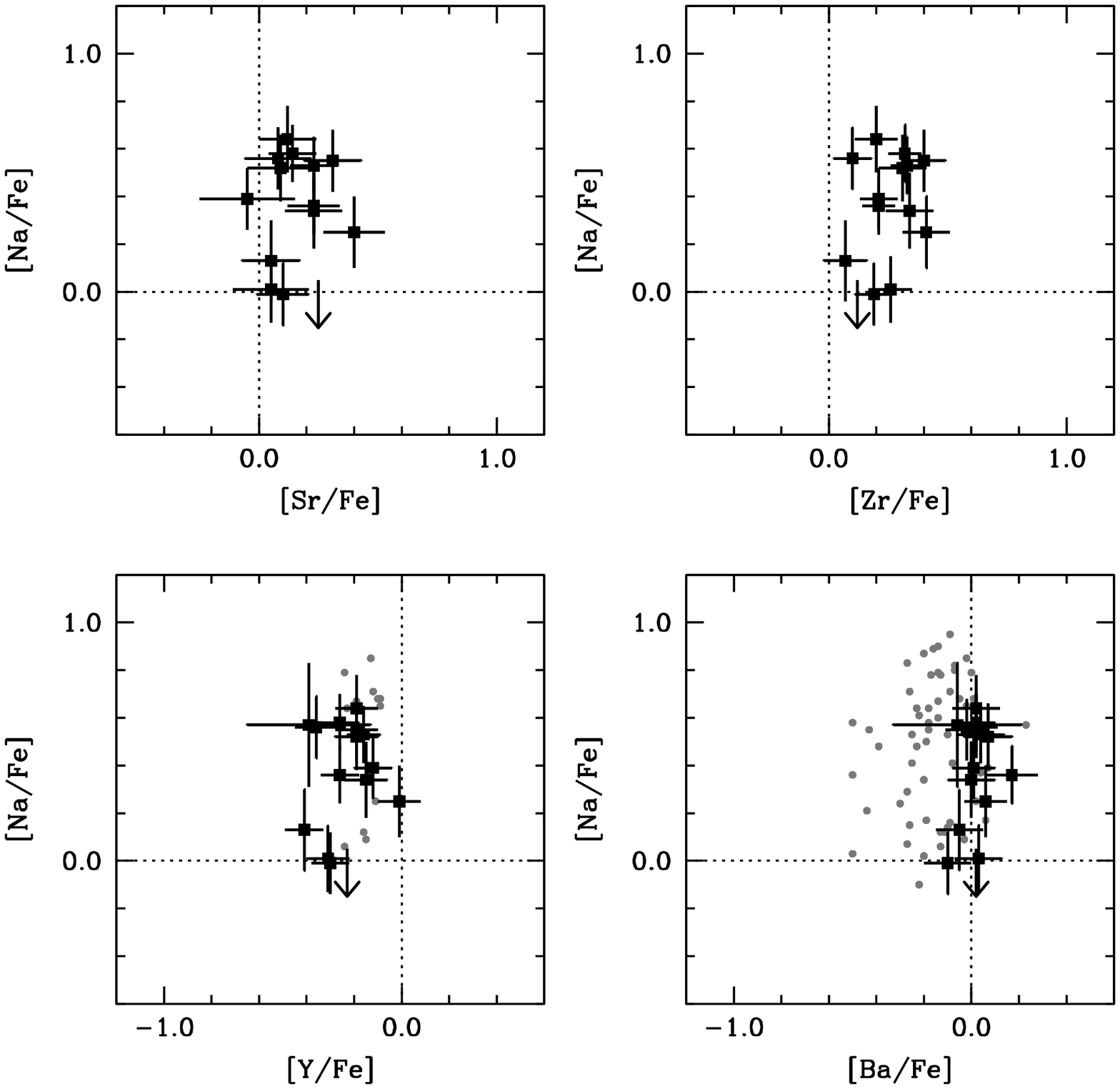}
\hspace*{0.0in}
\includegraphics[angle=0,width=3.4in]{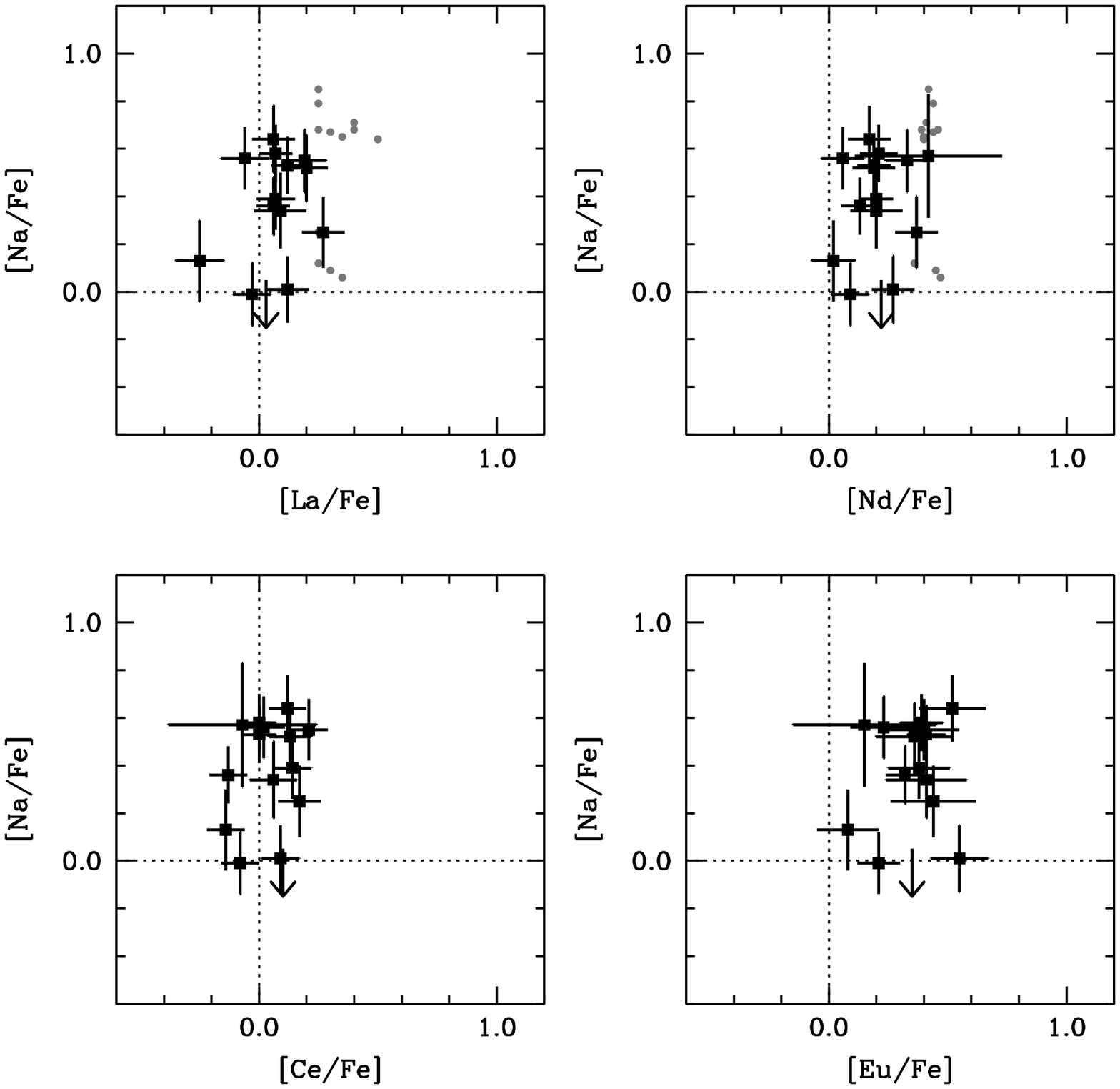} \\
\caption{
\label{naheavyplot}
Relationships among heavy elements and [Na/Fe] in \ngc.
Black points mark our derived abundance ratios, and
gray circles mark those from \citet{carretta14}.
Dotted lines mark the solar values.
No correlations are apparent.
}
\end{figure*}

Figure~\ref{naheavyplot} illustrates 
[Na/Fe] and eight of the heavy-element
ratios shown in Figure~\ref{rprodispplot}.
No correlations are apparent, and
the $p$-values indicate that
none of these correlations (including [Na/Fe] versus [Sm/Fe], not shown)
is even 1$\sigma$ significant.
The heavy element dispersion
is not related to the light element dispersion.

We have shown that the abundances of
the \ncap\ elements in \ngc\ 
correlate closely with one another
but not the lighter elements.
Next, we examine whether these correlations 
reflect true abundance dispersions or 
are an artifact of our analysis.
Atoms in the same state (e.g., ions)
would respond similarly to an inappropriate model atmosphere,
and the resulting abundance distributions would 
be smeared out to lower and higher ratios
in a correlated fashion.
Table~\ref{atmvar} lists the 1$\sigma$ uncertainty
in each [X/Fe] ratio
that could be attributed to 
uncertainties in the model atmosphere parameters.
For example, these values
are 0.11 and 0.12~dex
for [La/Fe] and [Eu/Fe].
The standard deviations of the
[La/Fe] and [Eu/Fe] ratios are each 0.13~dex.
The comparable magnitudes of these values
suggest that the correlated heavy element abundance ratios
may not have a cosmic origin.

To demonstrate this point, we
perform the following test.
Using the values presented in Table~\ref{atmvar},
we adjust the \teff\ values for each star
to force the individual [Eu/Fe] ratios 
to equal the mean [Eu/Fe] ratio.
We then apply these \teff\ adjustments
to the individual [La/Fe] ratios to
derive revised [La/Fe] ratios for each star.
The standard deviation of the
revised [La/Fe] ratios is 30~per cent smaller 
than the standard deviation of the
uncorrected ratios.
Other rare earth elements exhibit similar responses.
We conclude from this test that 
random errors in \teff\ 
can account for a substantial portion of
the dispersion,
and we infer that random errors in the
other model atmosphere parameters
may have similar impact.

All of the heavy-element ratios shown to correlate
in Figure~\ref{rprodispplot} are derived from ionized atoms,
so other elements
detected as ions
(Sc, Ti, V, and Cr)
might exhibit similar characteristics.
However, Table~\ref{atmvar} indicates that the 
[Sc~\textsc{ii}/Fe], [Ti~\textsc{ii}/Fe],
[V~\textsc{ii}/Fe], and [Cr~\textsc{ii}/Fe]
should not be expected to behave like, e.g.,
[La~\textsc{ii}/Fe] or [Eu~\textsc{ii}/Fe].
For example,
$\delta$[Ti~\textsc{ii}/Fe]/$\delta$\vt~$= -$0.04,
while 
$\delta$[La~\textsc{ii}/Fe]/$\delta$\vt~$= +$0.05.
Responses to $\delta$\teff\ and $\delta$\logg\ 
are also dissimilar.
Even though both ratios are derived from
lines of ionized atoms,
the line strengths and
excitation potentials 
are different.
The excitation potentials of Sc~\textsc{ii}, Ti~\textsc{ii},
V~\textsc{ii}, and Cr~\textsc{ii}
lines used range from 1.08 to 4.07~eV with a median of 1.50~eV,
while the excitation potentials of the
heavy rare earth elements' lines range from
0.00 to 1.38~eV with a median of 0.32~eV.
This difference is significant.
The
[Sc~\textsc{ii}/Fe], [Ti~\textsc{ii}/Fe],
[V~\textsc{ii}/Fe], and [Cr~\textsc{ii}/Fe]
ratios are not a good control group for 
[La~\textsc{ii}/Fe] or [Eu~\textsc{ii}/Fe],
as was assumed by \citet{roederer11a}.

We conclude that there is no compelling evidence of
a cosmic origin for the
correlations among pairs of \ncap-element ratios
in \ngc.
The detected correlations likely result from
random errors in the model atmosphere parameters.
We suggest that the correlations identified among
literature data by \citet{roederer11a}
may also be the result of similar random errors
in the model atmosphere parameters.
The \rpro\ dispersion in M15, however, is too large
($>$~0.7~dex)
to be explained entirely by this phenomenon,
and we note that \citet{tsujimoto14} have
recently offered a theoretical explanation of its origin.

\subsection{The $^{232}$Th nuclear chronometer in \ngc}
\label{age}

We derive upper limits on the Th abundance
based on the non-detection of the
Th~\textsc{ii} line at 4019~\AA.~
This line originates from radioactive $^{232}$Th,
which can only be produced via \rpro\ nucleosynthesis.
This upper limit sets
a lower limit on the
age of the \rpro\ material in \ngc.
We compare the Th abundance to Eu, a well-measured, stable
element presumably produced by the same nucleosynthesis channel.
The lowest Th/Eu ratio found in \ngc,
in star \mbox{2-1664}, is
$\log \epsilon$(Th/Eu)~$< -$0.47~$\pm$~0.09.
If we assume the Th and Eu were produced 
in the ratio reported in Table~9 of
\citet{roederer09},
this implies an age greater than 6.4~Gyr.
This value is not very constraining, but 
it implies that the \rpro\ material in \ngc\
is not young.

Alternatively, we could
assume that \ngc\ is old
\citep{melbourne00,marinfranch09}
and treat the initial Th/Eu production ratio as a free parameter.
An assumed age of 12~$\pm$~1.5~Gyr implies an initial
production ratio of 
$\log \epsilon$(Th/Eu)~$= -$0.21~$\pm$~0.09.
This implies that the \rpro\ material
was not produced in an ``actinide boost'' event
(cf.\ \citealt{hill02,schatz02}),
which would require an initial production ratio of
$\log \epsilon$(Th/Eu)~$= +$0.11~$\pm$~0.07
to yield an age consistent with the
prototype actinide boost halo star \mbox{CS~31082--001}.

\begin{figure}
\centering
\includegraphics[angle=0,width=3.1in]{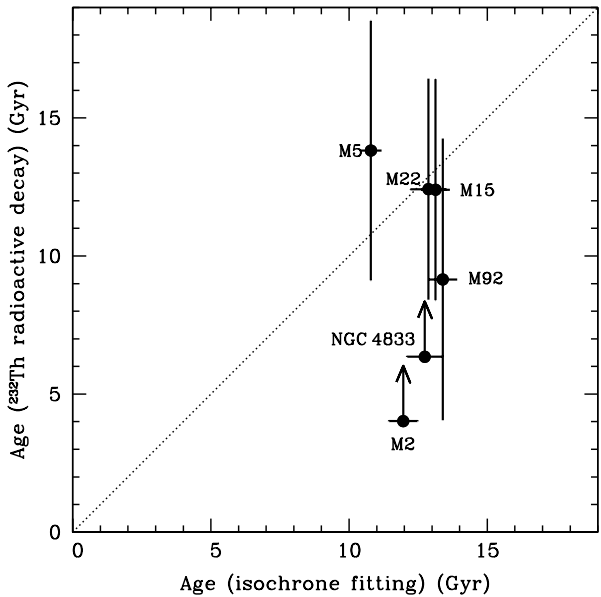}
\caption{
\label{ageplot}
Comparison of globular cluster ages derived from 
isochrone fitting and
radioactive decay of $^{232}$Th.
The relative ages presented by \citet{marinfranch09}
are based on stellar evolution models by \citet{dotter07}
and have been normalized to an absolute age of 13.0~Gyr.
The dotted line marks the 1:1 correspondence.
}
\end{figure}

To the best of our knowledge, no
actinide boost has yet been observed
in any globular cluster star.
$^{232}$Th has been detected in globular clusters
M5 \citep{lai11},
M15 \citep{sneden00},
M22 (Roederer et al.\ \citeyear{roederer11c}), and
M92 \citep{johnson01}.
We also report an unpublished upper limit on $^{232}$Th in M2,
$\log \epsilon$(Th/Eu)~$< -$0.42,
based on spectra obtained by I.U.R.\ 
and published in \citet{yong14}.
Ages calculated from the
$\log \epsilon$(Th/Eu) ratios are consistent with
ages deduced
by isochrone fitting \citep{marinfranch09},
as shown in Figure~\ref{ageplot}.\footnote{
\citet{yong08a,yong08b} detected Th in M4 and M5,
but the high [Th/Fe] ratio in M5 derived by \citeauthor{yong08b}
was not confirmed by \citet{lai11}.
We refrain from drawing any conclusions
from the [Th/Fe] ratios presented by \citeauthor{yong08b}} 
The uncertainties are substantial, but
we would expect to find values in the lower right corner 
of Figure~\ref{ageplot}
if an actinide boost is present in any of these six clusters
at the level found in \mbox{CS~31082--001}.
No detections are found here.

The astrophysical site(s) responsible for
\rpro\ nucleosynthesis are still debated.
Non-detection of the actinide boost
in globular clusters may provide
a new environmental test 
of \rpro\ nucleosynthesis models.
The sample of clusters where $^{232}$Th
has been studied is still small, however,
so new data are of great importance.

\section{Conclusions}
\label{conclusions}

We have collected new high-resolution spectroscopic
observations of 15~giants in the metal-poor globular cluster
\ngc.  
We derive stellar parameters for these stars
and examine up to 
354~lines of
48~species of
44~elements in each star.
We detect 43~species of 39~elements
and present upper limits derived from the non-detections
of the others.
Overall, the composition of stars in 
\ngc\ appears relatively normal
for a metal-poor globular cluster.

We derive a mean metallicity of 
$\langle$[Fe/H]$\rangle = -$2.25~$\pm$~0.02 ($\sigma =$~0.09)
from Fe~\textsc{i} lines and 
$\langle$[Fe/H]$\rangle = -$2.19~$\pm$~0.013 ($\sigma =$~0.05)
from Fe~\textsc{ii} lines.
These uncertainties support the results of
\citet{carretta14} that there is no
internal Fe dispersion within \ngc.
Our derived mean metallicity is 
0.20~$\pm$~0.02~dex lower than that derived by \citeauthor{carretta14}\
for four stars in common.
We derive [Fe/H]~$= -$2.19~$\pm$~0.04 for \ngc\
on the differential globular cluster scale
established by \citet{koch08,koch11}
relative to the K-giant Arcturus,
which supports our lower metallicity.

Our data reveal the abundance variations
among [O/Fe], [Na/Fe], [Mg/Fe], [Al/Fe], and possibly [Si/Fe]
commonly found in globular clusters.
There are bi-modal distributions among the
[Na/Fe], [Mg/Fe], and [Al/Fe] ratios.
Our results reaffirm those of \citet{carretta14}.
We also reproduce a hint of an anti-correlation
between [Si/Fe] and [Mg/Fe],
but our data do not reveal statistically-significant
correlations between [Si/Fe] and either [Na/Fe] or [Al/Fe].
We suspect the discrepancy between our data and 
\citeauthor{carretta14}\ is a result of the smaller
sample size examined by us.

Quantitative measures of potential 
anti-correlations between
[K/Fe] and [Mg/Fe] and between
[Sc/Fe] and [Mg/Fe]
reveal that neither is significant.
Furthermore, 
neither [K/Fe] nor [Sc/Fe]
correlate with
[Na/Fe], [Al/Fe], or [Si/Fe]
at the 2$\sigma$ level.
Our data demonstrate that
\ngc\ does not possess 
the extreme relations among [Mg/Fe], [K/Fe], and [Sc/Fe]
found in the globular clusters \mbox{NGC~2419} and \mbox{NGC~2808}.
We find no dispersion among any of the other iron-group elements
in \ngc.

We detect up to 20~\ncap\ elements in \ngc,
and we place upper limits on four others.
Abundances of the rare earth elements 
are consistent with \rpro\ nucleosynthesis.
Sr, Y, Zr, Mo, and Ba
are enhanced relative to the scaled solar \rpro\
distribution, 
but low levels of Yb, Hf, and Pb 
suggest that this is not due to enrichment by the \spro.
Instead, the weak and main 
components of the \rpro\ may be responsible.
The heavy-element abundance distribution in \ngc\
closely resembles that found in M5, M15, M92, 
\mbox{NGC~2419}, and the \rpro-only stellar groups in M2 and M22.
There is no correlation between the \ncap\ elements
and the light-element variations in \ngc,
which reaffirms and expands upon results obtained by \citet{carretta14}.
Our analysis of correlations between [La/Fe], [Eu/Fe], and other
\ncap\ elements in \ngc\
reveals that they are likely due to random
errors in the stellar parameters,
and we conclude that there is no compelling evidence for
cosmic dispersion among the heavy elements in \ngc.
This cluster does not possess wide star-to-star variations
like those observed in 
$\omega$~Cen, M2, M15, M22, or \mbox{NGC~1851}.

Upper limits derived from the non-detection of the
Th~\textsc{ii} line at 4019~\AA\ 
indicate that the \rpro\ material in \ngc\
was not formed in an actinide boost event.
No other clusters that have been studied
show evidence of an actinide boost, either.
The potential non-detection of the
actinide boost in globular clusters
presents a new environmental test of
\rpro\ nucleosynthesis models
that should be investigated further.

\section*{Acknowledgments}

I.U.R.\ thanks 
E.\ Carretta for sending equivalent width measurements and
D.\ Yong for advice during the early stages of the analysis.
We appreciate the helpful suggestions and rapid response
of the referee.
This research has made use of NASA's 
Astrophysics Data System bibliographic services, 
the arXiv preprint server operated by Cornell University, 
the SIMBAD and VizieR databases hosted by the
Strasbourg Astronomical Data Center,  
the Atomic Spectra Database \citep{kramida13} hosted by
the National Institute of Standards and Technology,
and the Two Micron All Sky Survey, which is a joint project of the 
University of Massachusetts and the Infrared Processing and Analysis 
Center/California Institute of Technology, 
funded by the National Aeronautics and Space Administration 
and the National Science Foundation.
\textsc{iraf} is distributed by the National Optical Astronomy Observatories,
which are operated by the Association of Universities for Research
in Astronomy, Inc., under cooperative agreement with the National
Science Foundation.

\label{lastpage}

\end{document}